\documentclass[final,5p,twocolumn]{elsarticle}

\usepackage{amssymb}
\usepackage{amsmath}
\usepackage[caption=false]{subfig}
\usepackage{multirow}
\usepackage{algorithm}
\usepackage{enumerate}
\usepackage{algpseudocode}
\usepackage{float}
\usepackage[colorlinks=true,linkcolor=black, citecolor=blue, urlcolor=blue]{hyperref}
\usepackage{tikz,pgfplots}
\usepackage{breqn}
\usepackage{bbm}
\usepackage{cleveref}
\usepackage{mathtools}
\usepackage{balance}
\pgfplotsset{%
    compat=newest, 
    tick label style={font=\footnotesize},
    label style={font=\small},
    legend style={font=\small},
    axis x line = center,
    axis y line = center,
    every axis/.style={pin distance=1ex},
    trim axis left
    } 

\newcommand{\iu}{\mathrm{i}\mkern1mu}
\DeclareMathAlphabet\mathbfcal{OMS}{cmsy}{b}{n}

\makeatletter
\def\ps@pprintTitle{%
  \let\@oddhead\@empty
  \let\@evenhead\@empty
  \def\@oddfoot{\reset@font\hfil\thepage\hfil}
  \let\@evenfoot\@oddfoot
}
\makeatother

\journal{ }

\begin{document}

\begin{frontmatter}

\title{Malliavin-Mancino estimators implemented with non-uniform fast Fourier transforms}

\author[uct-sta]{Patrick Chang}
\ead{chnpat005@myuct.ac.za}
\author[uct-sta]{Etienne Pienaar}
\ead{etienne.pienaar@uct.ac.za}
\author[uct-sta]{Tim Gebbie}
\ead{tim.gebbie@uct.ac.za}
\address[uct-sta]{Department of Statistical Sciences, University of Cape Town, Rondebosch 7700, South Africa}

\begin{abstract}
    We implement and test kernel averaging Non-Uniform Fast Fourier Transform (NUFFT) methods to enhance the performance of correlation and covariance estimation on asynchronously sampled event-data using the Malliavin-Mancino Fourier estimator. The methods are benchmarked for Dirichlet and Fej\'{e}r Fourier basis kernels. We consider test cases formed from Geometric Brownian motions to replicate synchronous and asynchronous data for benchmarking purposes. We consider three standard averaging kernels to convolve the event-data for synchronisation via over-sampling for use with the Fast Fourier Transform (FFT): the Gaussian kernel, the Kaiser-Bessel kernel, and the exponential of semi-circle kernel. First, this allows us to demonstrate the performance of the estimator with different combinations of basis kernels and averaging kernels. Second, we investigate and compare the impact of the averaging scales explicit in each averaging kernel and its relationship between the time-scale averaging implicit in the Malliavin-Mancino estimator. Third, we demonstrate the relationship between time-scale averaging based on the number of Fourier coefficients used in the estimator to a theoretical model of the Epps effect. We briefly demonstrate the methods on Trade-and-Quote (TAQ) data from the Johannesburg Stock Exchange to make an initial visualisation of the correlation dynamics for various time-scales under market microstructure.
\end{abstract}


\end{frontmatter}


\section{Introduction} \label{sec:intro}

Data-informed approaches to modelling the relationships between fast asynchronous streaming event-data features requires efficient algorithms to compute the dependency or similarity across data features. This can be useful to relate collections of similar features to similar but potentially useful information on the appropriate decision time-scale. When the dependency structure can be approximated by an averaged realised correlation or covariance matrix then the problem of estimation from asynchronous event data can be significantly simplified. Then the problem of correlation and covariance estimation over asynchronous event data can be addressed using the Malliavin-Mancino estimator \cite{MM2002,MM2009,MRS2017}.

This has several advantages over {\it ad-hoc} averaging and interpolation methods built on the underlying assumptions of continuity, such as the approach taken in the well understood Hayashi-Yoshida estimator \cite{PCRBTG2019}. However, the Malliavin-Mancino estimator is built on numerically evaluating Fourier transforms and their inverses. This has a computational cost. Quickly extracting realised correlations or covariances on a given time-scale for large feature sets of distinct asynchronous events without biased interpolation is key to avoiding spurious correlations that can lead to ineffective decision making under uncertainty. 

This paper directly addresses two key issues: First, that of performance as measured by computational speed. Second, the implicit dependence of time-scale in the estimation of realised covariances and correlations on asynchronous event data using the Malliavin-Mancino estimator. The key contribution is to mitigate the first problem using non-uniform fast Fourier transforms to compute the Fourier coefficients in the Malliavin-Mancino estimator, and to provide clarity into the second idea using insights from the non-uniform fast Fourier transform.

Performance is a key requirement in two related use cases, namely simulation and real-time estimation. Being able to carry out large scale Monte-Carlo simulations over many features and many time-scales where one needs to iterate and recompute the correlation matrix over event data. In a real-time environment where decisions are being made on streaming event-data, the use of fast methods can reduce the time-scales of effective data-sampling. For example, the minimum effective sampling rate of correlation based state detection is bounded by the compute time of the correlation matrix. A speed improvement on the compute time of the realised covariance, or realised correlation matrix potentially allows more time for learning algorithm convergence and identification. This can be of particular importance for learning algorithms that require many updates to identify a reliable optimal relationship between actions and system states given an objective, such as Q-learning based implementations of reinforcement learning for trading \cite{DH2017PRL,HGW2016QF,HW2014ARL}.

Concretely, we extend an approach to performance enhancement based on the fast Fourier transform \cite{CT1965} in the context of the Malliavin-Mancino estimator \cite{MM2002,MM2009} by using non-uniform fast Fourier transform methods \cite{BMK2018,DR1993,GL2004}. This combines the performance advantage of fast Fourier transforms while providing intuition into the time-scale averaging. This follows from the basic idea behind the non-uniform fast Fourier transform: convolving the data onto a uniform grid (dependent on the number of Fourier coefficients required) through a choice of averaging kernel. Furthermore, the averaging kernel has an explicit averaging scale which provides avenues for controlling speed and accuracy.

We hope to follow Ren\`{o} \cite{RENO2001} and Precup and Iori \cite{PI2007} by using the choice of the number of Fourier coefficients $N$ as the method of tuning the estimation to different time-scales. To implement this with confidence using NUFFT methods, we need to understand the relative dependencies between kernel averaging (proxied by the tolerances) and time-scale averaging (proxied by the number of Fourier coefficients) under simulation to evaluate their impact on the estimated correlations. Moreover, we need to ensure the NUFFT estimates recover the same estimates as the original implementation.

To explore this idea we consider three different averaging kernels: (i) the Gaussian kernel \cite{GL2004} (see \cref{eq:Der:5}), (ii) the Kaiser-Bessel kernel \cite{PS2003} (see \cref{eq:Der:8,eq:Der:9}), and (iii) the exponential of semi-circle kernel \cite{BMK2018} (see \cref{eq:Der:10,eq:Der:11}). In conjunction with these choices of averaging kernels, we consider two different choices of Fourier basis kernels: (i) the Dirichlet, and (ii) the Fej\'{e}r basis kernels. Combinations of these are compared with different length and breadth data-sets and for different numbers of Fourier coefficients. This allows us to better understand the relative algorithm performance by comparing algorithm compute times with data-size and various tolerance levels (see \Cref{fig:ONplots,fig:Errorplots}). 

These combinations of kernel choices are benchmarked against three vanilla algorithms that implement the Malliavin-Mancino estimator: (i) the benchmark ``for-loop'' implementation first provided by Mancino, Recchioni and Sanfelici \cite{MRS2017}, (ii) a vectorised implementation with speed enhancements assuming real-valued data \cite{HGW2016QF,MALHERBE2007,TGDWCMDH2005}, and (iii) a zero-padded Fast Fourier implementation \cite{DHTGDW2017,TGDWCMDH2005} that allows the use of the fast Fourier transform on asynchronous data without the need to apply an averaging kernel, but using an underlying missing data approach to implement lossless interpolation. 

Here an important observation is that using the zero-padded FFT to compute the Malliavin-Mancino estimator can only work for uniformly sampled data that has missing data points and fails for truly asynchronous data (see \Cref{fig:AccSynDS} and \Cref{subsubsec:FFTZP}). This is the key motivation for the necessary requirement of using a non-uniform FFT in the setting of speeding up the compute time of the Malliavin-Mancino estimator using the fast Fourier transform method for asynchronous event data. The zero-padded FFT biases the data, while the non-uniform FFT does not if correctly used. It is for this reason that we promote the idea of using the NUFFT in conjunction with the Malliavin-Mancino estimator if the data is asynchronous, discrete and event driven. 

The paper is organised as follows: \Cref{sec:algo} we outline the various implementation methods for the Malliavin-Mancino estimator. \Cref{sec:benchmarking} we benchmark the various algorithms to understand the factors impacting speed. Moreover, we determine the conditions required for the NUFFT implementation to recover the correct estimates. \Cref{sec:scale} we demonstrate the link between the number of Fourier coefficients and the implicit time-scale investigated along with its relation to a theoretical model of the Epps effect. We then carry-out EDA on real world TAQ data to investigate the correlation dynamics under market microstructure. We finally conclude in \Cref{sec:conclude} to summarise our findings. 

\section{Algorithm Outline} \label{sec:algo}
\subsection{Malliavin-Mancino estimators} \label{subsec:MM}

Malliavin and Mancino \cite{MM2002, MM2009} proposed an estimator that is constructed in the frequency domain. It expresses the Fourier coefficients of the volatility process using the Fourier coefficients of the price process $p_i(t) = \ln(S_i(t))$, where $S_i(t)$ is the generic asset price at time $t$. By re-scaling the trading times from $[0, T]$ to $[0, 2\pi]$ (see \cref{algo:rescale}) and using the \textit{Bohr} convolution product (see Theorem 2.1 of \cite{MM2009}) we have that for all $k \in \mathbb{Z}$ and $N$ samples:
\begin{equation} \label{eq:Der:1}
    \begin{aligned}
      \mathcal{F}(\Sigma^{ij})(k) = \lim_{N \rightarrow \infty} \frac{2 \pi}{2N+1} \sum_{|s| \leq N} \mathcal{F}(dp_i)(s) \mathcal{F}(dp_j)(k-s).
    \end{aligned}
\end{equation}

Here $\mathcal{F}(\ast)(\star)$ is the $\star^{\text{th}}$ Fourier coefficient of the $\ast$ process.
Now using previous tick interpolation to avoid a downward bias in the estimator \cite{BR2002} and a simple function approximation for the Fourier coefficients (see \cite{PCRBTG2019,MALHERBE2007,MM2009}), we obtain the Dirichlet representation of the integrated volatility/co-volatility estimator:\footnote{We try follow the notation of \cite{MM2002,MM2009,MRS2017,MS2011} where $\iu \in \mathbb{C}$ in the exponential defining the Fourier transform is such that $\operatorname{Re}(\iu)=0$ and $\operatorname{Im}(\iu)=1$. It should not be confused with integer indices $i$, for example on the times $t^i_h$.}

\begin{equation} \label{eq:Der:2}
    \hat{\Sigma}^{ij}_{n,N} = \frac{1}{2N+1} \sum_{\substack{|s|\leq N \\ h=1,\ell=1}}^{n_i-1,n_j-1} e^{\iu s(t^j_{\ell} - t^i_h)} \delta_{i}(I_h) \delta_{j}(I_{\ell}),
\end{equation}
where $(t^i_h)_{h=1,...,n_i}$ and $(t^j_{\ell})_{\ell=1,...,n_j}$ are the observation times for asset $i$ and $j$, and the price fluctuations are:
$$
\delta_{i}(I_h) = p_i(t_{h+1}^i) - p_i(t_{h}^i), \quad \delta_{j}(I_{\ell}) = p_j(t_{\ell+1}^j) - p_j(t_{\ell}^j),
$$
for the $i^{\text{th}}$ and $j^{\text{th}}$ asset respectively. Note that $n_i$ is the sample dimension for price $p_i$ and $n_j$ that of the price $p_j$, which \textit{a priori} can be different.

An alternate version of the Fourier estimator is the Fej\'{e}r representation:

\begin{equation} \label{eq:Der:3}
  \hat{\Sigma}^{ij}_{n,N} = \frac{1}{N+1}  \sum_{\substack{|s|\leq N \\ h=1,\ell=1}}^{n_i-1,n_j-1} \left( 1 - \frac{|s|}{N} \right) e^{\iu s(t^j_{\ell} - t^i_h)} \delta_{i}(I_h)  \delta_{j}(I_{\ell}),
\end{equation}
which is more stable under the presence of market microstructure noise \cite{MM2009}.

The various implementation methods follow the same general structure (outlined in \cref{algo:MM}). First, we re-scale the trading times from $[0, T]$ to $[0, 2\pi]$ (see \cref{algo:rescale}) and compute the Nyquist frequency\footnote{Mancino et al. \cite{MRS2017} picks $N$ such that MSE is minimised.} (see \cref{algo:nyquist}). Second, we compute the non-normalised Fourier coefficients $\mathcal{F}(dp_i)(k)$ $k \in \{ -N, ..., N\}$ for all assets. Finally, we compute either the Dirichlet or Fej\'{e}r representation of the estimator. The difference between the implementation methods are in the computation of the Fourier coefficients.

\subsection{Implementation methods} \label{subsec:methods}

The computationally intensive step in the Malliavin-Mancino estimator is the computation of the Fourier coefficients for each asset defined as:
\begin{equation}\label{eq:Der:FC}
    \mathcal{F}(dp_i)(k) = \frac{1}{2 \pi} \sum_{h=1}^{n_i - 1} \delta_{i}(I_h) e^{-\iu k t^i_h},
\end{equation}
for $k \in \{-N,...,N\}$, and $i = 1,...,D$ features. We outline the various methods to evaluate \cref{eq:Der:FC} along with the use-case,\footnote{The use-case refers to the ability to evaluate synchronous or asynchronous time-series data.} benefits, pitfalls and general algorithm complexity.\footnote{The complexity is given only for the synchronous case when $n = n_1 = ... = n_D$, $N=\frac{n}{2}$, and $t^i_{h+1} - t^i_h = \Delta t$, $\forall i = 1,...,D$ and $\forall h = 1, ..., n-1$.}

\subsubsection{Benchmark for-loop implementation} \label{subsubsec:MS}

The Mancino et al. implementation (see \cref{algo:MS}) is from the appendix of \cite{MRS2017} and uses a for-loop construction. The evaluation relies on looping through $\{-N,..., N\}$ to compute the $k^{\text{th}}$ Fourier mode. The implementation does not rely on any techniques to improve performance and will act as a benchmark to compare against other methods. The method can be used for all synchronous and asynchronous cases. The complexity is $O(n^2)$, the same as Discrete Fourier Transforms (DFTs).


\subsubsection{Vectorised implementation} \label{subsubsec:legacy}

The legacy code implementation (see \cref{algo:legacy}) is based on a MATLAB implementation \cite{DHTGDW2017, TGDWCMDH2005}. The difference compared to the Mancino et al. implementation is that all the Fourier modes are evaluated in parallel by vectorising the computation. Here we further improve upon the legacy code by exploiting techniques found in \citet{MRS2017}. Concretely, we exploit the Hermitian symmetry $\mathcal{F}(dp_i)(k) = \overline{\mathcal{F}(dp_i)(-k)}$ where $\overline{\mathcal{F}}$ denotes the conjugate function of $\mathcal{F}$. This is possible because the source strengths $\delta_{i}(I_h)$ are all real-valued. Therefore, we only need to evaluate $k \in \{1, ..., N\}$ and obtain the conjugates for these Fourier modes. Finally, $\mathcal{F}(dp_i)(0) = \sum_{h=1}^{n_i-1} \delta_{i}(I_h) / 2 \pi$ must be computed to complete the range of Fourier modes required for the convolution. The method can be used for all synchronous and asynchronous cases with a complexity of $O(n^2)$. The key concern with this method is the memory usage constraints that it can face. After inspecting \cref{algo:legacy} we see that a large matrix of size $(n \times N)$ is required for the vectorisation which can adversely affect performance by either: (i) pre-maturely ending the computation due to insufficient memory or heap-size constraints, or (ii) slow down performance due to an over-reliance on virtual-memory management. Therefore, memory management is crucial for effective performance enhancements of large data-sets.

\subsubsection{The fast Fourier transform} \label{subsubsec:FFT}

The FFT implementation used here is the current state-of-the-art FFTW package \cite{FFTW05} based on the Cooley-Tukey algorithm \cite{CT1965} to compute the Fourier modes. This implementation also exploits the Hermitian symmetry, making this the fastest implementation known to the authors. This implementation has a well understood complexity of $O(n \log n)$. The key constraint of this method is its restriction to strictly synchronous data, so the evaluation becomes a simple DFT.

\subsubsection{The zero-padded fast Fourier transform} \label{subsubsec:FFTZP}

The Zero-padded FFT (ZFFT) implementation extends the FFT implementation by zero padding missing observations. Therefore, allowing the computation of the asynchronous case under a missing data representation (see \cref{algo:ZPFFT}). The implementation computes the minimum sampling interval $\Delta t$ and creates a new over-sampled grid with intervals $\Delta t$. The observations are then placed at the nearest neighbour of the over-sampled grid.\footnote{It is also recommended that this be implement to preserve the filtration structure of time-series events by moving to the nearest right neighbour so that information in the future is shifted, at worst, further into the future, but never into the past relative to a particular time to avoid temporal contamination.} The FFT algorithm is then applied to the new over-sampled grid. The implementation retains a complexity of $O(n \log n)$ but is slower than the FFT implementation since it does not exploit the Hermitian symmetry and requires the additional step of creating an over-sampled grid. This is our benchmark asynchronous approach to the fast Fourier transform. 

\begin{figure}[h!]
    \centering
    \includegraphics[width=0.5\textwidth]{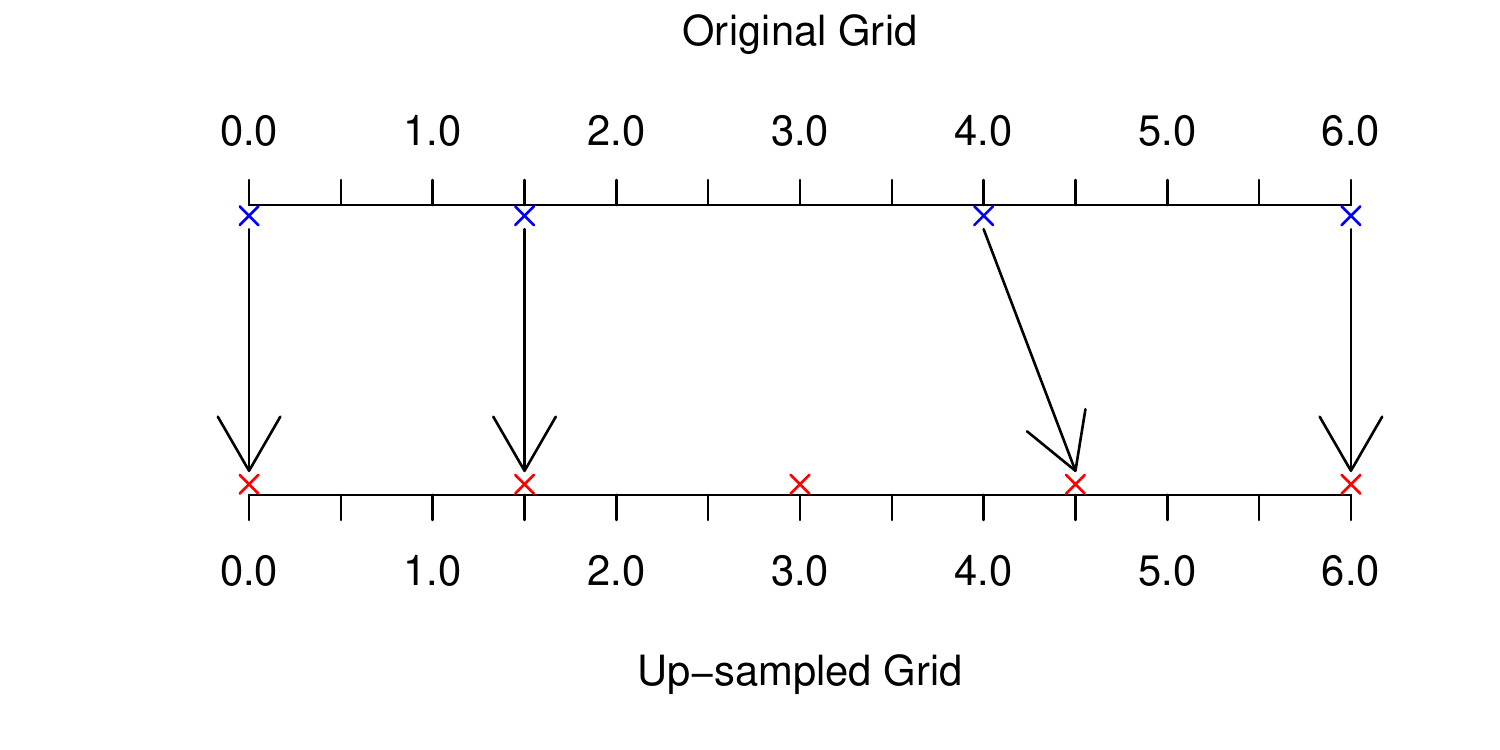}
    \caption{A toy example to show how the zero-padding works for the zero-padded FFT implementation. Here the minimum sampling interval is $\Delta t = 1.5$. A new uniform over-sampled grid is created and observations are placed on the nearest neighbouring point of the over-sampled grid.}
    \label{fig:FFTZP_demontration}
\end{figure}

\Cref{fig:FFTZP_demontration} demonstrates two points: (i) how the zero-padded implementation works, and (ii) why the implementation does not work for the asynchronous case using an arrival time representation. The original grid has equal spacing $\Delta t$ when asynchrony is induced using a missing data representation. Therefore, the over-sampled grid will be at the same time points as the original grid with a value of zero when there is a missing observation. Meaning that there is no shifting of time points, allowing the correct recovery of \cref{eq:Der:FC}. However, the original grid does not have equal spacing $\Delta t$ when asynchrony is induced using an arrival time representation. Resulting in the time points being shifted (seen in the third arrow from the left in \Cref{fig:FFTZP_demontration}) and the incorrect recovery of \cref{eq:Der:FC}.

\subsubsection{Non-uniform fast Fourier transform} \label{subsubsec:NUFFT}

The Non-Uniform FFT (NUFFT) implementation of the Malliavin-Mancino estimator is the main contribution of this paper. We want a fast algorithm to evaluate \cref{eq:Der:FC} when $(t^i_h)_{h=1,...,n_i}$ are non-uniformly spaced in $[0, 2\pi]$. This can be achieved by using the 1-dimensional ``type 1'' NUFFT \cite{BMK2018,GL2004} (also known as the adjoint NUFFT \cite{PS2003}). We adopt the popular NUFFT algorithm \cite{BMK2018,GL2004,PS2003}: (i) convolve the non-uniform source points onto an over-sampled uniform grid, (ii) apply the FFT on the uniform up-sampled grid, and (iii) deconvolve the effects of the convolution in the Fourier space.

The convolution is achieved with a kernel $\varphi (x)$.\footnote{The choice of kernel has a fascinating history and has a significant impact on the speed of NUFFTs. We refer the reader to \cite{BMK2018} for further details.} We consider the three most popular kernels: the Gaussian kernel using the fast Gaussian gridding implementation from \cite{GL2004}, the Kaiser-Bessel kernel using the implementation approach of \cite{PS2003}, and the exponential of semi-circle used by the state-of-the-art FINUFFT package \cite{BMK2018}. 

To set the theoretical scene: let $M = 2N + 1$ be the number of Fourier modes we want returned, $\sigma$ be the over-sampling ratio (most studies have settled on $\sigma = 2$ \cite{BMK2018}), $\xi_\ell$ be the $\ell^{\text{th}}$ location on the over-sampled grid with $\ell \in \{0, ..., M_r-1 = \sigma M-1\}$ and $\omega$ is the spreading width with $M_{sp}$ as the spreading in each direction. 
The Gaussian kernel and its Fourier transform is defined as:
\begin{equation} \label{eq:Der:5} 
  \varphi_{_{G}}(x) = e^{-x^2 / 4 \tau} \quad \mbox{and} \quad \hat{\varphi}_{_{G}}(k) = \sqrt{2 \pi} e^{-k^2 \tau}.
\end{equation}
%
Here $\tau$ is defined as

\begin{equation} \label{eq:Der:7}
  \tau = \frac{1}{M^2} \frac{\pi}{\sigma(\sigma-0.5)} M_{sp}.
\end{equation}
The Kaiser-Bessel pair is defined as:
\begin{equation} \label{eq:Der:8}
     \varphi_{_{KB}}(x) = \frac{1}{\pi}
\begin{cases}
        \frac{\sinh\left(b \sqrt{M_{sp}^2 - M_r^2 x^2} \right)}{\sqrt{M_{sp}^2 - M_r^2 x^2}} & |x| \leq \frac{M_{sp}}{M_r}, \\
      \frac{\sin \left( b \sqrt{M_r^2 x^2 - M_{sp}^2} \right)}{\sqrt{M_r^2 x^2 - M_{sp}^2}} & \text{otherwise},
\end{cases}
\end{equation}
and 
\begin{equation} \label{eq:Der:9}
  \hat{\varphi}_{_{KB}}(k) = \frac{1}{M_r} I_0 \left( m \sqrt{b^2 - \left( 2 \pi k / M_r \right)^2} \right),
\end{equation}
where $b = \pi \left( 2 - \frac{1}{\sigma} \right)$ and $I_0(\cdot)$ is the modified zero-order Bessel function \cite{PS2003}. Finally, the exponential of semicircle pair is defined as:
\begin{equation} \label{eq:Der:10}
  {\phi}_{_{ES}}(x) = 
  \begin{cases}
        e^{\beta \left( \sqrt{1-x^2} - 1 \right)}  & |x| \leq 1, \\
        0 & \text{otherwise},
\end{cases}
\end{equation}
and 
\begin{equation} \label{eq:Der:11}
  \hat{{\phi}}_{_{ES}}(k) = \int_{-\infty}^{\infty} {\phi}_{_{ES}}(x) e^{\iu kx} dx,
\end{equation}
where $\beta = 2.3 \omega$. The kernel is re-scaled to have support between $[-\alpha, \alpha]$ with $\alpha = \pi \omega / M_r$. Thus the re-scaled kernel is then $\varphi_{_{ES}}(x) = {\phi}_{_{ES}}(x/\alpha)$ and $\hat{{\varphi}}_{_{ES}}(k) = \alpha \hat{{\phi}}_{_{ES}}(\alpha k)$.
The exponential of semicircle kernel has no known analytic Fourier transform; therefore, numerical integration is used to obtain $\hat{\varphi}_{_{ES}}(k)$. See \cite{BMK2018} for more details on their implementation. 

We focus our attention on the implementation for the various kernels which have different periodicity. The Gaussian and exponential of semi-circle are $2 \pi$-periodic with domain on $[0, 2 \pi]$ \cite{BMK2018,GL2004}, while the  Kaiser-Bessel kernel is $1$-periodic with domain on $[0, 1]$. Therefore $\xi_{\ell}$, $t^i_h$ $\in [0, 2\pi]$ for the Gaussian and exponential of semi-circle kernel, and $\xi_{\ell}$, $t^i_h$ $\in [0, 1]$\footnote{\citet{PS2003} have domain on $[-\frac{1}{2}, \frac{1}{2}]$, but we change it to $[0, 1]$ for simpler implementation. The actual domain is not important provided the periodicity is correct, this is because all that matters for the convolution is the distances between $t^i_h$ and $\xi_{\ell}$.} for the Kaiser-Bessel kernel. 

Now let $p$ be the periodicity, then its periodisation is
\begin{equation} \label{eq:Der:14}
  \tilde{\varphi}(x) = \sum_{r = -\infty}^\infty \varphi(x - rp).
\end{equation}
Hence the source strength on the over-sampled grid is given by the periodic discrete convolution

\begin{equation} \label{eq:Der:15}
  f_{\varphi,dp_i}(\xi_{\ell}) = \sum_{h=1}^{n_i-1} \delta_{i}(I_h) \tilde{\varphi}(\xi_{\ell} - t^i_h), \quad \text{for} \ \ell = 0, ..., M_r-1.
\end{equation}
The full derivation to obtain \cref{eq:Der:15} can be found in either \cite{BMK2018,GL2004,PS2003}. The second step is to now evaluate the DFT of the over-sampled grid using the standard FFT

\begin{equation} \label{eq:Der:16}
\begin{aligned}
  F_{\varphi}(dp_i)(k) = \sum_{\ell=0}^{M_r-1} f_{\varphi,dp_i}(\xi_{\ell}) e^{-2 \pi \iu k \ell / M_r}, \\ \text{for} \ k = -M_r/2, ..., M_r/2-1.
\end{aligned}
\end{equation}
The final step, as a consequence of the convolution theorem is to correct the effects of the convolution and retain the $M$ central frequencies \cite{BMK2018}

\begin{equation} \label{eq:Der:17}
  F(dp_i)(k) = F_{\varphi}(dp_i)(k) / \hat{\varphi}(k), \quad \text{for} \ k = -N, ..., N.
\end{equation}

The Fourier coefficient $F(dp_i)(k)$ in \cref{eq:Der:17} is the evaluation of $\mathcal{F}(dp_i)(k)$ in \cref{eq:Der:FC} using non-uniform fast Fourier techniques. The level of numerical accuracy between \cref{eq:Der:17} and \cref{eq:Der:FC} can be measured as the relative $\ell^2$-norm in the output vector defined as:
\begin{equation} \label{eq:Der:18}
  \epsilon = \frac{\lVert \boldsymbol{F}(dp_i) - {\mathbfcal{F}}(dp_i) \rVert_2}{\lVert {\mathbfcal{F}}(dp_i) \rVert_2}.
\end{equation}
Moreover, the desired level of accuracy can be controlled by the amount of spreading in each direction $M_{sp}$ (in terms of number of grid points). We found that setting $M_{sp} = \lfloor \frac{-\ln(\epsilon) (\sigma-1/2)}{(\pi (\sigma-1))} + \frac{1}{2} \rfloor$ for the Gaussian kernel, $M_{sp} =  \lfloor \frac{1}{2} (\lceil \log_{10}(\frac{1}{\epsilon}) \rceil + 2) \rfloor$ for the Kaiser-Bessel kernel, and $M_{sp} =  \lfloor \frac{1}{2} (\lceil \log_{10}(\frac{1}{\epsilon}) \rceil + 2) \rfloor + 2$ for the exponential of semi-circle kernel allow us to achieve the desired relative error level.\footnote{We tuned the $M_{sp}$ such that it always strictly achieves the desired error level. We note that our choice of $M_{sp}$ is stricter than that in the literature. Specifically, \cite{BMK2018} set $\omega = \lceil \log_{10}(\frac{1}{\epsilon}) \rceil + 1$. \href{https://github.com/CHNPAT005/PCEPTG-MM-NUFFT/blob/master/Test/NUFFT\%20error}{NUFFT error} is the test script to check that the desired relative $\ell^2$ error is strictly achieved and can be found in the GitHub resource \cite{PCEPTG2020CODE}.} 

\begin{figure}[h!]
    \centering
    \includegraphics[width=0.45\textwidth]{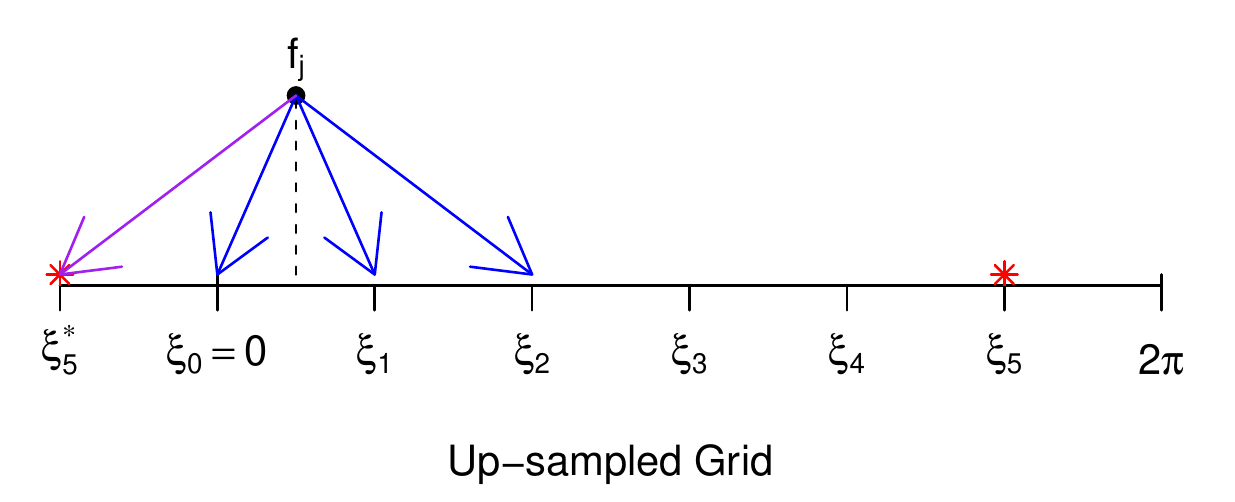}
    \caption{The figure is a toy example to show how the spreading works for a single source point $t^i_h$. The up-sampled grid is denoted by $\xi_{\ell}$, and the source strength is spread to the nearest grid points as $\delta_{i}(I_h) \varphi(\xi_{\ell} - t^i_h)$. The figure aims to show that we only need to evaluate $\varphi(\xi_{\ell} - t^i_h)$ instead of $\tilde{\varphi}(\xi_{\ell} - t^i_h)$. The grid points $\xi_5^*$ and $\xi_5$ (denoted by a red star) is the same point due to the periodicity, but the distance between $\xi_5 - t^i_h$ is large resulting in $\varphi(\xi_5 - t^i_h) \approx 0$. $\tilde{\varphi}(\xi_5 - t^i_h)$ fixes this by accounting for the periodicity. We can reduce the unnecessary computation of \cref{eq:Der:14} by contributing $\delta_{i}(I_h)$ to $f_{\varphi,dp_i}(\xi_5)$ with $\delta_{i}(I_h) \varphi(\xi_5^* - t^i_h)$.}
    \label{fig:NUFFT_spread}
\end{figure}

At first glance, \cref{eq:Der:15} seems a lot more expensive than it actually is. This is based on two observations: first, the kernels in equation \cref{eq:Der:5,eq:Der:8,eq:Der:10} are sharply peaked in a manner such that the contribution of $\delta_{i}(I_h)$ to grid points outside the kernel width is zero (the kernels have small numerical support). Second, the evaluation of $\tilde{\varphi}(\cdot)$ is unnecessary; we only need to evaluate $\varphi(\cdot)$ (see \Cref{fig:NUFFT_spread}). This is because the purpose of $\tilde{\varphi}(\cdot)$ is to account for the periodicity when spreading near the end points of the over-sampled grid. Using these observations, we can efficiently implement \cref{eq:Der:15} by looping through the source points. Find the nearest up-sampled grid point $\xi_{\ell^*}$ that is less than or equal to $t^i_h$. Spread to the $s \in \{-M_{sp},.., M_{sp}\}$ nearest grid points $\xi_{\ell^* - s}$ with $\delta_{i}(I_h) \varphi(t^i_h - \xi_{\ell^*} - \frac{sp}{M_r})$, subject to the condition that when $\ell^*-s<0$ the index becomes $\ell^*-s+M_r$, and when $\ell^*-s \geq M_r$ the index becomes $\ell^*-s-M_r$ to account for the correct indices due to the periodicity.

The method can be used for all synchronous and asynchronous cases and has a complexity of $O\left(M_r \log M_r + n \lvert \log \left( \epsilon \right) \rvert \right)$ \cite{BMK2018}.


\subsection{Insights from NUFFTs}\label{subsec:Avescale}

The use of non-uniform FFT methods presents not only a speed advantage, but provides insights in: (i) the implicit time-scale averaging (controlled by $N$) in the Malliavin-Mancino estimator, and (ii) interpolation of financial data.

First, the Malliavin-Mancino estimator aims to represent the Fourier coefficients of the volatility process as a function of the Fourier coefficients of the price process. Therefore, investigation into different time scales of the volatility process is limited to the sampling rate of the price process. The highest sampling rate present in the data is $N_0$, then the Nyquist frequency is $0.5 N_0 = N$ --- the highest component frequency we can investigate without introducing aliasing. Meaning we are band-limited to frequencies $\leq N$. 

To reconstruct the volatility process at the Nyquist frequency, we require at least $2N$ samples. This condition is satisfied by construction of the \textit{Bohr} convolution product with Fourier modes ranging from $\{-N, ..., N\}$ --- resulting in a sampling frequency $M=2N+1$ samples. 

The relation between the number of Fourier modes and the sampling interval is simply $\frac{T}{M} = \Delta t$. Therefore, we can investigate different time scales by investigating different frequency ranges. This is due to a consequence of the sampling theorem --- which in essence states that in order to perfectly reconstruct a certain frequency, one needs at least twice the amount of samples. By reducing the number of Fourier modes (investigating larger time scales), we are reducing the number of samples and thereby aliasing the larger frequencies. With this in mind, we are able to investigate different time scales by perfectly reconstructing frequencies $< 0.5M$ through the cost of aliasing frequencies $\geq 0.5M$ --- a result used by \cite{PI2007,RENO2001}.

\begin{figure}[h!]
    \centering
    \includegraphics[width=0.45\textwidth]{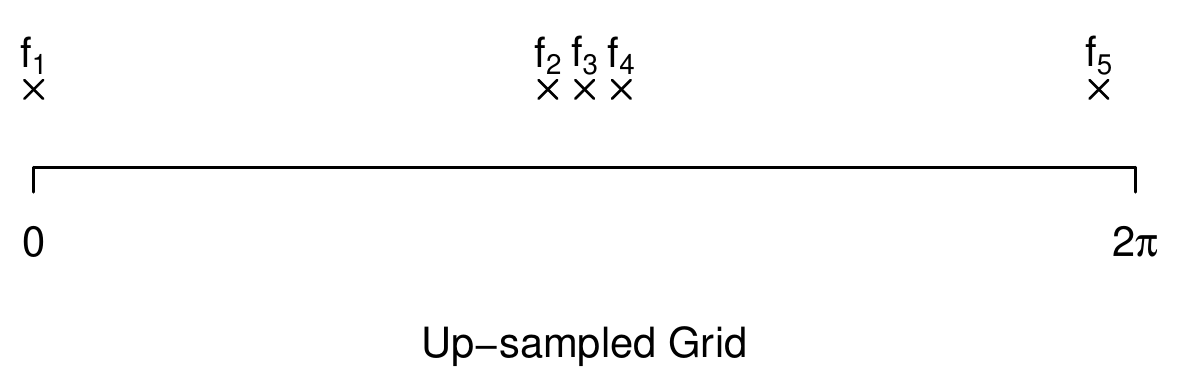}
    \caption{A toy example is used to show how the choice of $N$ has an impact on averaging and the time scale. For fixed $M_{sp} = 6$, when $N$ is small ({\it e.g.} $N=5$) $M_r$ is also small ($M_r = 22$). Meaning the $M_r$ grid points will be more spaced out and each $M_r$ grid point will have contributions from most of the source strengths $\delta_{i}(I_h)$; therefore, investigating smaller time scales by averaging. On the other hand, when $N$ is large ({\it e.g.} $N=40$) $M_r$ is also large ($M_r = 162$). Meaning the $M_r$ grid points will be more closely spaced and only some of the $M_r$ grid points will have contributions from separate source points and majority of the $M_r$ grid points will have no contributions. This illustrates the intuitive idea of how changing $N$ allows one to investigate different time scales using the ideas from NUFFTs.}
    \label{fig:NUFFT_ave}
\end{figure}

The insight of NUFFT methods is that the relation between $N$ and the time scale is demonstrated more intuitively (see \Cref{fig:NUFFT_ave}). For fixed $M_{sp}$, when $N$ is small $M_r$ is also small. Therefore, the $M_r$ grid points will be more spread out and each grid point will have contributions from multiple source strengths averaged based on the choice of kernel $\varphi(\cdot)$. While for the case when $N$ is large $M_r$ will also be large. Meaning the $M_r$ grid points are more tightly packed and fewer grid points will have contributions from separate source strengths --- essentially there is less averaging.

Second, the interpolation is explicit in NUFFT methods. This is interesting because interpolation of financial data can result in estimates being biased --- such as linear interpolation \cite{BR2002} or interpolation based on underlying continuity assumptions such as the Hayashi-Yoshida estimator \cite{PCRBTG2019}. We argue these methods are flawed because they do not account for the effects of interpolation --- whereas NUFFT methods account for this by deconvolving the interpolation effects in the Fourier space.

Before moving on, we highlight that one of the main characteristics of the Malliavin-Mancino estimators is that it does not require the manipulation of the original data in the computation of \cref{eq:Der:FC}. However, there is an explicit averaging step in the evaluation of the Fourier coefficients using NUFFT methods. Therefore, we need to carefully examine if the use of \cref{eq:Der:17} rather than \cref{eq:Der:FC} in \cref{eq:Der:1} will affect the resulting estimates (see \Cref{subsec:accuracy}).

\section{Algorithm Performance and Benchmarking} \label{sec:benchmarking}

The benchmarking is done using Monte Carlo simulations.\footnote{All the seeds for replication of the work are provided in the respective script files from our GitHub resource \cite{PCEPTG2020CODE}} We compare the relative performance of the algorithms and investigate the various factors influencing speed and accuracy. We use the Geometric Brownian Motion (GBM) which satisfies the following SDEs:
\begin{equation} \label{eq:BM:1}
  \frac{dS_i(t)}{S_i(t)} = \mu_i dt + \sigma_i dW_i(t), \ \ \ \ i = 1, ..., D,
\end{equation}
with $\mathrm{Corr}(dW_i, dW_j) = \rho^{ij}$.
The GBM is simulated using the Euler–Maruyama scheme (which is strong order 0.5 in the sense of \cite{kloeden2013numerical}) with equal spacing $\Delta t$ between the observations (see \cref{algo:GBM}) and are at the same time across the features. This is known as the synchronous case. Asynchrony is then induced from the synchronous case using two approaches: (i) the missing data representation, and (ii) the arrival time representation. The missing data representation is achieved by randomly sampling and removing a certain percentage of observations. The arrival time representation is achieved by sampling the synchronous price path using an exponential inter-arrival time with rate $\lambda_i$.

\begin{figure*}[p]
\centering
    \subfloat{\label{fig:ONplots:a}\includegraphics[width=0.48\textwidth]{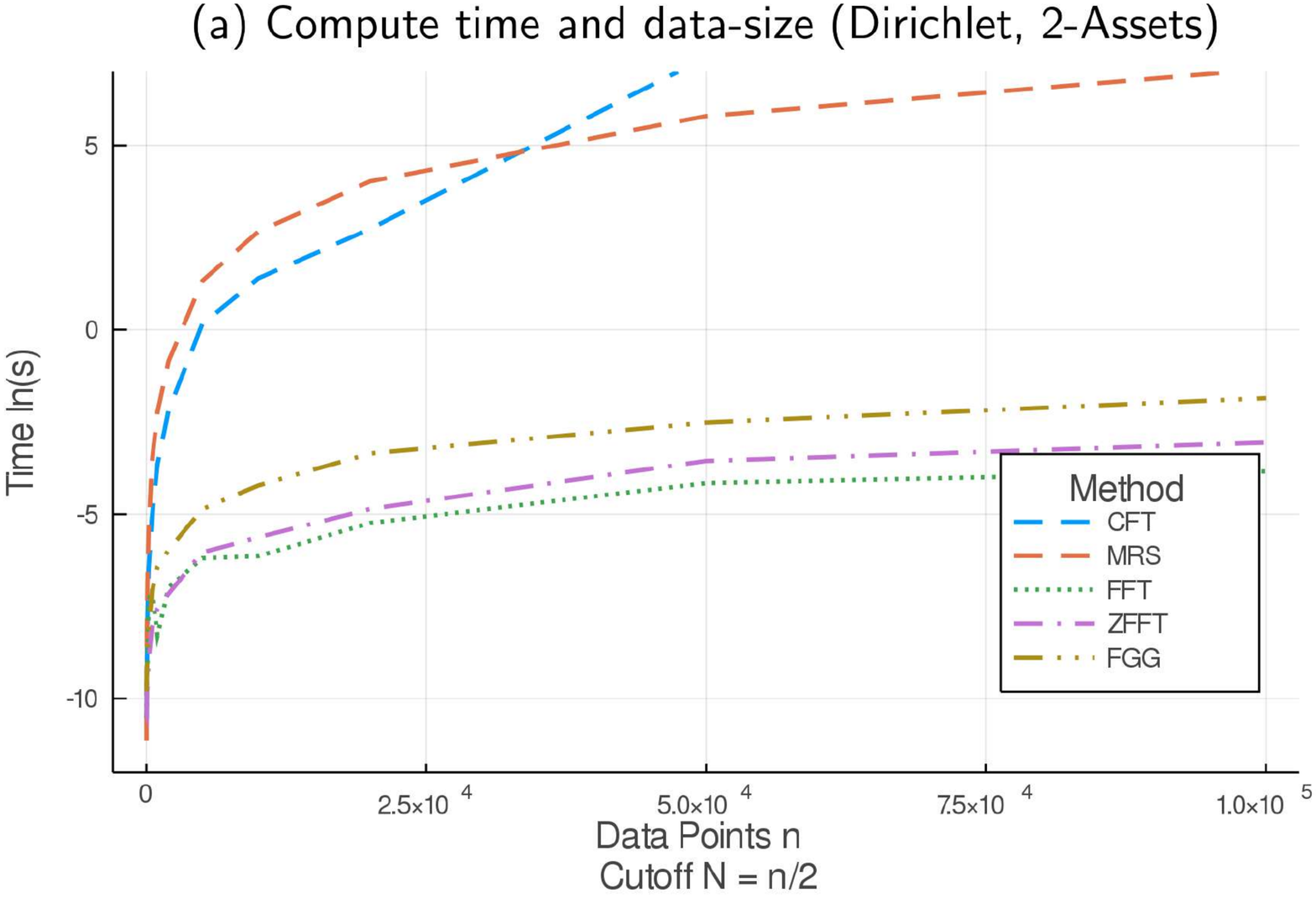}}
    \subfloat{\label{fig:ONplots:b}\includegraphics[width=0.48\textwidth]{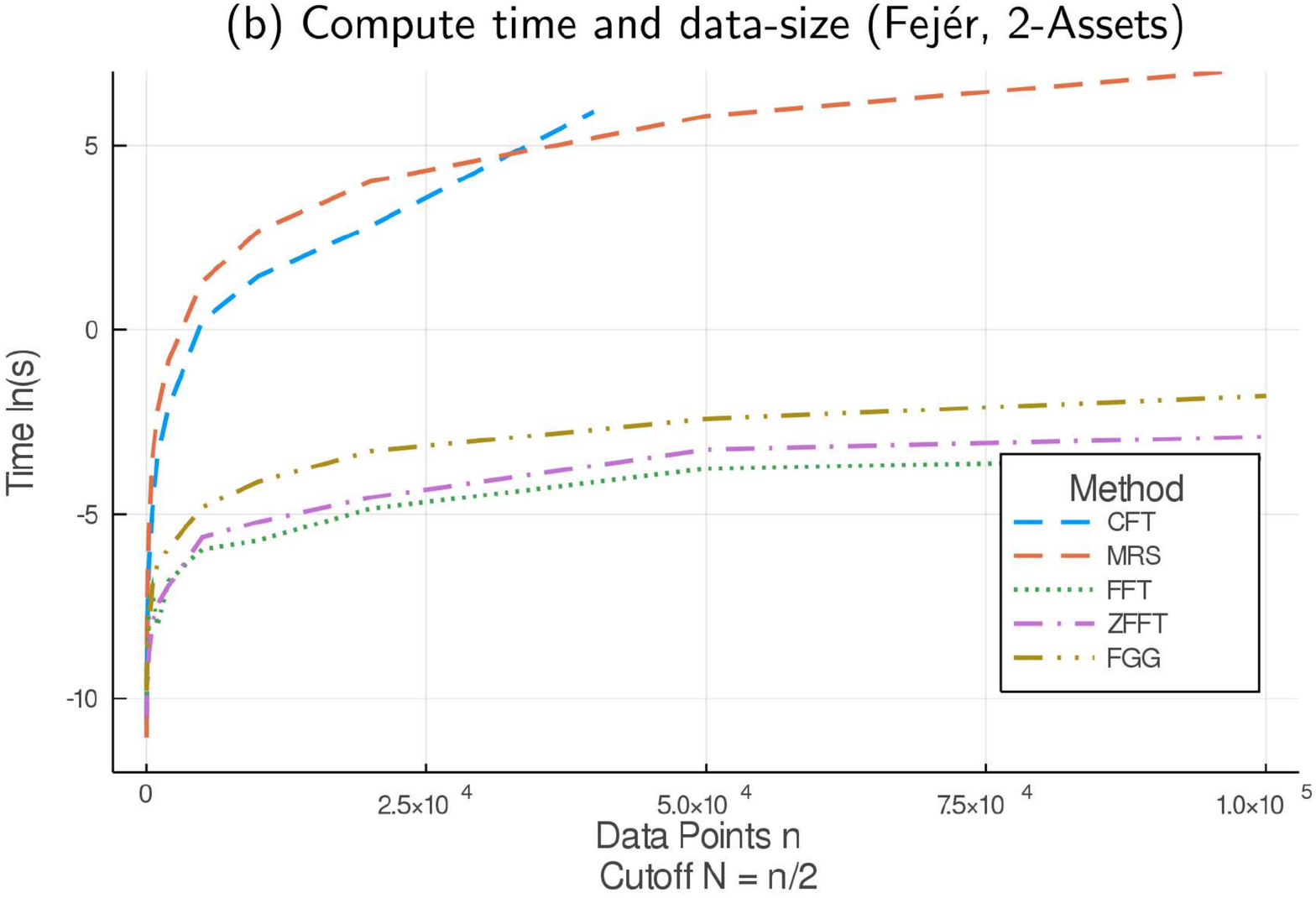}} \\
    \subfloat{\label{fig:ONplots:c}\includegraphics[width=0.48\textwidth]{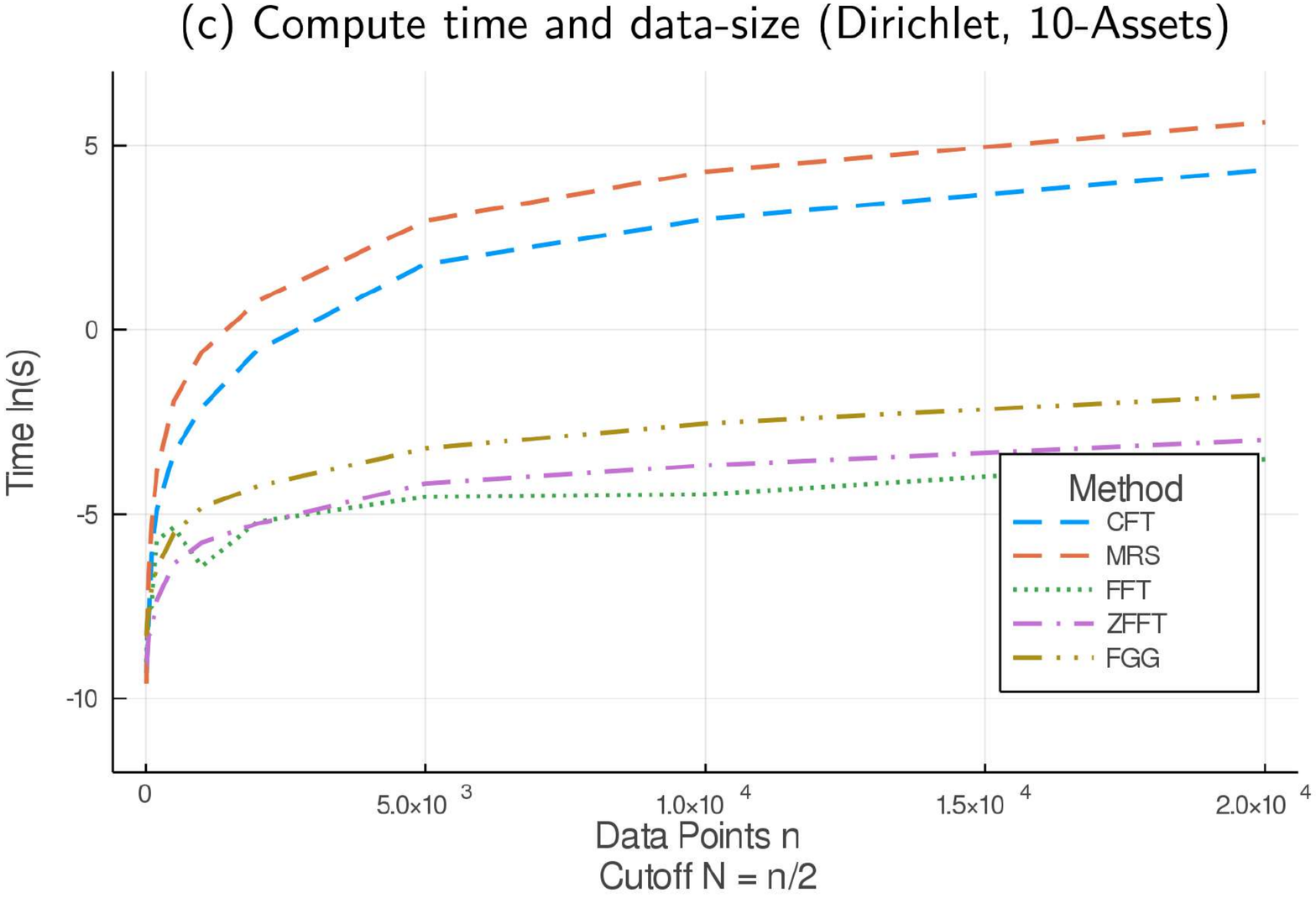}}
    \subfloat{\label{fig:ONplots:d}\includegraphics[width=0.48\textwidth]{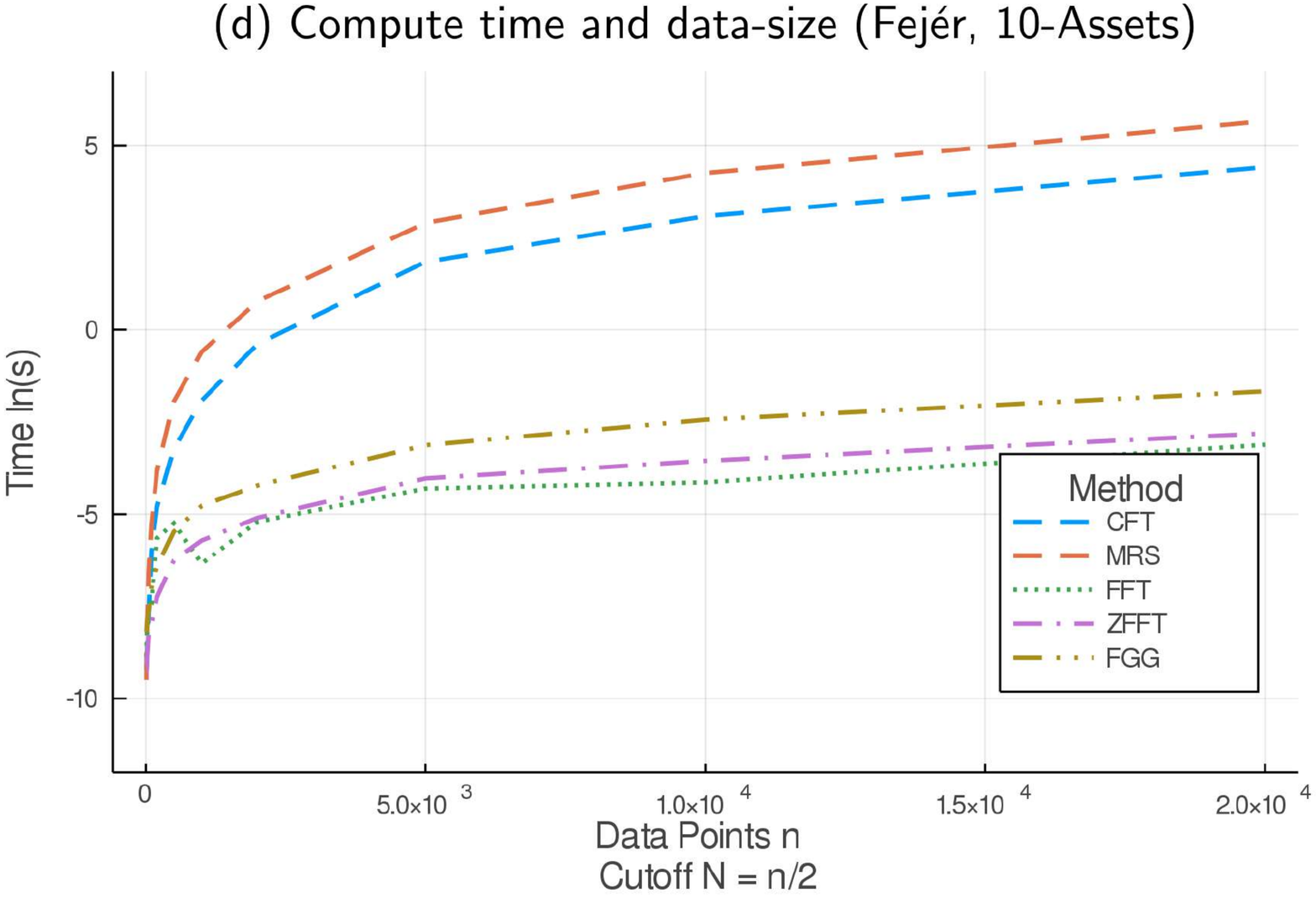}}  \\
    \subfloat{\label{fig:ONplots:e}\includegraphics[width=0.48\textwidth]{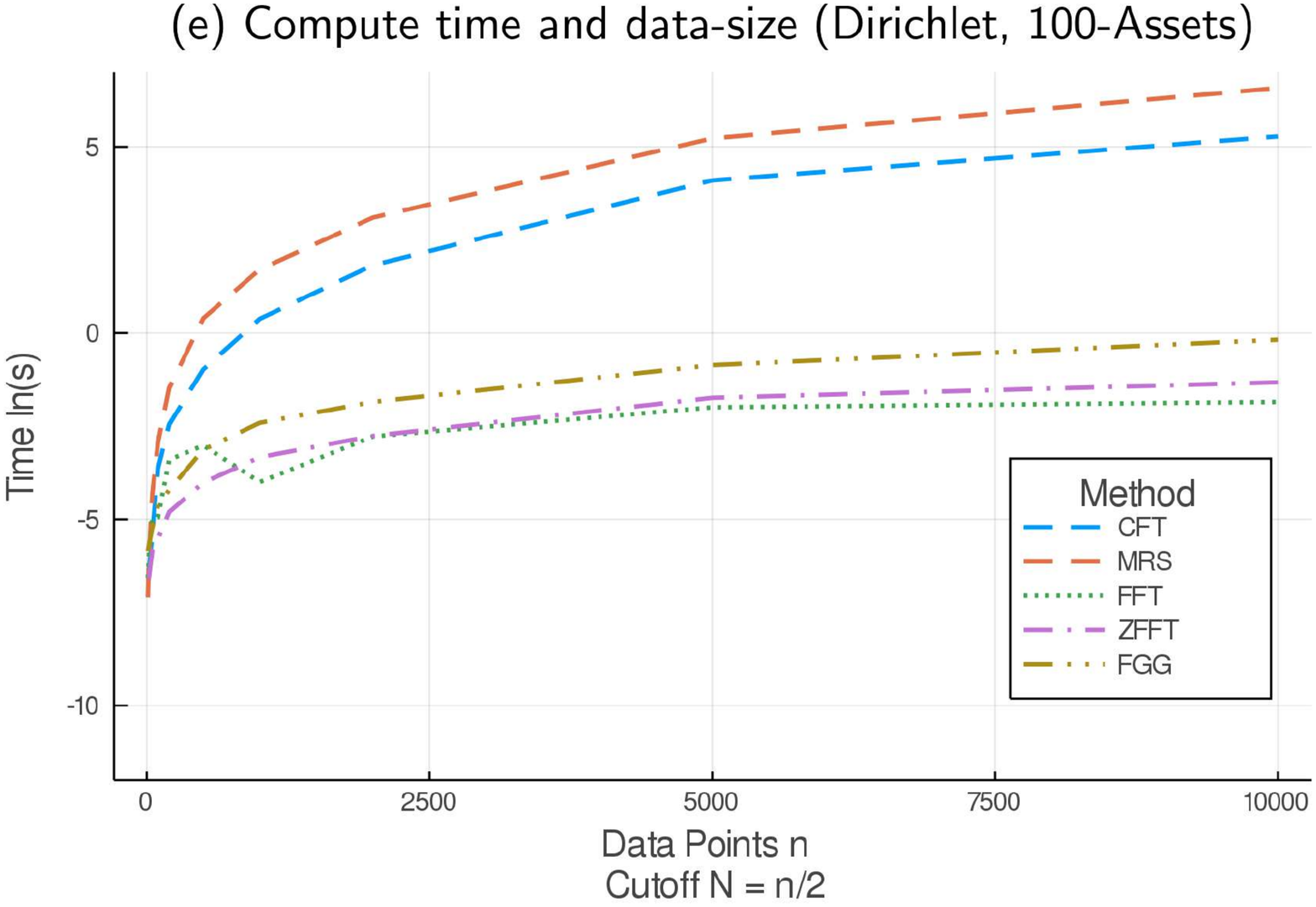}}
    \subfloat{\label{fig:ONplots:f}\includegraphics[width=0.48\textwidth]{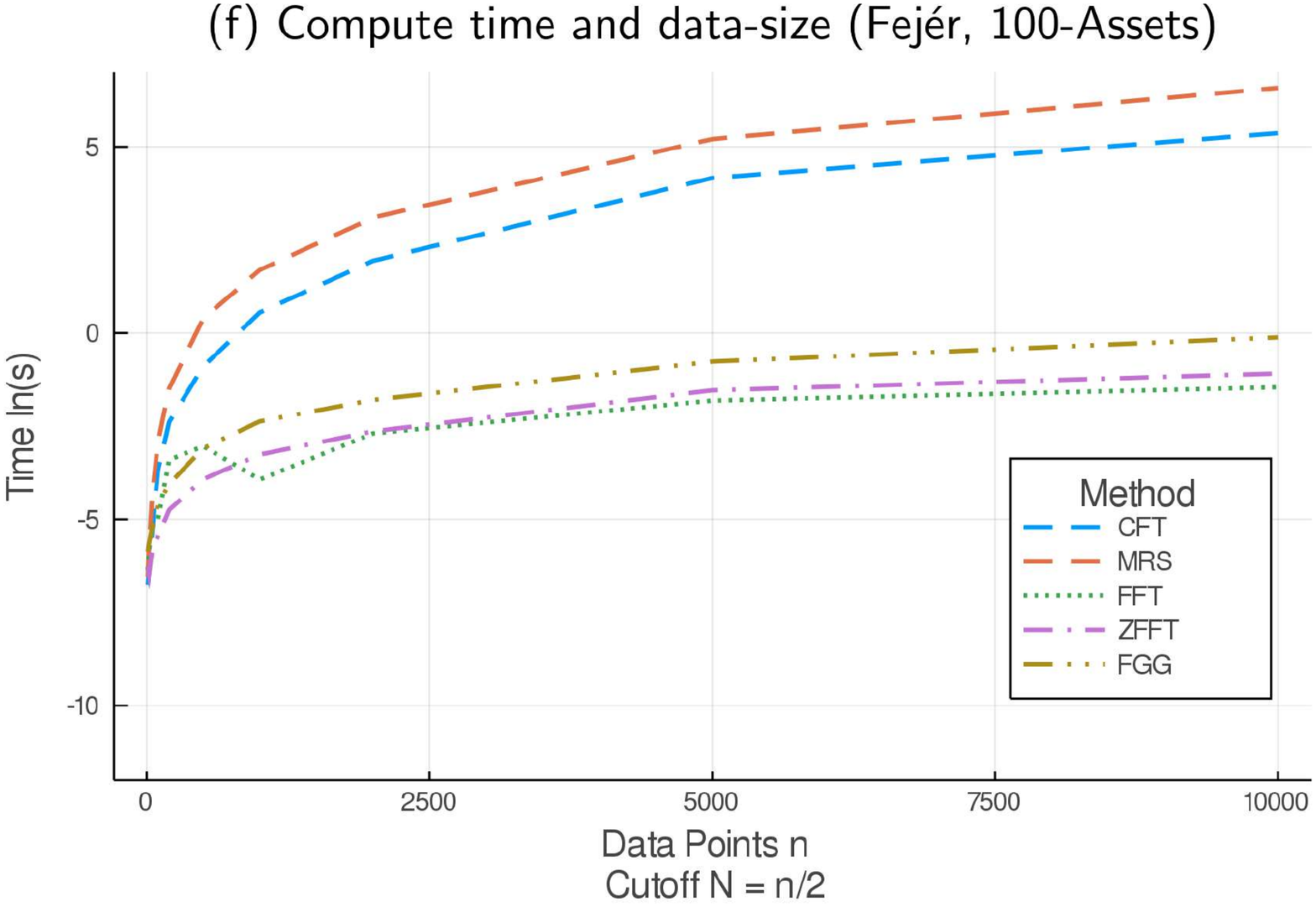}}
\caption{We investigate the algorithm complexity between traditional implementation methods against fast Fourier methods. We plot the logarithm of compute time (measured in seconds) as a function of the number of data points $n$ for various implementation methods. The $n$ is the number of price observations simulated using a Geometric Brownian Motion with \cref{algo:GBM}. We investigate the synchronous case since the Nyquist cutoff scales linearly with the number of data points when the data points are uniform in time. Furthermore, we obtain run times for the Dirichlet and Fej\'{e}r kernel basis for $D=2, 10$ and $100$ features to investigate the impact breadth has on the run time. The traditional methods investigate are the vectorised implementation (CFT - blue dashes) and the ``for-loop'' implementation (MRS - orange dashes). The fast Fourier methods investigated are the FFT (FFT - green dots), zero-padded FFT (ZFFT - purple dash-dots) and the NUFFT implementation using the fast Gaussian gridding (FGG - dark green dash-dot-dots) with the default $\epsilon = 10^{-12}$. The figures demonstrate the efficacy of fast Fourier methods in reducing compute time for both basis kernels of the Malliavin-Mancino estimator. The figures can be recovered using the Julia script files \href{https://github.com/CHNPAT005/PCEPTG-MM-NUFFT/blob/master/Scripts/Timing/Dirichlet\%20Timing}{Dirichlet Timing} and \href{https://github.com/CHNPAT005/PCEPTG-MM-NUFFT/blob/master/Scripts/Timing/Fejer\%20Timing}{Fejer Timing} on the GitHub resource \cite{PCEPTG2020CODE}.}
\label{fig:ONplots}
\end{figure*}

\begin{table*}[htb]
\centering
\begin{tabular}{|l|l|l|l|l|l|}
\hline
Method           & MRS  & KB & ES & FGG & FINUFFT   \\ \hline
Time (s) & 1176s & 2.161s & 0.190s  & 0.119s & 0.0331s \\ \hline
\end{tabular}
\caption{The measured compute times are given in seconds for various algorithms using 2 features with $10^5$ data points. The methods considered are: the ``for-loop'' implementation (MRS), the fast Gaussian gridding (FGG), Kaiser-Bessel (KB), exponential of semi-circle using our naive implementation (ES) and the implementation from FINUFFT (FINUFFT). The NUFFT methods are computed using the default $\epsilon = 10^{-12}$. The times are extracted from \Cref{fig:ONplots,fig:Errorplots}. }
\label{tab:time}
\end{table*}

\begin{table*}[htb]
\centering
\begin{tabular}{|lll|l|ll|ll|}
\hline
      &       &       &             & \multicolumn{2}{c|}{Dirichlet {[}sec{]}} & \multicolumn{2}{c|}{Fej\'{e}r {[}sec{]}} \\
$n_1$ & $n_2$ & $N$   & $1/\lambda_2$ & FGG                 & MRS                & FGG                          & MRS                        \\ \hline
319   & 336   & 48855 & 30          & 0.047676            & 3.83602            & 0.050635                     & 3.93206                    \\
319   & 272   & 36164 & 40          & 0.027140            & 2.51448            & 0.032651                     & 2.64331                    \\
319   & 214   & 7128  & 50          & 0.004719            & 0.40825            & 0.005266                     & 0.44622                    \\
319   & 169   & 38917 & 60          & 0.021548            & 2.19374            & 0.028445                     & 2.40181                    \\
319   & 168   & 25006 & 70          & 0.013686            & 1.39089            & 0.016143                     & 1.50772                    \\
319   & 106   & 2281  & 80          & 0.000995            & 0.09976            & 0.001036                     & 0.10474                    \\
319   & 119   & 3325  & 90          & 0.001844            & 0.14592            & 0.001923                     & 0.15903                    \\
319   & 98    & 2227  & 100         & 0.000765            & 0.09277            & 0.000805                     & 0.09451                    \\ \hline
\end{tabular}
\caption{The Dirichlet and Fej\'{e}r computation times (measured in seconds) are given for varying degree of asynchrony. Here a synchronous GBM with $10^4$ data points is sampled with $1/\lambda_1 = 30$ for the first feature, while $1/\lambda_2$ ranges from 30 to 100 in increments of 10 seconds. The table reports the exact $n_i$ and $N$ from the sampling. The fast Gaussian gridding implementation of the NUFFT (FGG) outperforms the for-loop implementation (MRS); as $1/\lambda_2$ increases, $n_2$ and $N$ decrease, resulting in a faster compute time. However, increasing $1/\lambda_2$ does not guarantee that $\Delta t_0$ will be larger, which can lead to a larger $N$ and therefore a longer compute time.}
\label{tab:asyntime}
\end{table*}

\subsection{Benchmark Timing} \label{subsec:time}

The common factors affecting the computation time for all the algorithms are: (i) the number of data points $n_1,...,n_D$, (ii) the number of Fourier coefficients $M=2N+1$, and (iii) the number of features $D$. The parameter specific to the non-uniform FFT method is the tolerance $\epsilon$ which determines the spreading width $\omega$. 

First, we investigate the common factors affecting computation time for the various algorithms.\footnote{The benchmarking is done using a 2.5GHz base clock speed Quad-Core Intel i7-4870HQ with 16GB of 1600MHz DDR3L (CL=11) RAM on MacOS version 10.15.1 with JuliaPro version 1.2.0. GCC8 is used as a requirement for the Julia interface to FINUFFT provided by \cite{K2018}.} To this end, we investigate the computation time as a function of the number of data points for a synchronous GBM ($n=n_1=...=n_D$). The synchronous GBM is used because the Nyquist frequency is $N = n/2$ for the synchronous case. Therefore, the number of Fourier coefficients scale linearly with the number of data points. 

\Cref{fig:ONplots} we compare the following algorithms: the for-loop implementation (MRS), the vectorised implementation (CFT), the FFT implementation (FFT), the zero-padded FFT implementation (ZFFT) and the fast Gaussian gridding implementation of the NUFFT (FGG) using the default $\epsilon = 10^{-12}$. The $O(n)$ plots are plotted with compute time\footnote{The compute time is the minimum estimate over 10 replications. As the minimum is a robust estimator for the location parameter of the time distribution \cite{CR2016}.} on the log scale for better comparison, and include the Dirichlet and Fej\'{e}r representation for $D = 2, 10$ and $100$ features.\footnote{The induced correlation matrix for $D=10$ and $100$ are created using a uniform random matrix and re-scaled appropriately. The function can be found in \href{https://github.com/CHNPAT005/PCEPTG-MM-NUFFT/blob/master/Functions/Monte\%20Carlo\%20Simulation\%20Algorithms/RandCovMat}{gencovmatrix} provided in our GitHub resources \cite{PCEPTG2020CODE}. We use a uniform random matrix, such a choice will only produce positive correlations but is computationally convenient and has no influence estimates of the compute times.} Taking a closer look at \Cref{fig:ONplots} we notice several things. 

First, the for-loop and vectorised implementation take on the same general shape but the vectorised implementation is faster due to the exploitation of the Hermitian symmetry. \Cref{fig:ONplots:a,fig:ONplots:b} demonstrate the limitation of the vectorisation: memory usage. This is because each element in the complex matrix (of size $n \times N$) requires 16 bytes to store the Complex 64-bit floating point number. For $5 \times 10^4$ data points we require 20GB of memory, therefore demanding the use of virtual memory which results in a deterioration of performance.

Second, the difference in speed between the naive methods compared to the fast Fourier transform methods is significant. Looking at \Cref{tab:time} for 2 features with $n=10^5$ data points, the for-loop implementation takes $1176$ seconds while the fast Gaussian gridding takes $0.119$ seconds --- $10,000$ times faster than the naive for-loop compute time.

Third, between the fast Fourier methods from fastest to slowest we have: FFT, zero-padded FFT, and FGG. This is because the FFT computes $N$ Fourier modes, the zero-padded FFT computes $M$ Fourier modes, and the FGG computes $M_r$ Fourier modes. On top of that, the FFT requires no steps before performing the FFT whereas the zero-padded FFT needs to zero-pad missing data while the FGG requires the convolution and deconvolution step.

Finally, the breadth of features $D$ can impact computation time depending on the choice of $N$. The case when $N$ is the same across all features is simple. We then only need to compute the $M$ Fourier coefficients for $D$ features. This is presented in \Cref{fig:ONplots}. When $N$ is the same across all features, we are presented with two advantages: (i) the time scale investigated will be the same for all the features, and (ii) if one uses the Fej\'{e}r basis kernel we can guarantee positive semi-definiteness in the covariance matrix \cite{MRS2017, MS2011}. The case when $N$ changes for the different features becomes more nuanced. For example, different features have different Nyquist frequencies in the arrival time representation. Here we need to compute $\frac{D(D-1)}{2}$ pairwise estimates for each $\hat{\Sigma}_{n,N}^{ij}$ entry. A potential problem arises when $N$ is independently obtained to investigate the co-movement between events for each feature pair \cite{PCRBTG2019}, or when the Dirichlet basis kernel is used. We are not guaranteed a positive semi-definite matrix which can present challenges. For example, when an invertible covariance matrix is a necessary requirement such as in the case of portfolio optimisation. This can be ameliorated by transforming the non-positive semi-definite covariance matrix estimate to the closest positive semi-definite matrix under some appropriate norm \cite{Lindskog2001}, or using extensions of these type of transformations \cite{Matoti2009}. However, doing so comes with an additional computational cost.

\begin{figure*}[p]
\centering
    \subfloat{\label{fig:Errorplots:a}\includegraphics[width=0.48\textwidth]{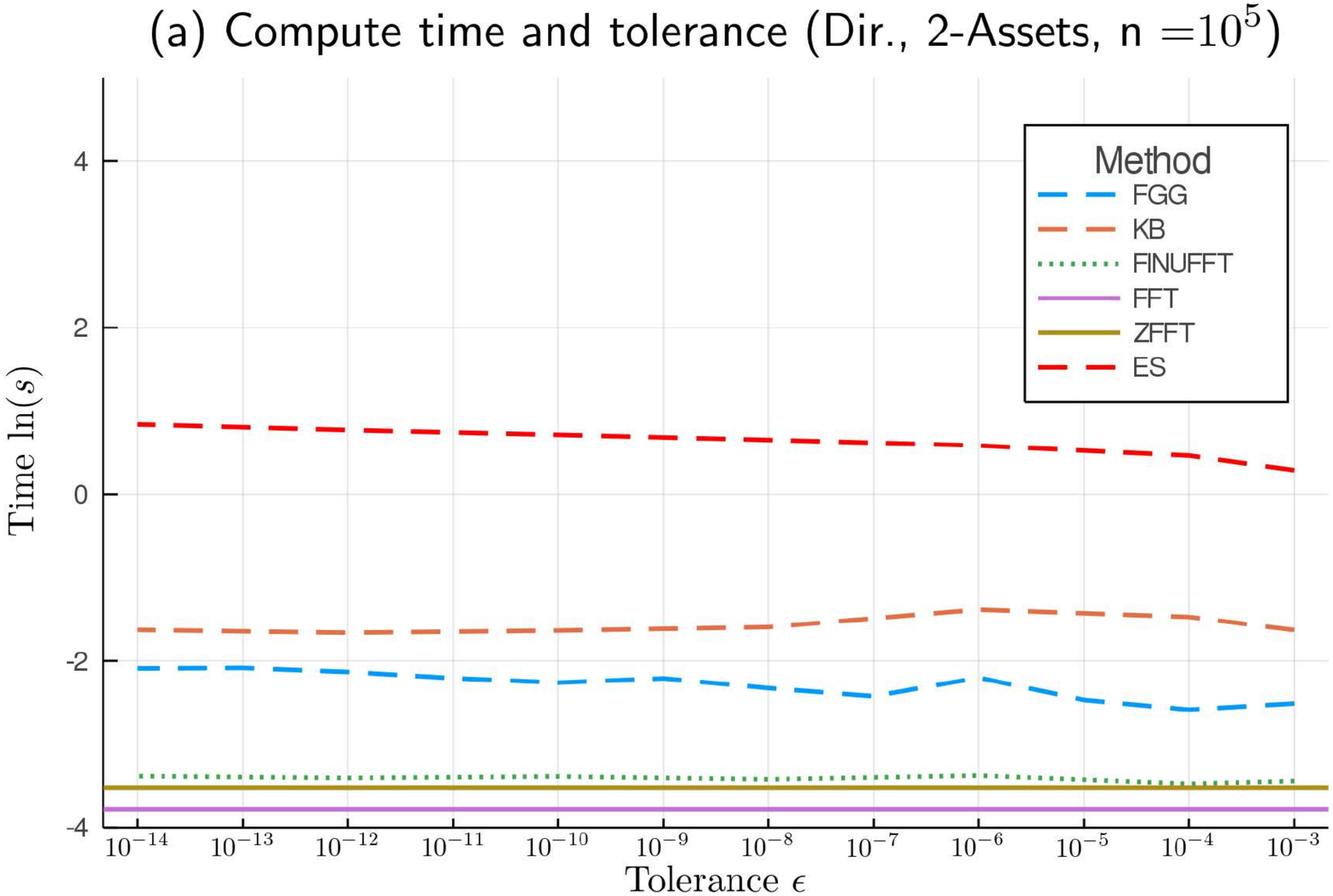}}
    \subfloat{\label{fig:Errorplots:b}\includegraphics[width=0.48\textwidth]{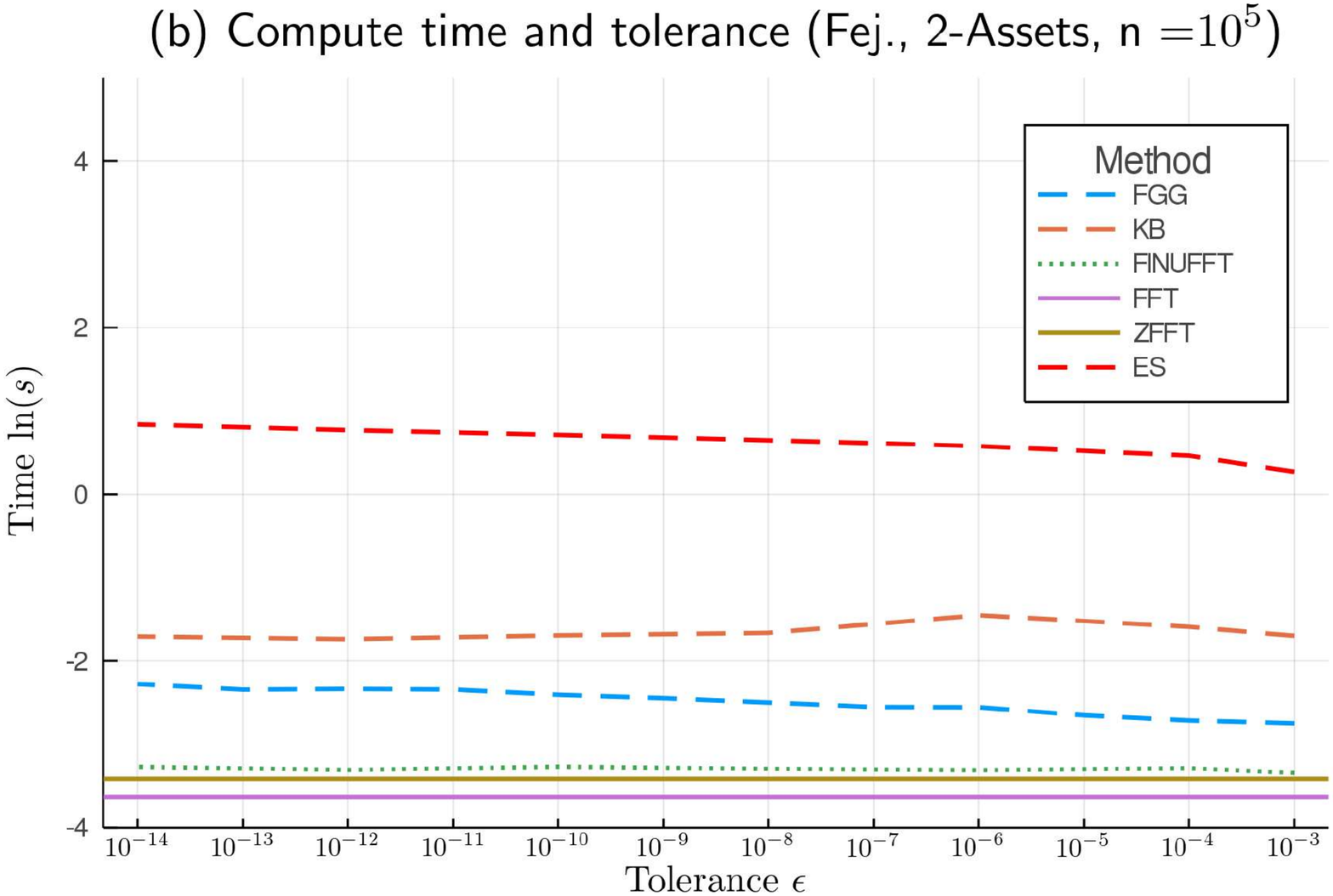}} \\
    \subfloat{\label{fig:Errorplots:c}\includegraphics[width=0.48\textwidth]{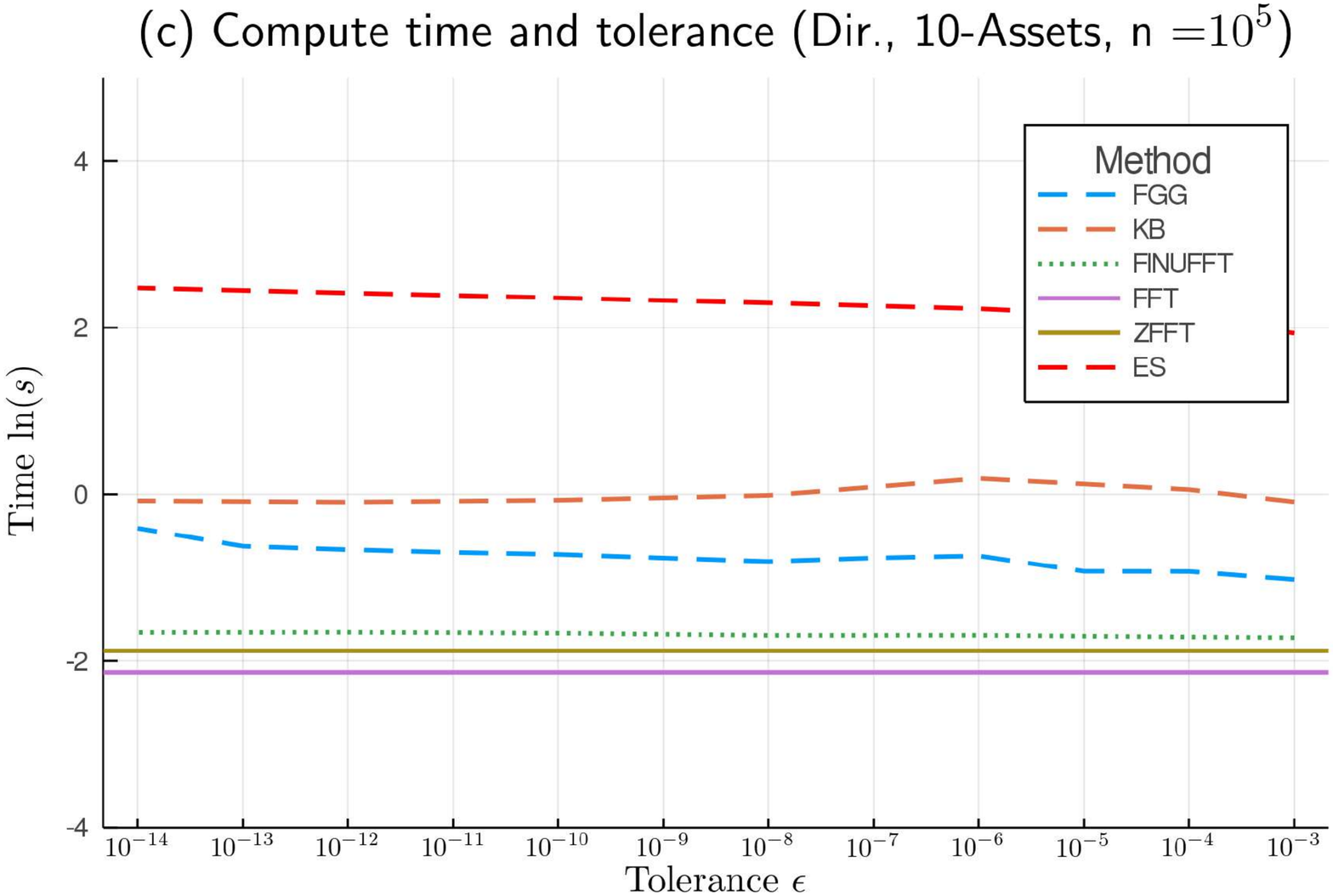}}
    \subfloat{\label{fig:Errorplots:d}\includegraphics[width=0.48\textwidth]{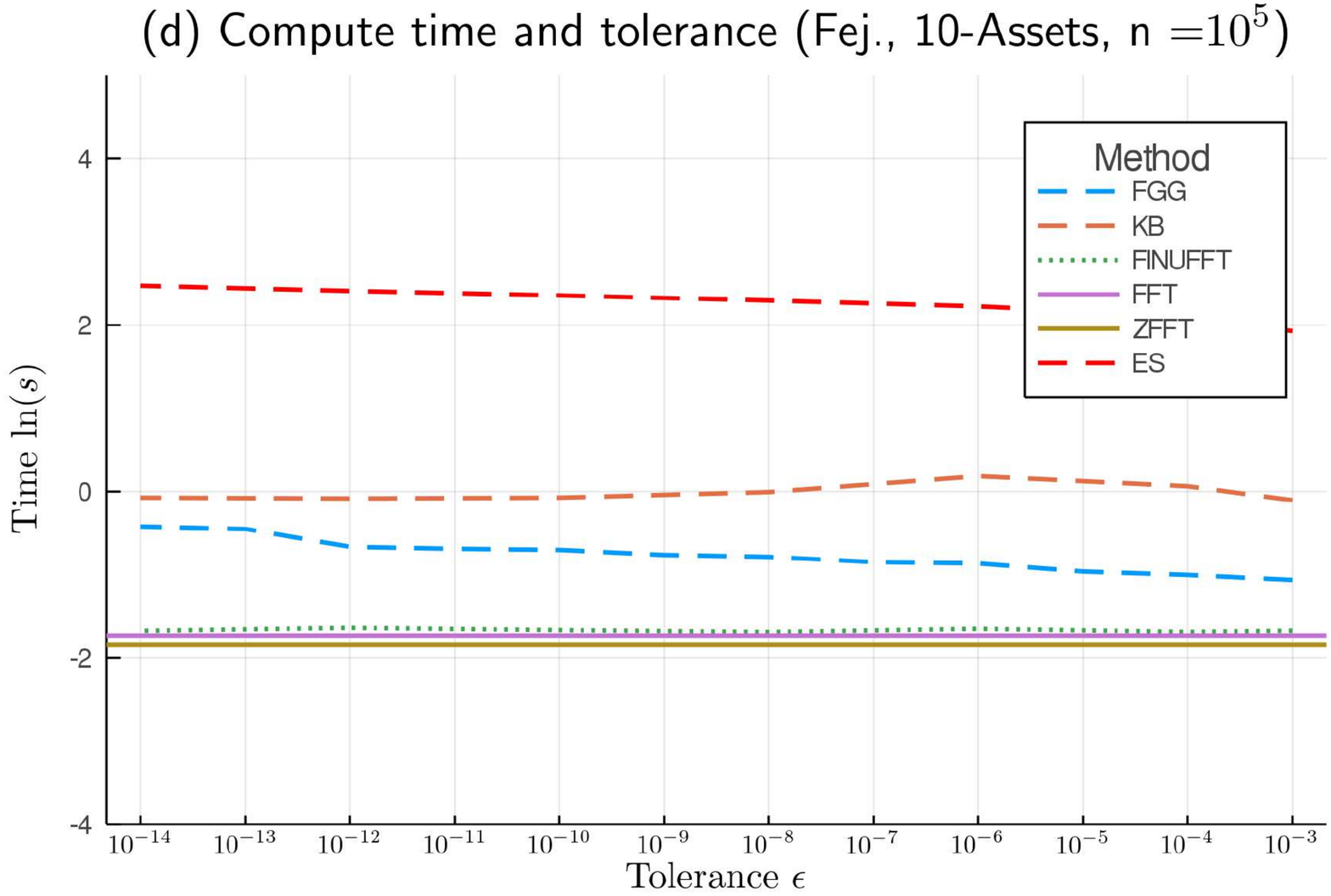}}  \\
    \subfloat{\label{fig:Errorplots:e}\includegraphics[width=0.48\textwidth]{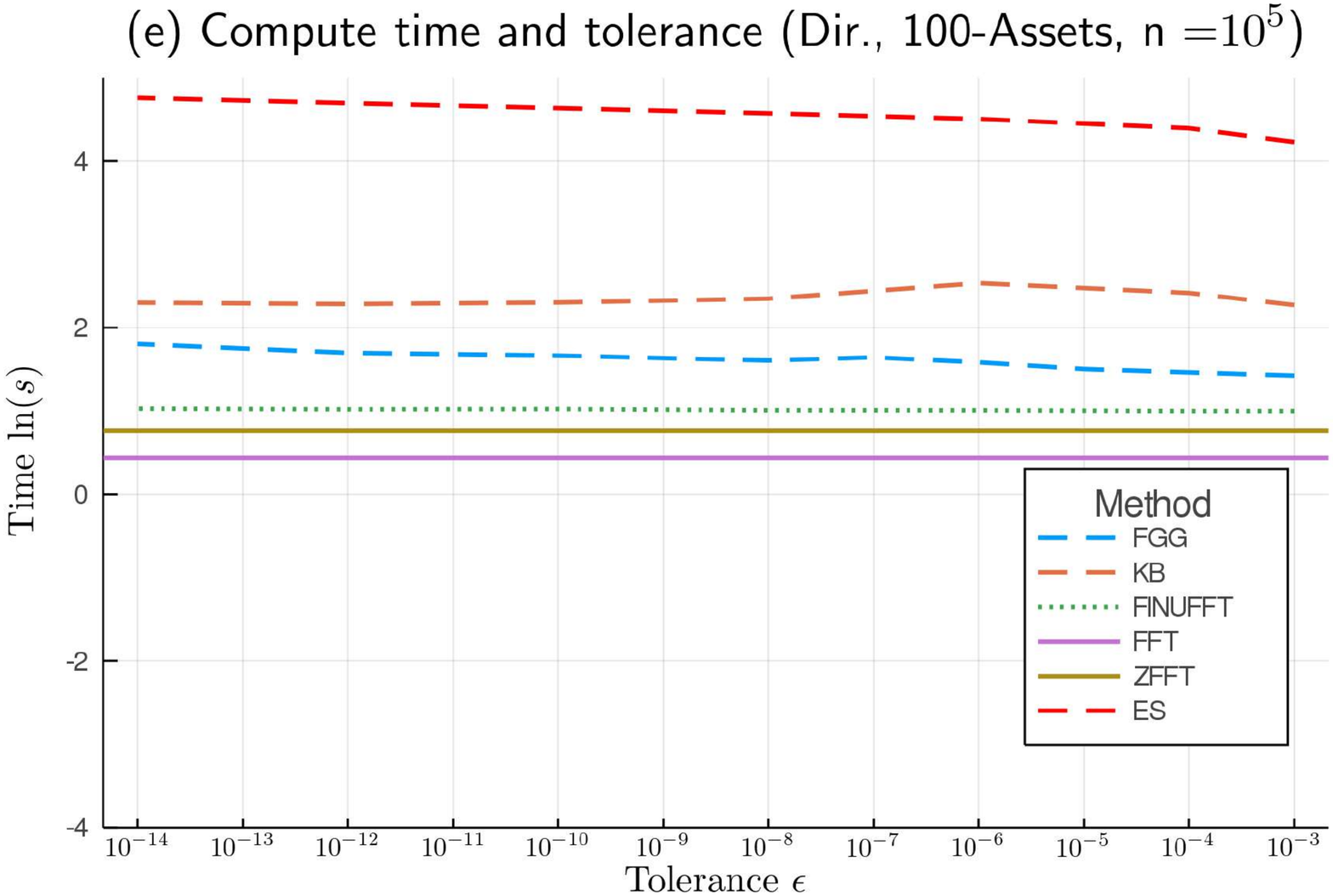}}
    \subfloat{\label{fig:Errorplots:f}\includegraphics[width=0.48\textwidth]{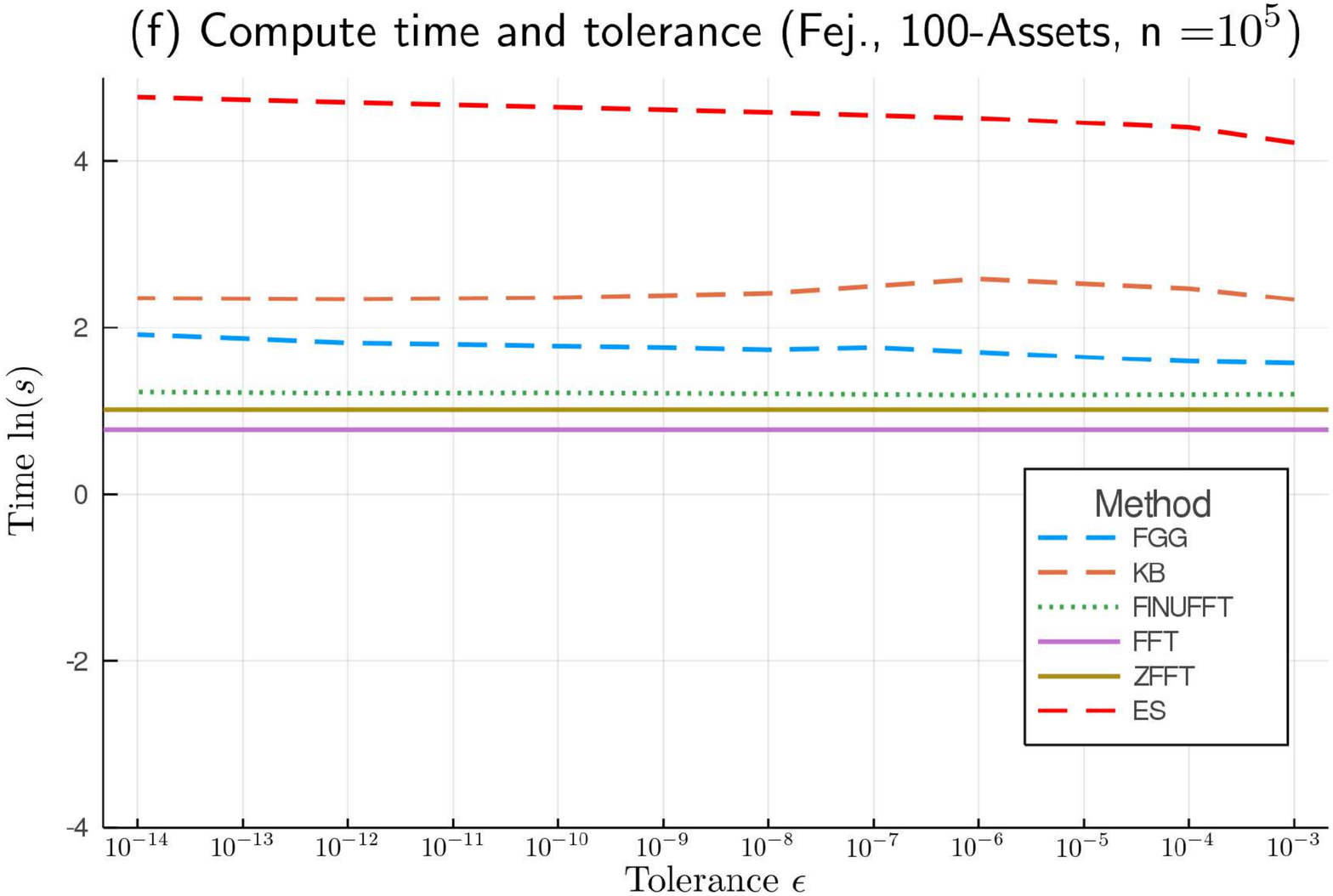}}
\caption{We investigate the algorithm complexity for various non-uniform fast Fourier transform methods by plotting the time (measured in seconds) as a function of the error tolerance $\epsilon$ which determines the spreading width $\omega$ for the various averaging kernels. The FFT (FFT - purple line) and zero-padded FFT (ZFFT - dark green line) is plotted as a baseline for comparison. We simulate a synchronous Geometric Brownian Motion using \cref{algo:GBM} with $n = 10^5$ data points for $D = 2, 10$ and $100$ respectively. The run times are obtained for the two basis kernels: Dirichlet (Dir.) and Fej\'{e}r (Fej.). The non-uniform FFT methods considered are the Gaussian kernel (FGG - blue dashes), the Kaiser-Bessel kernel (KB - orange dashes) and the exponential of semi-circle kernel with our naive implementation (ES - red dashes) and the implementation by FINUFFT (FINUFFT - green dots). The results are consistent with the results from \cite{BMK2018}, where the FINUFFT implementation is faster than the fast Gaussian gridding which is faster than the Kaiser-Bessel. The evaluations are all done ``on-the-fly'' without any pre-computation. The figures can be recovered using the Julia script file \href{https://github.com/CHNPAT005/PCEPTG-MM-NUFFT/blob/master/Scripts/Timing/Error\%20Timing}{Error Timing} on the GitHub resource \cite{PCEPTG2020CODE}.}
\label{fig:Errorplots}
\end{figure*}

Let us investigate the degree of asynchrony as a variable of study, specifically the affect on computation time under the arrival time representation. Here we only consider the case when $D=2$. Let $1/\lambda_1$ be the mean inter-arrival time used to sample the first feature, and $N_1$ be the corresponding Nyquist frequency\footnote{The Nyquist frequency here is $\lfloor \frac{T}{2 \Delta t_0} \rfloor$ where $\Delta t_0$ is the smallest distance between two consecutive prices \cite{MRS2017}.} from the feature; similarly $1/\lambda_2$ and $N_2$ for the second feature. Therefore, the $N$ used in \cref{eq:Der:1,eq:Der:2,eq:Der:3} is $N = \min\{N_1,N_2\}$ to avoid aliasing.

\Cref{tab:asyntime} reports the Dirichlet and Fej\'{e}r computation time (minimum estimate over 10 replications and measured in seconds) for the for-loop implementation (MRS) and the fast Gaussian gridding implementation of the NUFFT (FGG) using the default $\epsilon = 10^{-12}$. We simulate a GBM with $n=10^4$ data points. This process is then sampled using an exponential inter-arrival with rate $\lambda_i$. Here the first feature is sampled with an average inter-arrival ($1/\lambda_1$) of 30 seconds, while the second feature is sampled with an average inter-arrival ($1/\lambda_2$) ranging from 30 to 100 seconds in increments of 10. The exact number of observed data points $n_i \approx n/\lambda_i$ and Nyquist frequency $N$ from the sampling is also reported.\footnote{Using the Nyquist frequency under asynchrony will result in the estimate being biased. This is demonstrated in \Cref{subsec:accuracy}.}

\Cref{tab:asyntime} demonstrates two things: first, the FGG is significantly faster than the for-loop implementation. Second, as $1/\lambda_2$ increases we get a faster compute time (most of the time) because $n_2$ and $N$ decrease. However, this is not guaranteed because a larger $1/\lambda_2$ does not ensure $\Delta t_0$ (minimum $\Delta t$) will also be larger. Therefore, $N$ does not always decrease which can lead to longer compute times.

The last variable influencing compute time to investigate is the impact of tolerance $\epsilon$. This is explored by plotting the computation time as a function of $\epsilon$. Here we use the synchronous case ($n=n_1=...=n_D$) with $n=10^5$ data points.

\Cref{fig:Errorplots} we compare the following algorithms: the FFT implementation (FFT) and the zero-padded FFT implementation (ZFFT) as a baseline for comparison. The NUFFT methods include the fast Gaussian gridding with the Gaussian kernel (FGG), the Kaiser-Bessel kernel (KB), the exponential of semi-circle using our naive implementation (ES) and the FINUFFT package (FINUFFT). The $O(n)$ plots are plotted on the log scale as the minimum compute time estimate over 10 replications. The figures include the Dirichlet and Fej\'{e}r representation for $D = 2, 10$ and $100$ features.

Looking more closely, we see that the zero-padded FFT implementation needs to assign $n$ data points to the over-sampled grid, and the NUFFT methods need to assign $\omega n$ points to the over-sampled grid. Furthermore, the zero-padded FFT requires no evaluations whereas the NUFFT methods require $\omega n$ evaluations of $\delta_{i}(I_h) \varphi(\cdot)$. 

The key differences between the NUFFT algorithms is in how they reduce the number of evaluations required. The technique used in the fast Gaussian gridding is to reduce the number of exponential evaluations for $\varphi_{_G}(\cdot)$. This is achieved by separating the exponential into three components:
\begin{equation} \label{eq:Der:20}
  e^{-(x_j - 2\pi m / M_r)^2/ 4\tau} = \left[{e^{-x_j^2 / 4\tau} }\right] \left[{e^{x_j \pi / M_r \tau}}\right]^m \left[{e^{-(\pi m / M_r)^2 / \tau}}\right].
\end{equation}
By splitting the exponential this way, we only need two exponential evaluations per source point instead of $\omega$ exponential evaluations for each source point. Reducing the number of exponential evaluations from $\omega n$ to $\omega + 2n$. 
The advantage with using the Kaiser-Bessel kernel is that it is both smooth and has narrow support \cite{BMK2018}. This can be exploited to cut the number of kernel evaluations by reducing $\omega$ while maintaining a comparable level of accuracy. For example, the Gaussian kernel requires $\omega = 24$ for roughly 12 digit accuracy while the Kaiser-Bessel kernel requires $\omega = 13$ for the same 12 digit accuracy \cite{BMK2018}. 
Finally, the exponential of semi-circle has narrow support similar to that of the Kaiser-Bessel kernel but is simpler and faster to evaluate. The downfall is that there is no known analytic Fourier transform, thus incurring the additional cost of numerical integration to evaluate \cref{eq:Der:11}.

Our implementation of the exponential of semi-circle is naive compared to \cite{BMK2018}. We do not exploit the piecewise polynomial kernel approximation to accelerate the evaluation of \cref{eq:Der:10}. Furthermore, we use naive numerical integration to compute \cref{eq:Der:11} using the adaptive Gauss-Kronrod quadrature \href{https://github.com/JuliaMath/QuadGK.jl}{QuadGK} rather than the Gauss-Legendre quadrature with ``phase winding'' \cite{BMK2018}. The naive implementation of the exponential of semi-circle serves two purposes: (i) allowing the like-for-like comparison between the various kernels and their algorithms based on our implementation, and (ii) illustrating the importance of the implementation techniques used by \cite{BMK2018}. This is seen in \Cref{fig:Errorplots} where the exponential of semi-circle is significantly slower than the Gaussian or Kaiser-Bessel kernel without the implementation techniques due to the numerical integration required for \cref{eq:Der:11}. However, \cite{BMK2018} are able to reduce the compute time of the exponential of semi-circle to a similar time as our zero-padded FFT with the appropriate implementation techniques in place.

The results in \Cref{fig:Errorplots} are consistent with that of \cite{BMK2018}. Between the non-uniform FFT methods considered: the FINUFFT implementation of the exponential of semi-circle is the fastest, followed by the fast Gaussian gridding, the Kaiser-Bessel kernel evaluated ``on-the-fly'',\footnote{Without any pre-computations, preventing large RAM overhead.} and lastly the exponential of semi-circle using the naive implementation. 

\subsection{Benchmark Accuracy} \label{subsec:accuracy}

\begin{figure*}[p]
\centering
    \subfloat{\label{fig:AccSynDS:a}\includegraphics[width=0.48\textwidth]{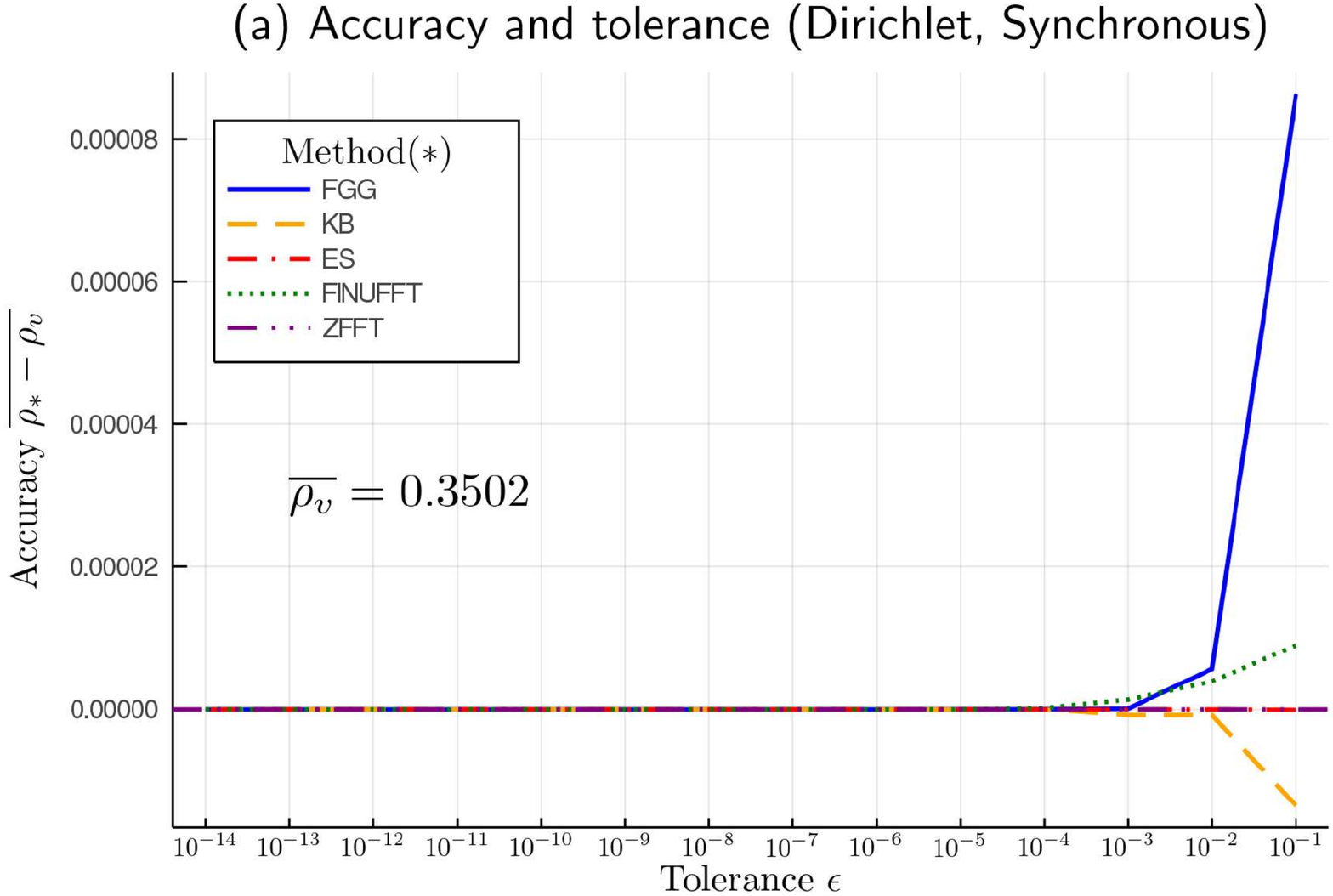}}
    \subfloat{\label{fig:AccSynDS:b}\includegraphics[width=0.48\textwidth]{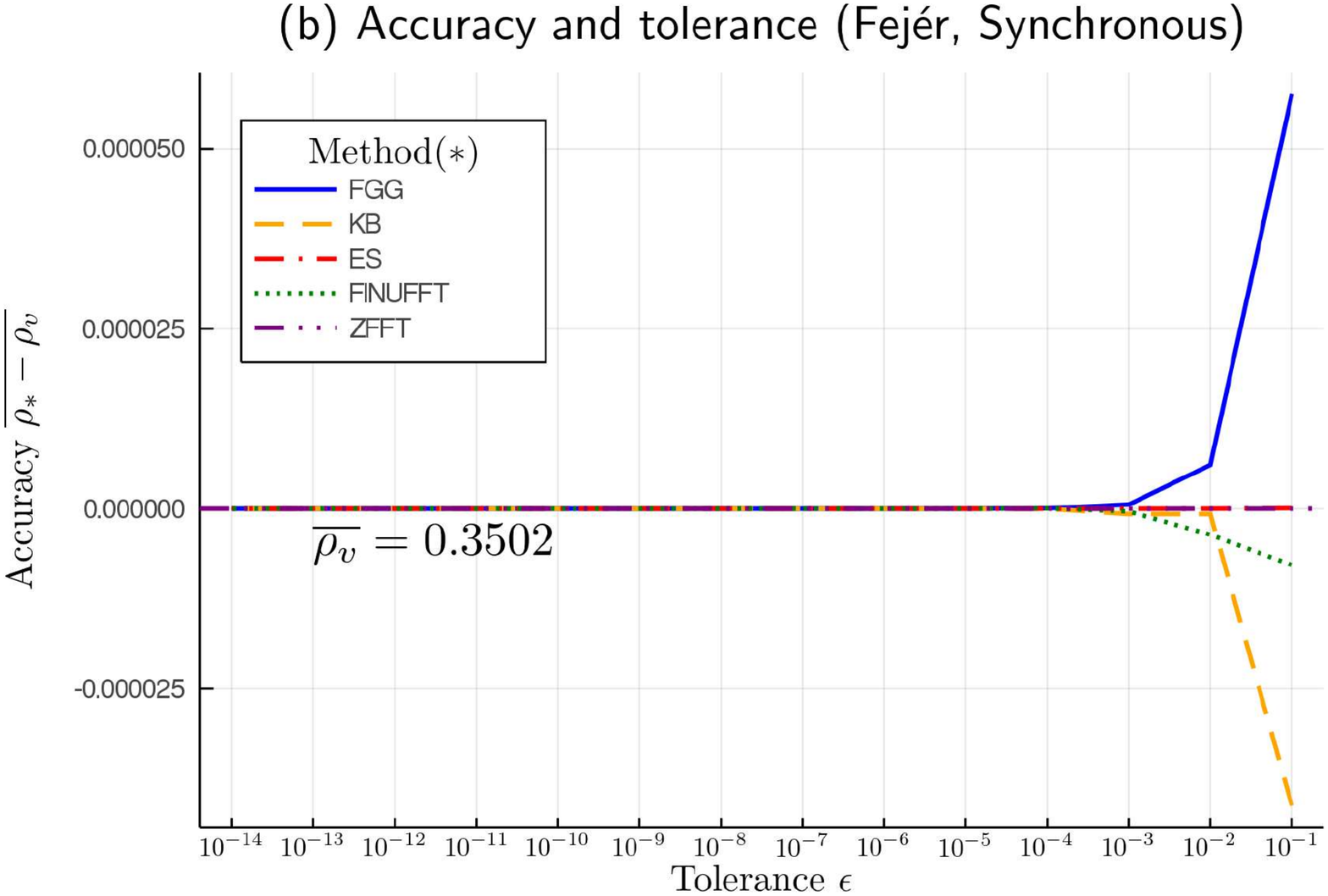}} \\
    \subfloat{\label{fig:AccSynDS:c}\includegraphics[width=0.48\textwidth]{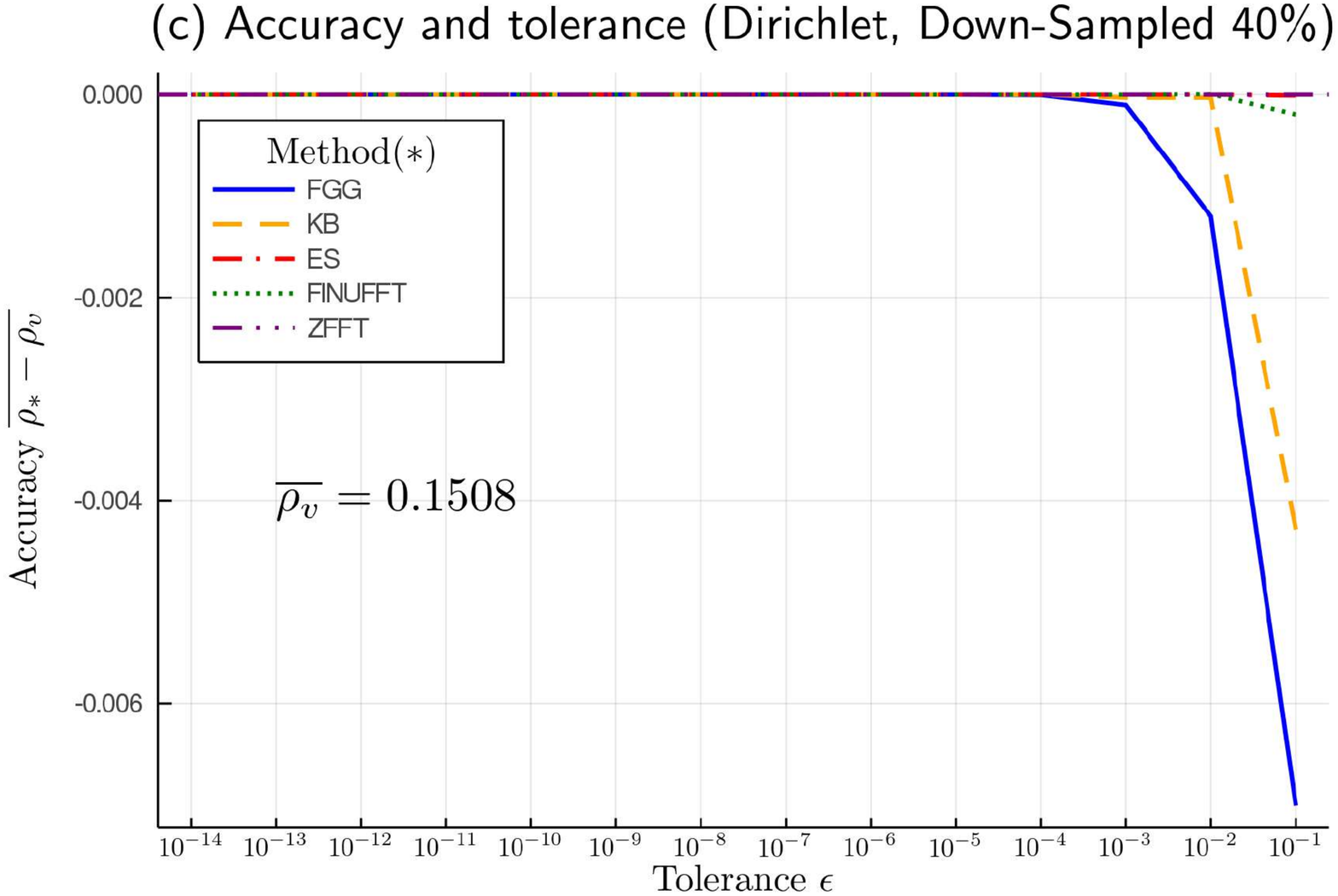}}
    \subfloat{\label{fig:AccSynDS:d}\includegraphics[width=0.48\textwidth]{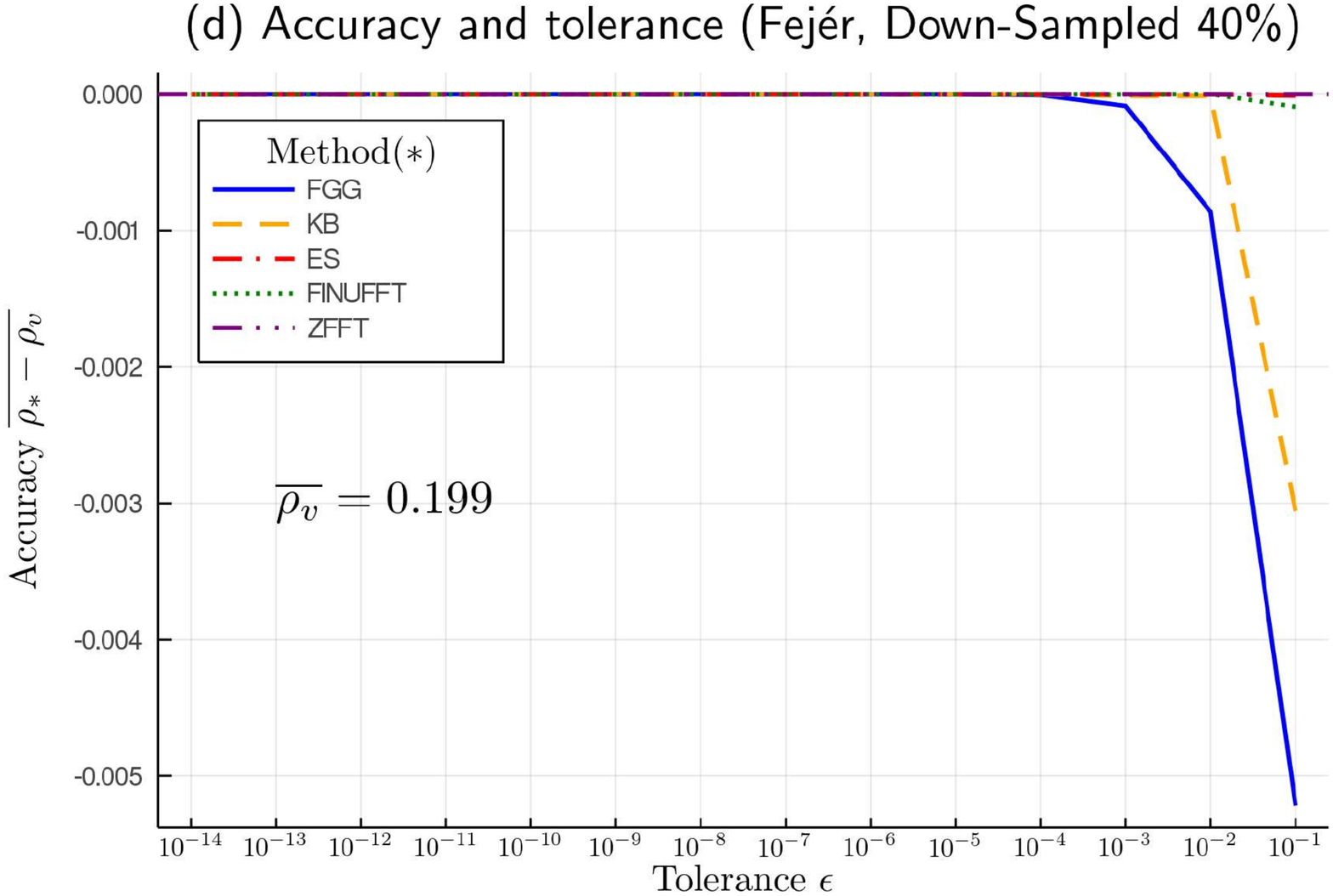}}  \\
    \subfloat{\label{fig:AccSynDS:e}\includegraphics[width=0.48\textwidth]{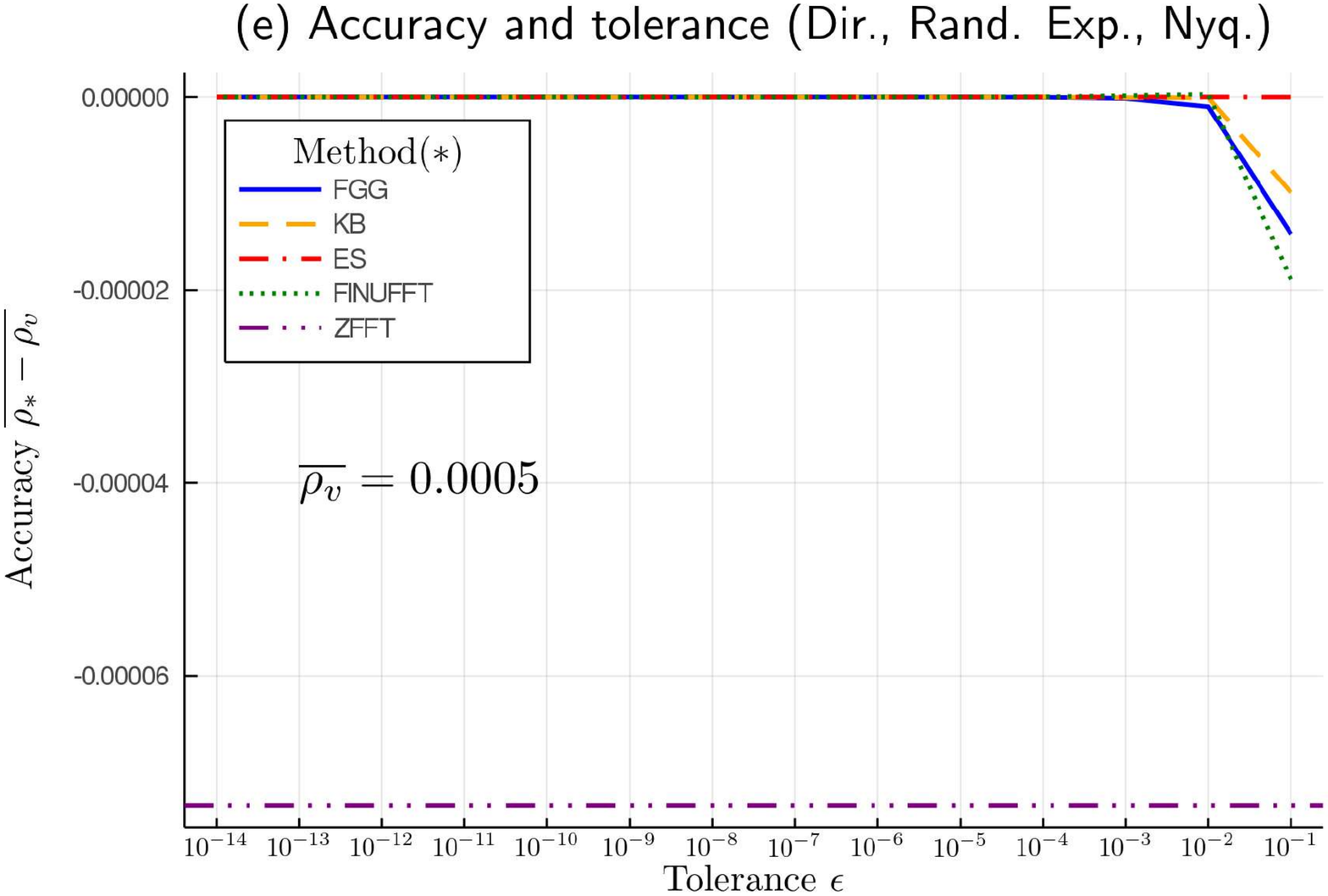}}
    \subfloat{\label{fig:AccSynDS:f}\includegraphics[width=0.48\textwidth]{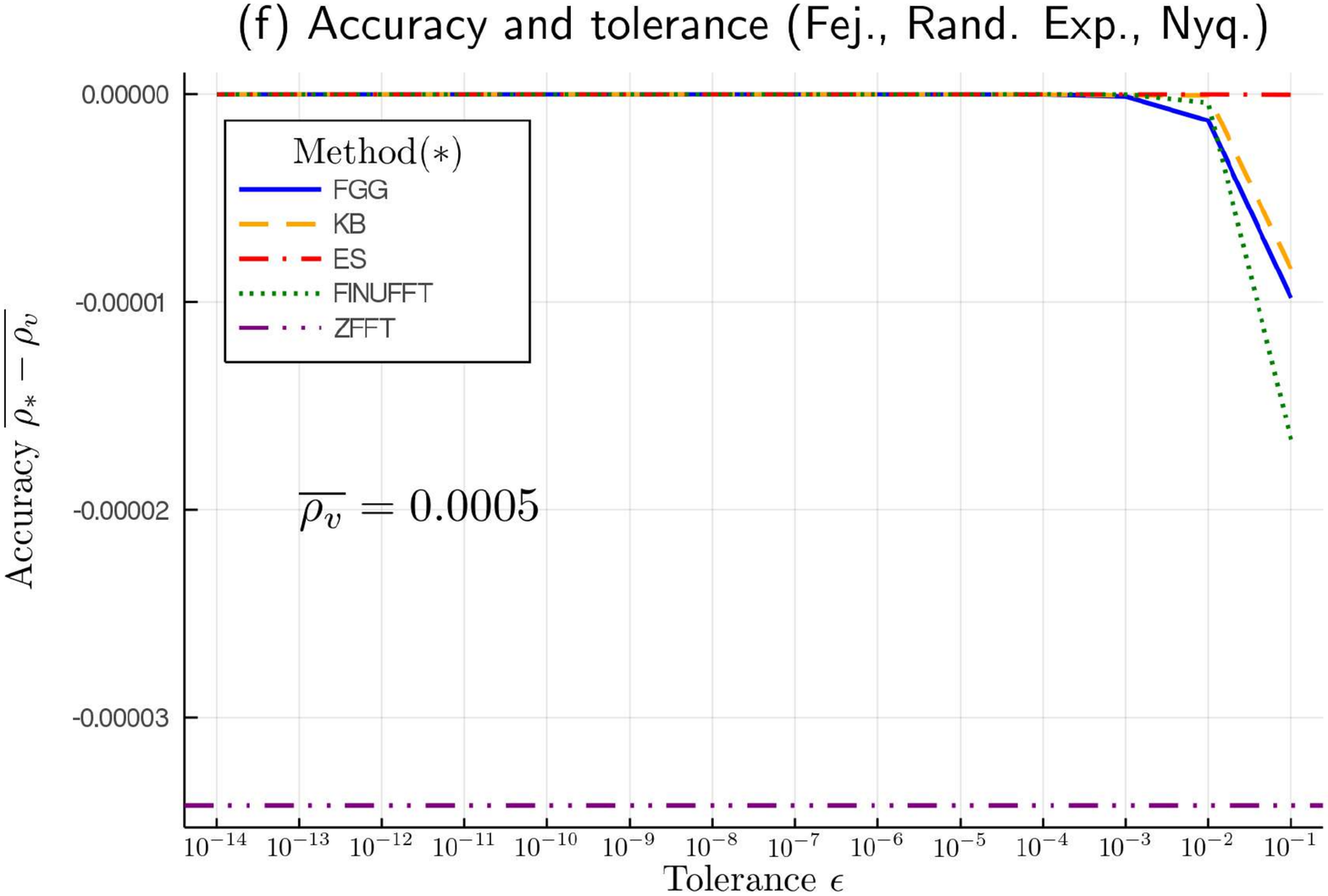}}
\caption{\scriptsize{We investigate the accuracy of various fast Fourier methods as a function of the error tolerance $\epsilon$ for various simulation settings. Accuracy is measured as the difference between the estimates of the various fast Fourier methods $\rho_*$ and the estimate using the vectorised implementation (CFT) $\rho_v$ averaged over the 100 replications. The average correlation estimate from the vectorised implementation is provided as an inset in each figure. The base-line price process is the synchronous Geometric Brownian Motion with $10^4$ data points simulated using \cref{algo:GBM}. The three simulation cases are: the synchronous case (a) and (b), the missing data representation (c) and (d), and the arrival time representation (e) and (f). The missing data representation is down-sampled by $40\%$ while the arrival time representation is sampled with an exponential inter-arrival time (Rand. Exp.) with mean 30 and 45 for the first and second price path respectively. The fast Fourier methods investigated are: the fast Gaussian gridding (FGG - blue line), the Kaiser-Bessel kernel (KB - orange dashes), the exponential of semi-circle with our naive implementation (ES - red dash-dots) and the FINUFFT implementation (FINUFFT - green dots) and finally, the zero-padded FFT (ZFFT - purple dash-dot-dot). The accuracy is tested for the two basis kernels: Dirichlet (Dir.) and Fej\'{e}r (Fej.). Furthermore, the Nyquist frequency (Nyq.) is used for all the correlation estimate, therefore it must be noted that for the arrival time representation, $N$ changes for each replication. We see firstly, provided the tolerance $\epsilon < 10^{-4}$, the NUFFT methods can accurately recover the estimates, secondly, the divergence from the CFT implementation when $\epsilon \geq 10^{-4}$ may simply be an artefact of random errors from the lack of precision requested, and finally, the zero-padded FFT fails for arrival time representation. The figures can be recovered using the Julia script file \href{https://github.com/CHNPAT005/PCEPTG-MM-NUFFT/blob/master/Scripts/Accuracy/AccSynDS}{AccSynDS} and \href{https://github.com/CHNPAT005/PCEPTG-MM-NUFFT/blob/master/Scripts/Accuracy/AccRE}{AccRE} on the GitHub resource \cite{PCEPTG2020CODE}.}}
\label{fig:AccSynDS}
\end{figure*}

\begin{figure*}[p]
\centering
    \subfloat{\label{fig:AccRE:a}\includegraphics[width=0.48\textwidth]{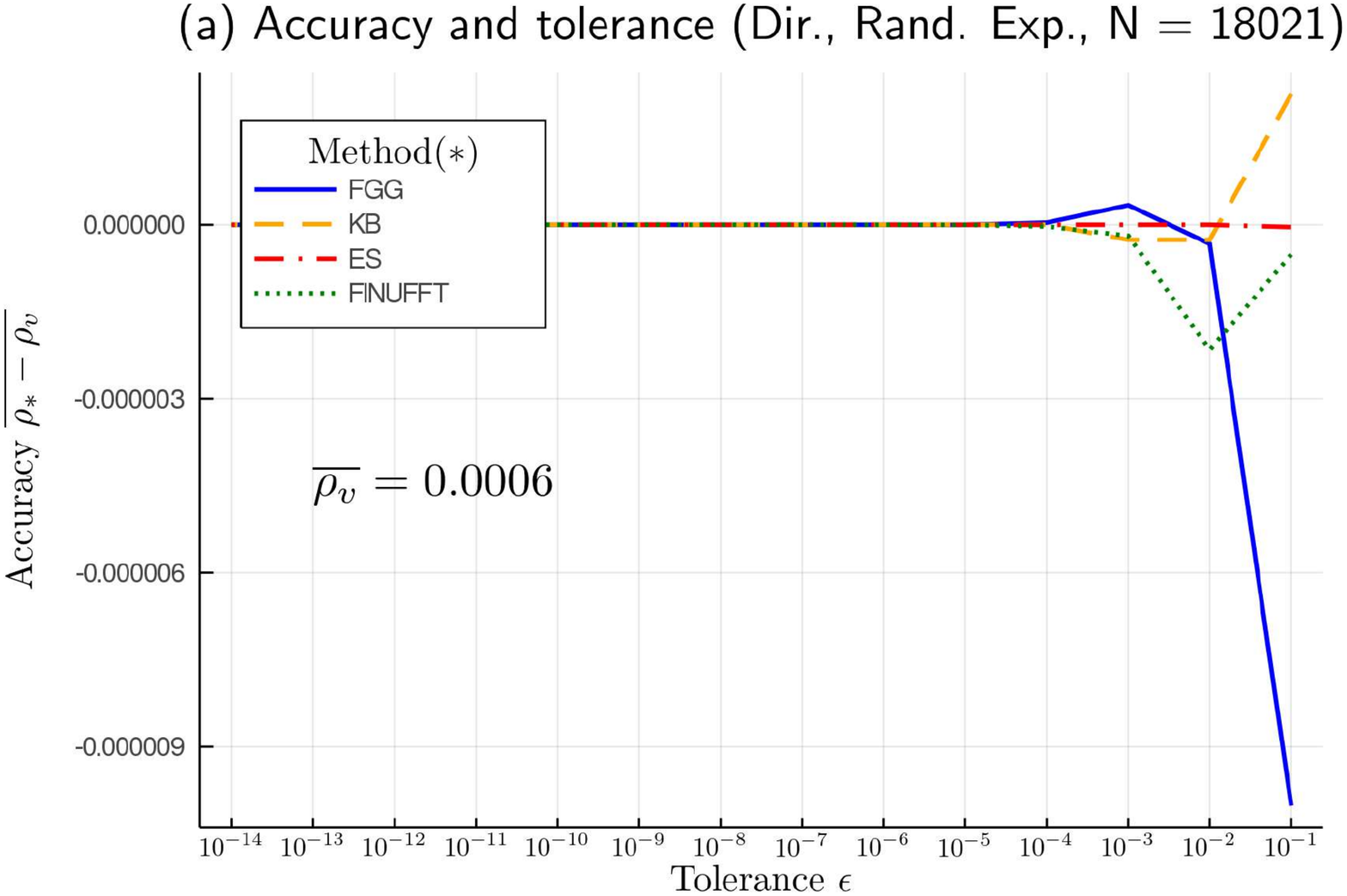}}
    \subfloat{\label{fig:AccRE:b}\includegraphics[width=0.48\textwidth]{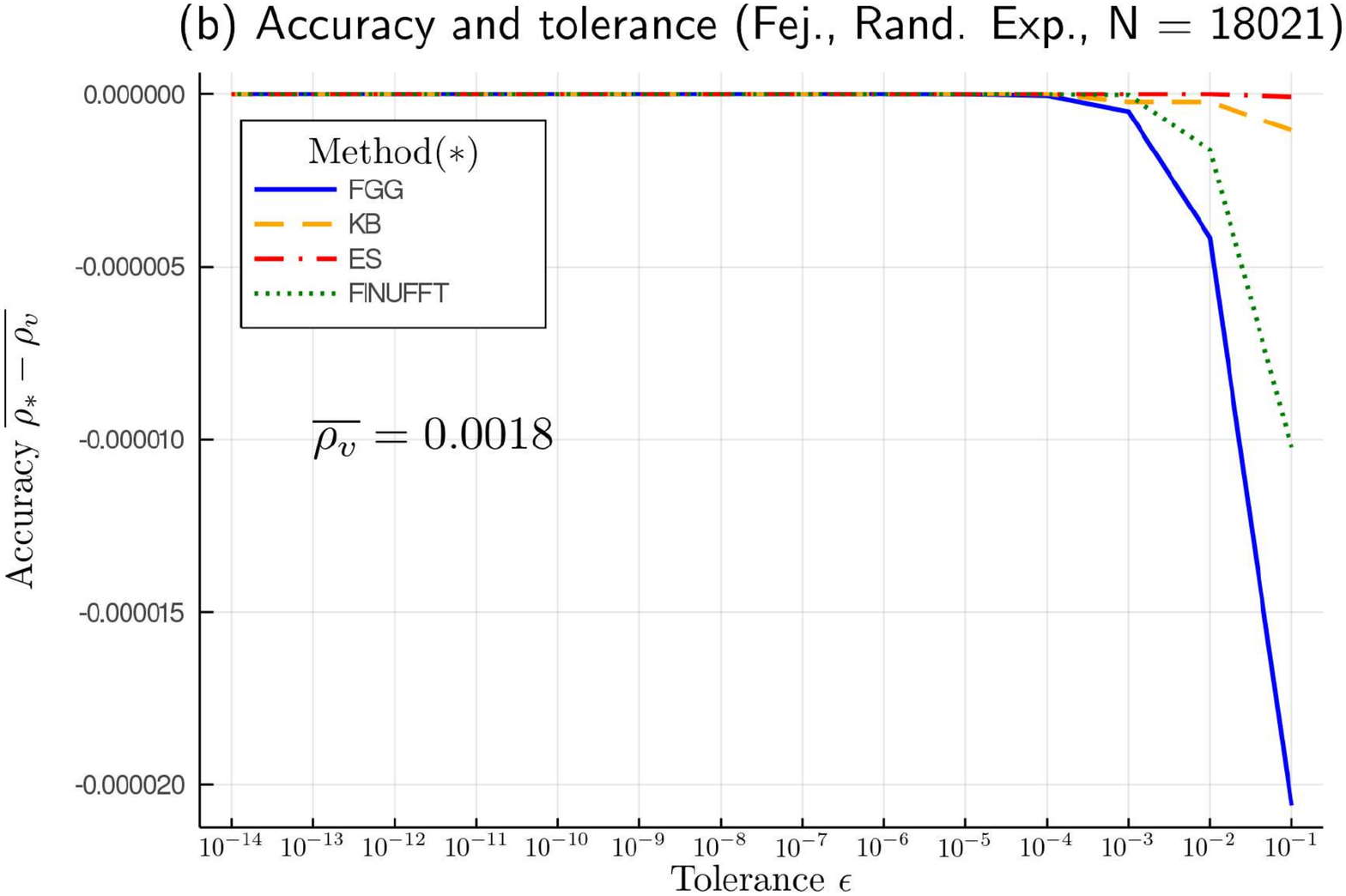}} \\
    \subfloat{\label{fig:AccRE:c}\includegraphics[width=0.48\textwidth]{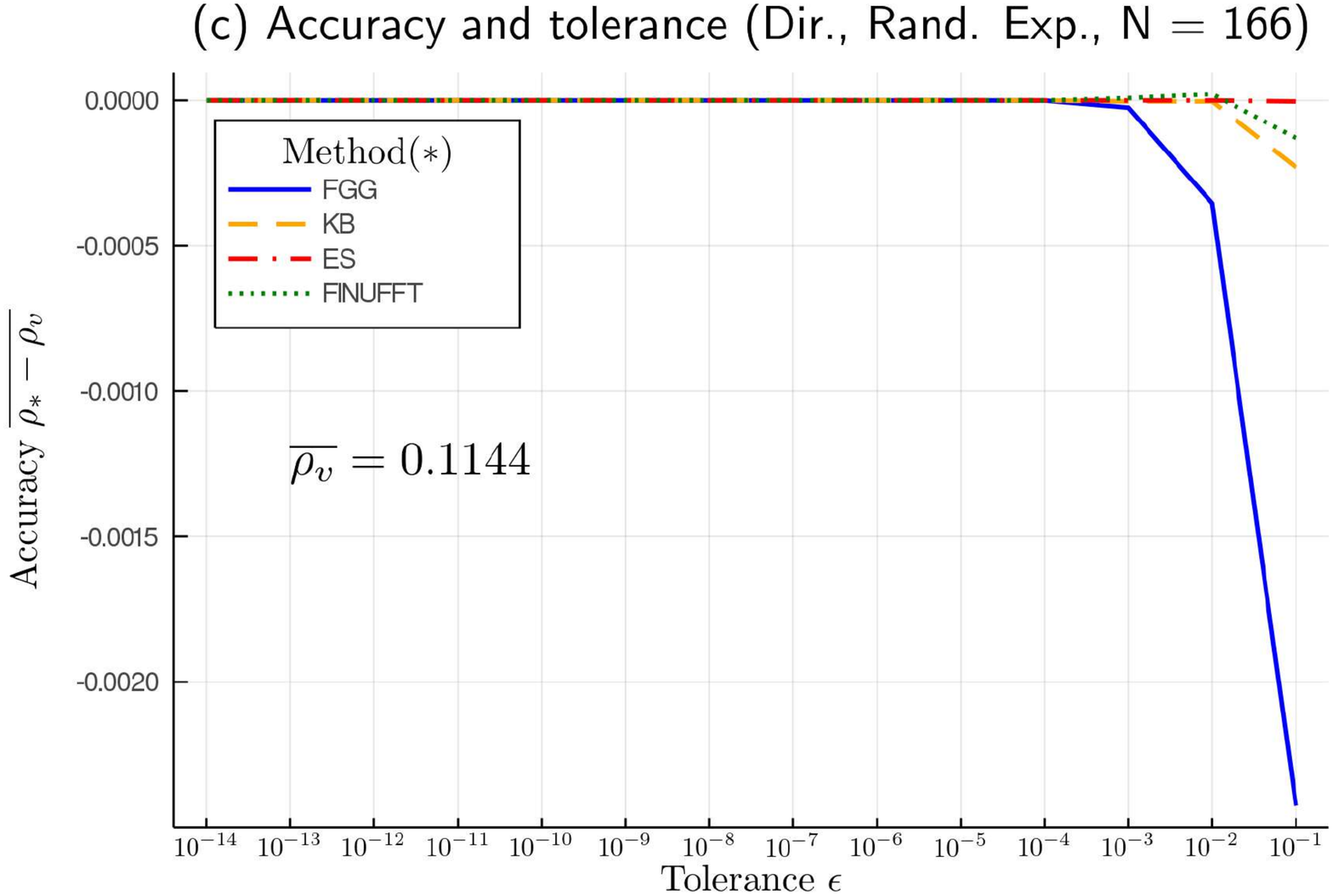}}
    \subfloat{\label{fig:AccRE:d}\includegraphics[width=0.48\textwidth]{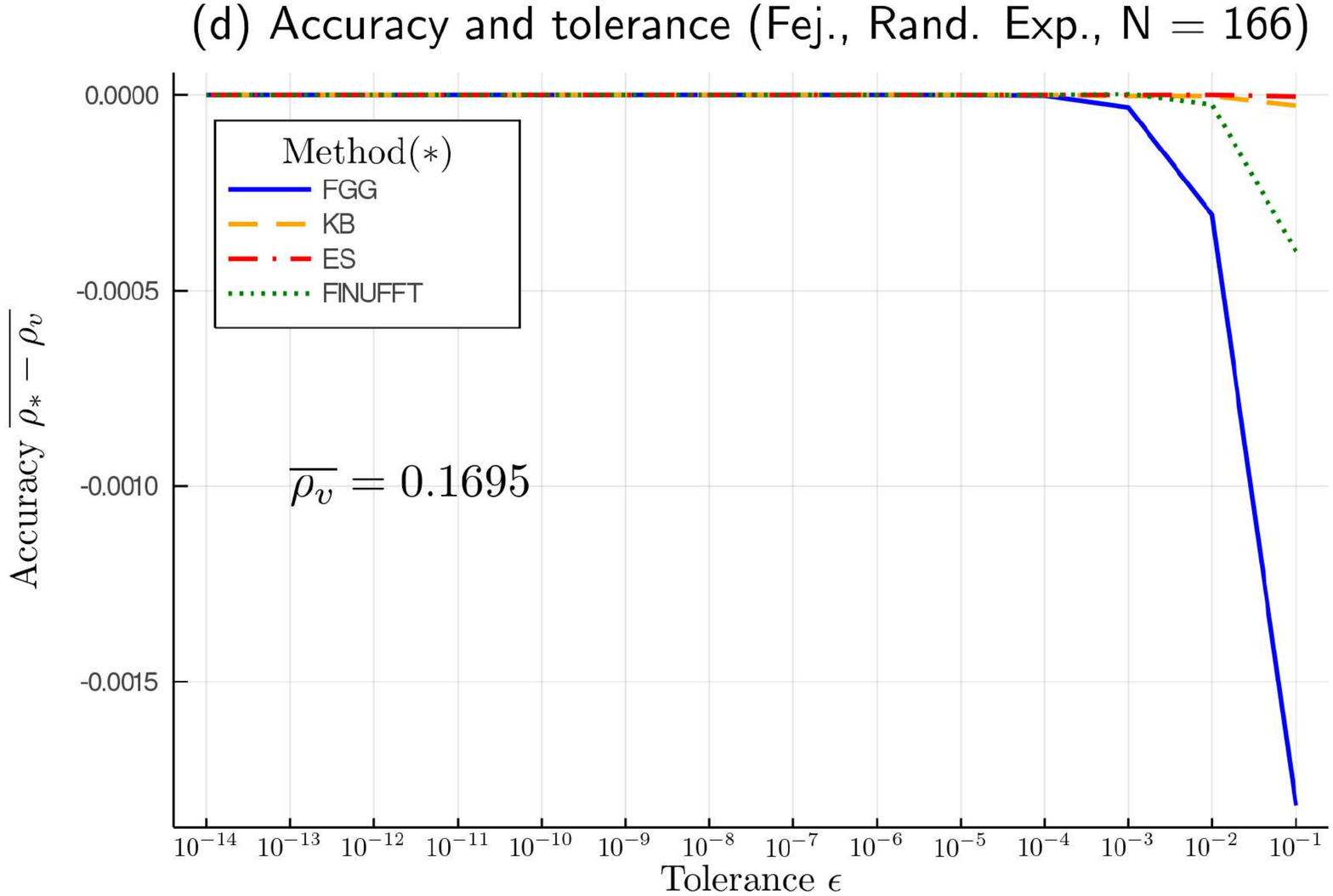}}  \\
    \subfloat{\label{fig:AccRE:e}\includegraphics[width=0.48\textwidth]{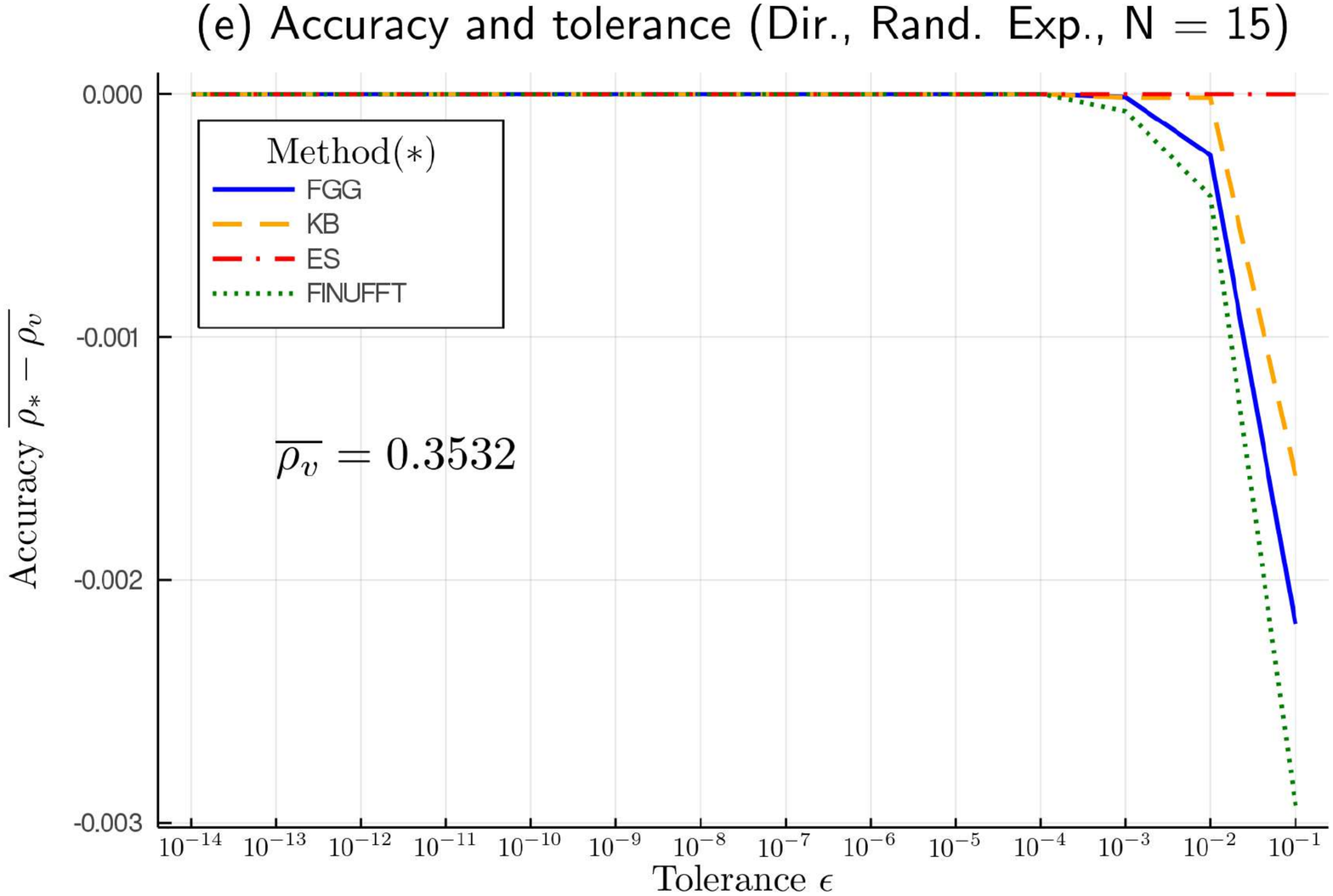}}
    \subfloat{\label{fig:AccRE:f}\includegraphics[width=0.48\textwidth]{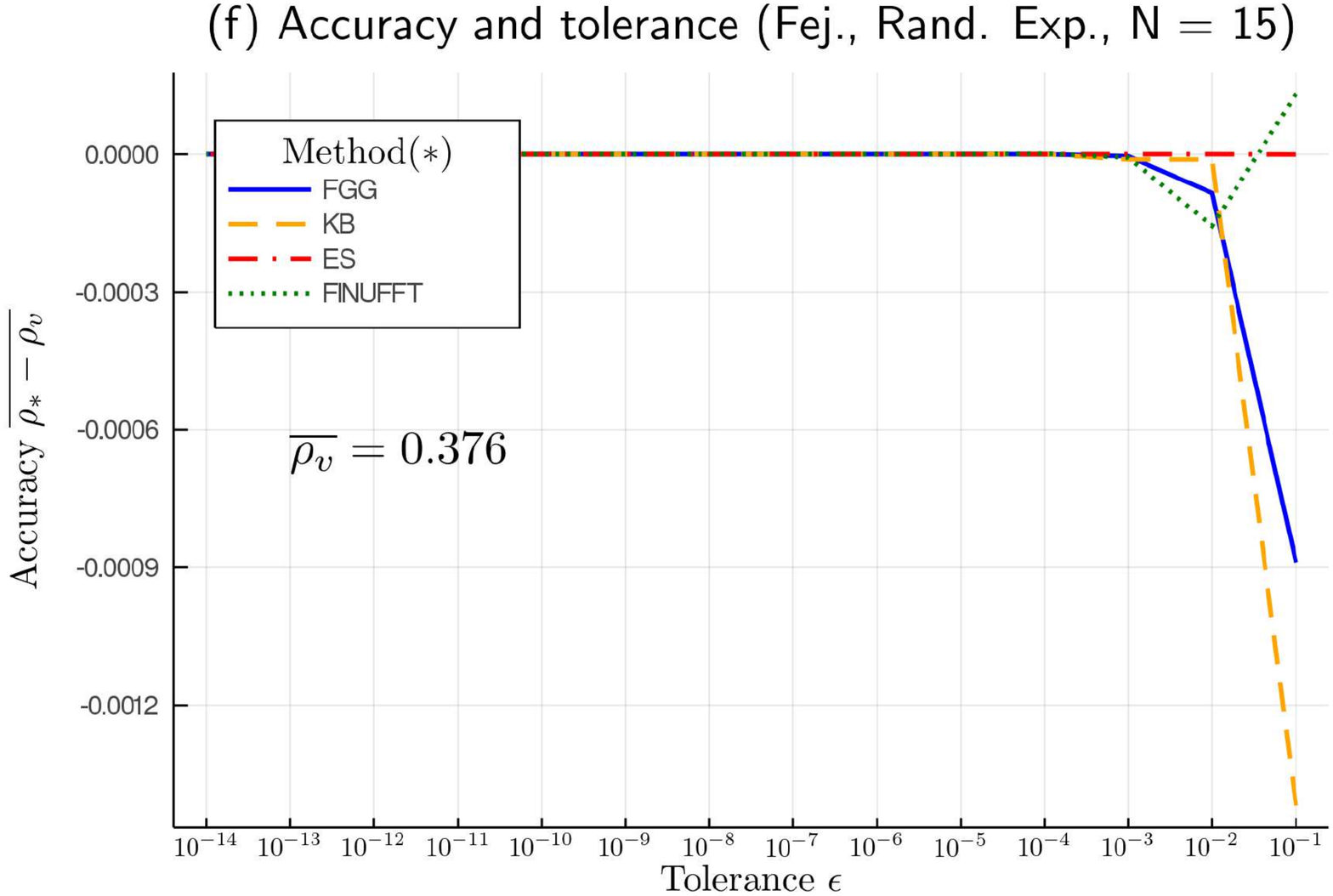}}
\caption{We investigate the inter-play between kernel averaging and time-scale averaging. We plot the accuracy of various fast Fourier methods as a function of the error tolerance $\epsilon$ for various choices of $N$. Accuracy is measured as the difference between the estimates of the various fast Fourier methods $\rho_*$ and the estimate using the vectorised implementation (CFT) $\rho_v$ averaged over the 100 replications. The average correlation estimate from the vectorised implementation is provided as an inset in each figure. The base-line price process is the synchronous Geometric Brownian Motion with $10^4$ data points simulated using \cref{algo:GBM}. The synchronous price paths are then sampled with an exponential inter-arrival time (Rand. Exp.) with mean 30 and 45 for the first and second price path respectively to create the arrival time representation of asynchrony. The fast Fourier methods investigated are: the fast Gaussian gridding (FGG - blue line), the Kaiser-Bessel kernel (KB - orange dashes), the exponential of semi-circle with our naive implementation (ES - red dash-dots) and the FINUFFT implementation (FINUFFT - green dots). The accuracy is tested for the two basis kernels: Dirichlet (Dir.) and Fej\'{e}r (Fej.). We see that there is no clear relation between the two types of averaging, rather the divergence for higher tolerance levels seems be an artefact of errors arising from the lack of precision requested. The figures can be recovered using the Julia script file \href{https://github.com/CHNPAT005/PCEPTG-MM-NUFFT/blob/master/Scripts/Accuracy/AccRE}{AccRE} on the GitHub resource \cite{PCEPTG2020CODE}.}
\label{fig:AccRE}
\end{figure*}

We have demonstrated the merit of fast Fourier techniques in terms of speed. We now investigate the accuracy of the fast Fourier methods and conditions when they fail. Moreover, we need to find out what level of numerical accuracy is required to ensure that the Fourier coefficients evaluated using NUFFT techniques \cref{eq:Der:17} can recover the same estimates \cref{eq:Der:2,eq:Der:3} using the direct evaluation of \cref{eq:Der:FC}. This is done by testing the fast Fourier methods on the synchronous case, the missing data representation, and the arrival time representation. We then look at the inter-relation between two types of averaging: (i) kernel averaging---the convolution step in the NUFFT algorithms, and (ii) time-scale averaging---the choice of $N$ in the Malliavin-Mancino estimator. Ensuring that the NUFFT methods can correctly recover estimates for different choices of $N$ allows us to quickly and accurately investigate various time scales. Finally, we compare the MSE and bias of the estimator under asynchronous sampling using the direct evaluation of \cref{eq:Der:FC} against NUFFT methods of evaluating the Fourier coefficients \cref{eq:Der:17}.

The setting for the following two experiments are as follows: a bivariate Geometric Brownian Motion with $n = 10^4$ is simulated using \cref{algo:GBM}. The daily parameters for the GBM are: $\mu_1 = 0.01$, $\mu_2 = 0.01$, $\sigma^2_1 = 0.1$, $\sigma^2_2 = 0.2$ and $\rho^{12} = 0.35$. We set $\Delta t = \frac{1}{86400}$, therefore each unit interval can be thought of as a second in Calendar time. From the synchronous case we create the missing data representation by randomly removing $40\%$ of data points from each path. The arrival time representation is achieved by sampling each price path with an exponential inter-arrival time with mean 30 and 45 for the first and second price paths respectively.

We measure accuracy as the difference between the estimates from the fast Fourier methods and the estimates from the vectorised implementation averaged over 100 replications. This allows us to directly see if the estimates obtained using the fast Fourier methods recover the same estimates using the direct evaluation.

\Cref{fig:AccSynDS} investigates the accuracy for three scenarios: the synchronous case, the missing data representation, and the arrival time representation. The fast Fourier methods investigated are: the zero-padded FFT (ZFFT), the fast Gaussian gridding (FGG), the Kaiser-Bessel kernel (KB), the exponential of semi-circle with our naive implementation (ES) and the FINUFFT implementation (FINUFFT). First, the NUFFT methods can accurately recover the estimates provided the tolerance $\epsilon < 10^{-4}$. Furthermore, we see that the ES kernel (red dashes) recover the correct estimates for $\epsilon = 10^{-1}$. This is not a property of the exponential of semi-circle kernel as the FINUFFT implementation diverges from $\epsilon = 10^{-4}$ (due to their more lenient choice of $\omega$). Rather, this is a result of our choice of $M_{sp}$ for the ES implementation to ensure the requested tolerance is always strictly met. Concretely, this means each source point must be spread in each direction for a minimum of $M_{sp}=4$ grid points for the Gaussian kernel, $M_{sp}=3$ grid points for the Kaiser-Bessel kernel, and $M_{sp}=3$ grid points for the exponential of semi-circle to recover the vectorised estimate.\footnote{The $M_{sp}$ requirement is calculated based on $\epsilon = 10^{-4}$ for the Gaussian and Kaiser-Bessel kernel and $\epsilon = 10^{-1}$ for the ES kernel.} Second, the non-uniform FFT methods diverge away from the vectorised implementation when tolerance $\epsilon \geq 10^{-4}$. There is no clear pattern in the divergence for the various kernels, therefore it seems the errors are a simple artefact arising from the lack of precision requested in $\boldsymbol{F}(dp_i)$. Finally, the zero-padded FFT recovers the correct estimate for the synchronous case and missing data representation. More importantly, it fails for the arrival time representation because of the shifting of time points (see \Cref{fig:FFTZP_demontration}). Non-uniform FFT methods overcome this through a convolution and deconvolution step to correct the effects of shifting the points to a uniform grid by trying to preserve the power spectrum. 

Previously in \Cref{fig:AccSynDS}, the arrival time representation had $N$ changing for each replication. \Cref{fig:AccRE} we fix three cases of $N$ and measure the accuracy using the arrival time representation (with the same parameters as before) to better understand the relationship between the kernel averaging and the time-scale averaging. The first $N$ is computed as the minimum Nyquist frequency across the 100 replications resulting in $N = 18,021$. The second $N$ is computed based on the smallest {\it average} sampling interval resulting in $N = \lfloor \frac{10,000}{30 \times 2} \rfloor = 166$. Finally, the last $N$ is chosen to be arbitrarily small subject to the condition that the corresponding $M_r$ is larger than $\omega$ for $\epsilon = 10^{-14}$.\footnote{This is to ensure the up-sampled grid is larger than the total spreading width.} We pick $N = 15$ for the final case. The zero-padded FFT is excluded because the implementation only computes the case when $N$ is the Nyquist frequency. We see that there is no clear relation between the two types of averaging. For any choice of $N$, we can recover the vectorised estimate provided the tolerance $\epsilon < 10^{-4}$. There is no clear pattern in the divergence for the various kernels and the lack of accuracy in the estimates are due to the lack of precision requested in $\boldsymbol{F}(dp_i)$.

\begin{figure*}[htb]
    \centering
    \subfloat{\includegraphics[width=0.48\textwidth]{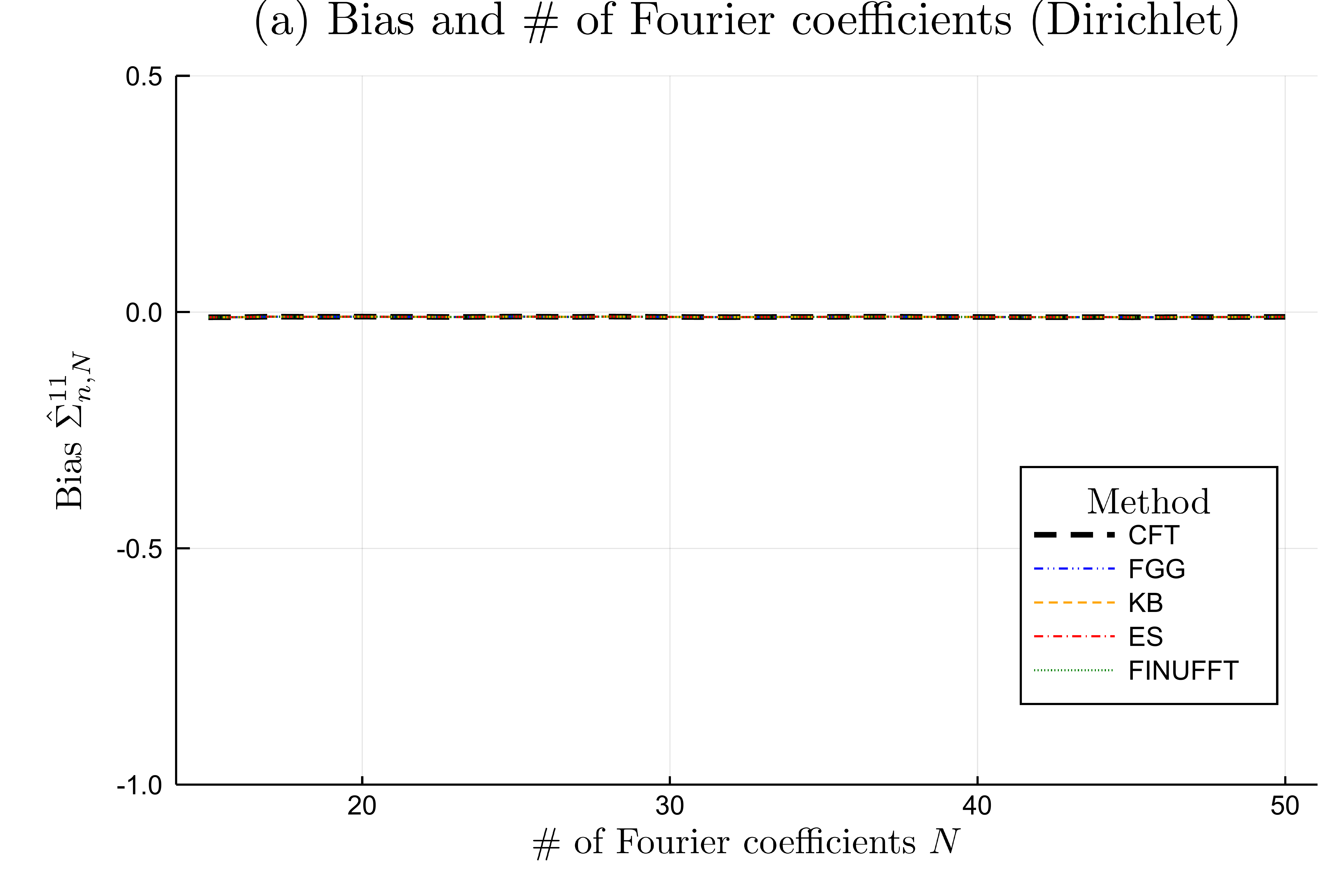}}
    \subfloat{\includegraphics[width=0.48\textwidth]{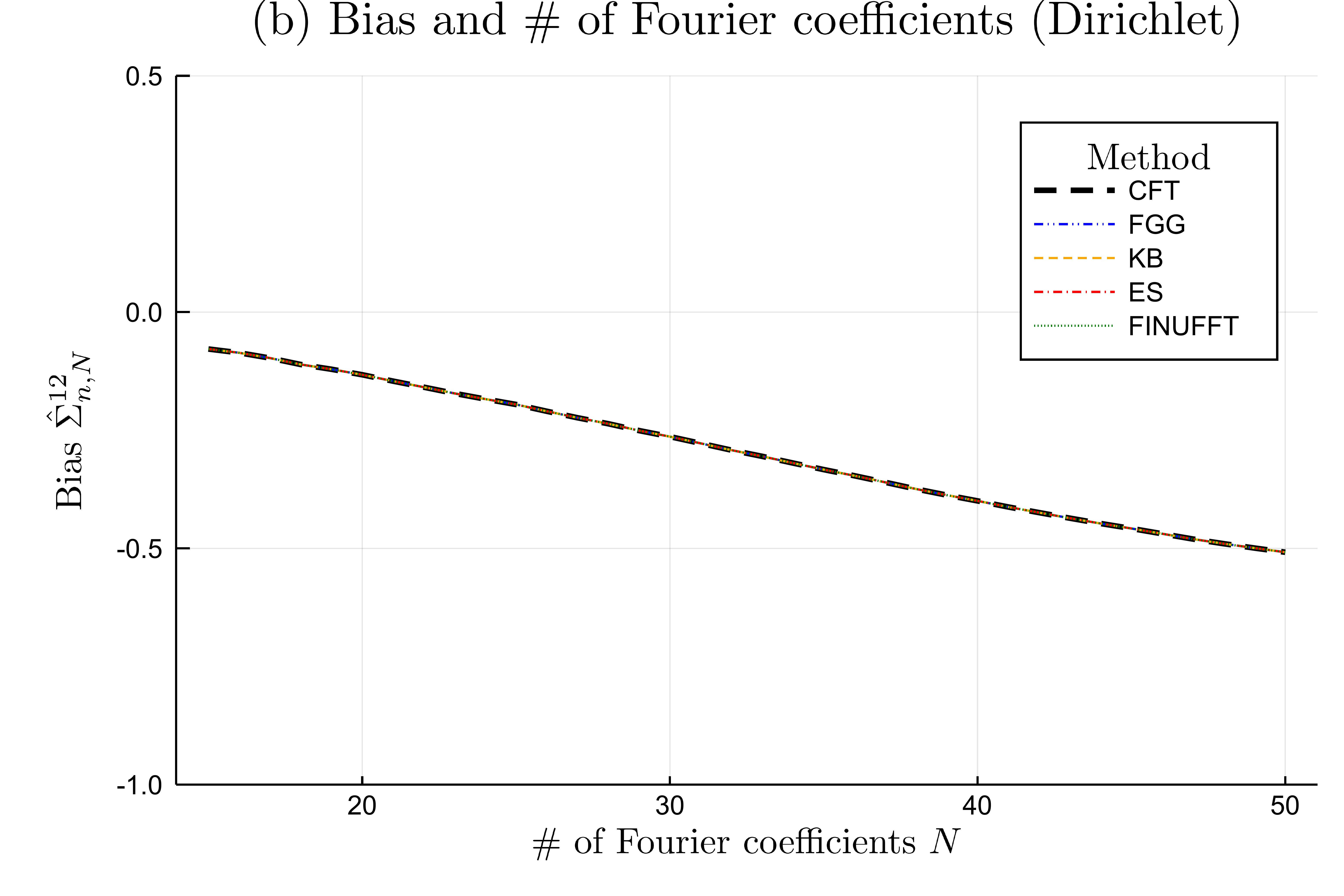}} \\
    \subfloat{\includegraphics[width=0.48\textwidth]{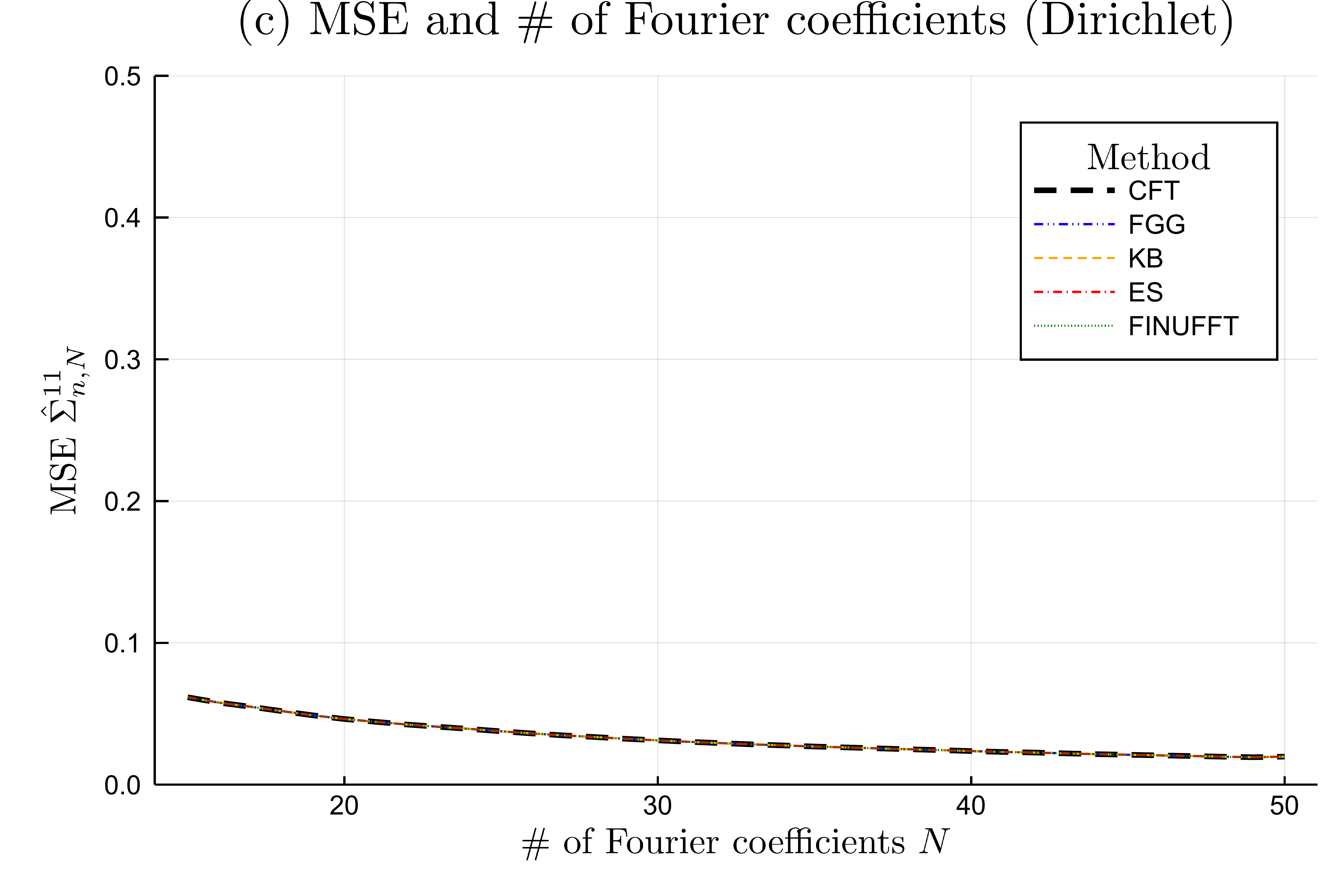}}
    \subfloat{\includegraphics[width=0.48\textwidth]{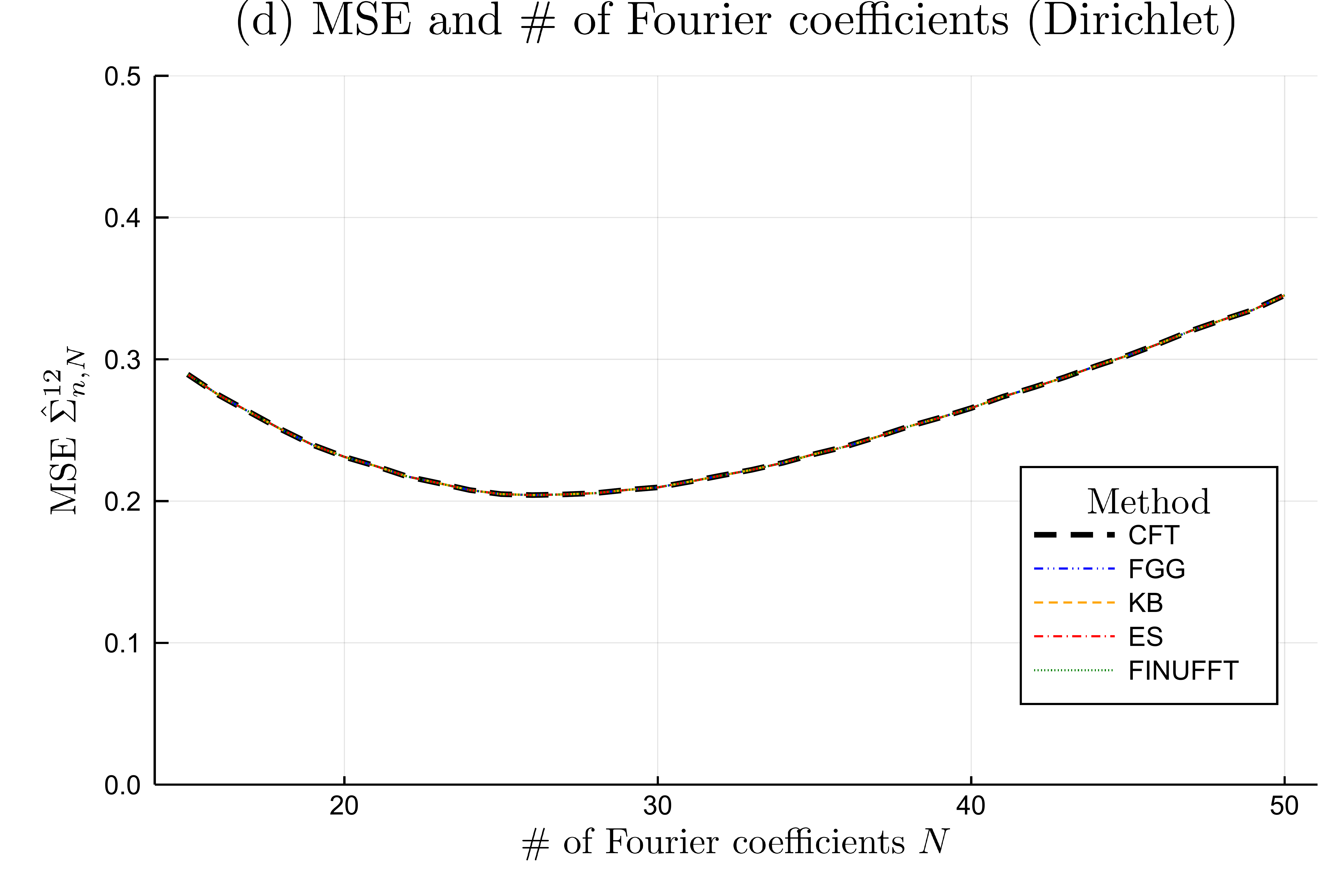}}
    \caption{Here we show the bias and MSE of the integrated covariance as a function of the number of Fourier coefficients using the Dirichlet representation (the Fej\'{e}r can be found in \Cref{fig:MSEBias_Fej}). The base-line price process is a synchronous GBM with $n=100$ data points. The missing data representation is induced here using the regular non-synchronous trading \cite{MRS2017}, where the second asset is observed at every second trade of asset one. The methods investigated are: the vectorised implementation (CFT - black dashes), the fast Gaussian gridding (FGG - blue dash-dot-dot), the Kaiser-Bessel kernel (KB - orange dashes), the exponential of semi-circle with our naive implementation (ES - red dash-dots) and the FINUFFT implementation (FINUFFT - green dots). The NUFFT methods are computed using the default $\epsilon = 10^{-12}$. We see that the fast Fourier methods recover the same bias and MSE results as the vectorised implementation and are consistent with the results obtained by \cite{MRS2017}. The figures can be recovered using the Julia script file \href{https://github.com/CHNPAT005/PCEPTG-MM-NUFFT/blob/master/Scripts/Accuracy/MSEBias}{MSEBias} on the GitHub resource \cite{PCEPTG2020CODE}.}
\label{fig:MSEBias}
\end{figure*}

\Cref{fig:MSEBias} compares the MSE and bias of the integrated covariance as a function of the number of Fourier coefficients for the vectorised implementation and the NUFFT methods (with default $\epsilon = 10^{-12}$). A bivariate GBM with $n=100$ is simulated with the same parameters as before (except $\Delta t = 1/n$). Asynchrony is induced using a special case of the missing data representation: the regular non-synchronous trading used by \cite{MRS2017}. Here the second asset is observed at every second trade of asset one. The methods investigated are: the vectorised implementation (CFT), the fast Gaussian gridding (FGG), the Kaiser-Bessel kernel (KB), the exponential of semi-circle with our naive implementation (ES) and the FINUFFT implementation (FINUFFT). We see that the NUFFT methods recover the same bias and MSE results as the vectorised implementation. When asynchrony is introduced the integrated volatility $\hat{\Sigma}_{n,N}^{11}$ (see \cref{eq:Der:2,eq:Der:3}) does not present a bias for all values of $N$, but the MSE is still large for small values of $N$ due to the variance of the estimator \cite{MRS2017}. However, we have an increase in bias for larger $N$ (smaller time-scales) with co-volatility $\hat{\Sigma}_{n,N}^{12}$. This is a result of the Epps effect (the decay in corrections as time-scales decrease). To remove this effect, a smaller $N$ (larger time-scales) must be chosen. 

Mancino et al. \cite{MRS2017} suggest picking $N$ by minimizing the MSE for an optimal bias and variance tradeoff.\footnote{We also performed a sensitivity analysis confirming that the NUFFT methods and vectorised implementation correctly recover the integrated covariance. This is confirmed through the linear relationship by plotting the estimate against the true integrated covariance for a range of values (see \Cref{fig:Sensitivity}).} The speed of convergence with respect to the degree of asynchrony of the Malliavin-Mancino estimator was then investigated with the MSE criterion in mind \cite{Chen2019,MRS2017,PHL2016}. Picking $N$ to minimise the MSE could be seen as inadvertently assuming that there is some latent model generating the data with some appropriate limiting properties, and that we need only be concerned about an estimators deviation from this implied latent model. This can potentially lead one to inadvertently average away empirically important sources of the Epps effect. Allowing $N$ to be chosen to isolate a particular time-scale by its implied choice of $\Delta t$ is weaker but it can allow us to disentangle genuine and statistical sources of the Epps effect (see \Cref{subsec:sim} for further details). 

\section{Correlations and time-scale averaging} \label{sec:scale}

\subsection{Simulated data}\label{subsec:sim}

We now consider the relationship between the time-scales and the Epps effect \cite{EPPS1979}. Concretely, we demonstrate how different time-scales can be investigated with the Malliavin-Mancino estimator through the choice of $N$. Specifically, by using $\Delta t = \frac{T}{2N + 1}$ (see \Cref{subsec:Avescale}). This follows the insights introduced by Ren\`{o} \cite{RENO2001}, and Precup and Iori \cite{PI2007} to investigate the Epps effect. Precup and Iori were able to demonstrate that the higher the level of asynchrony, the larger the drop in correlation for the Epps effect. This is demonstrated in \Cref{fig:AccSynDS} with the average correlation provided as insets for varying level of asynchrony. Additionally, Ren\`{o} was able to demonstrate the Epps effect as a function of sampling frequency under the arrival time representation of asynchrony. Demonstrated here in \Cref{fig:AccRE} with the average correlation provided as insets for various $N$.
\begin{figure*}[p]
    \centering
    \subfloat{\label{fig:avescale:a}\includegraphics[width=0.48\textwidth]{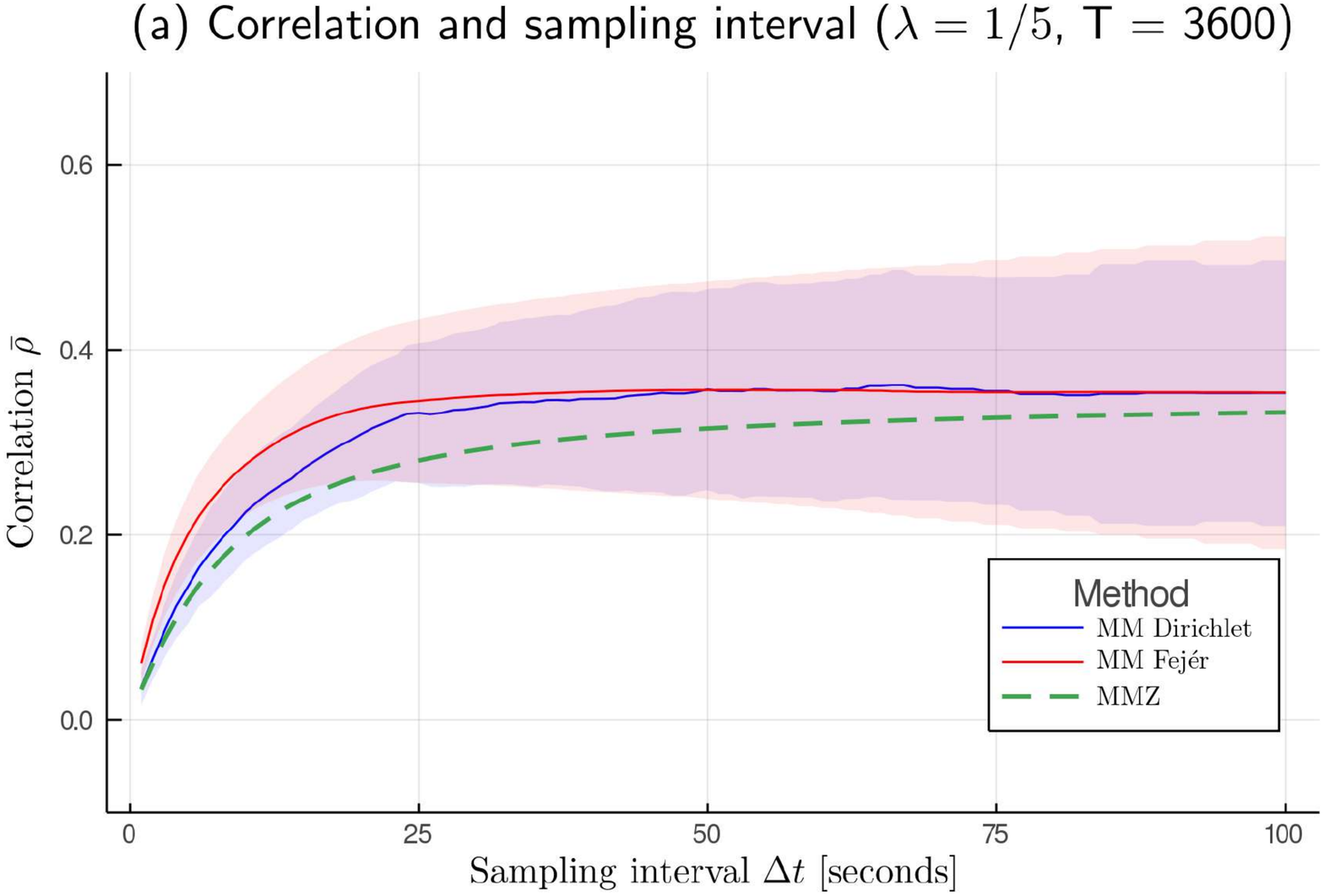}}
    \subfloat{\label{fig:avescale:b}\includegraphics[width=0.48\textwidth]{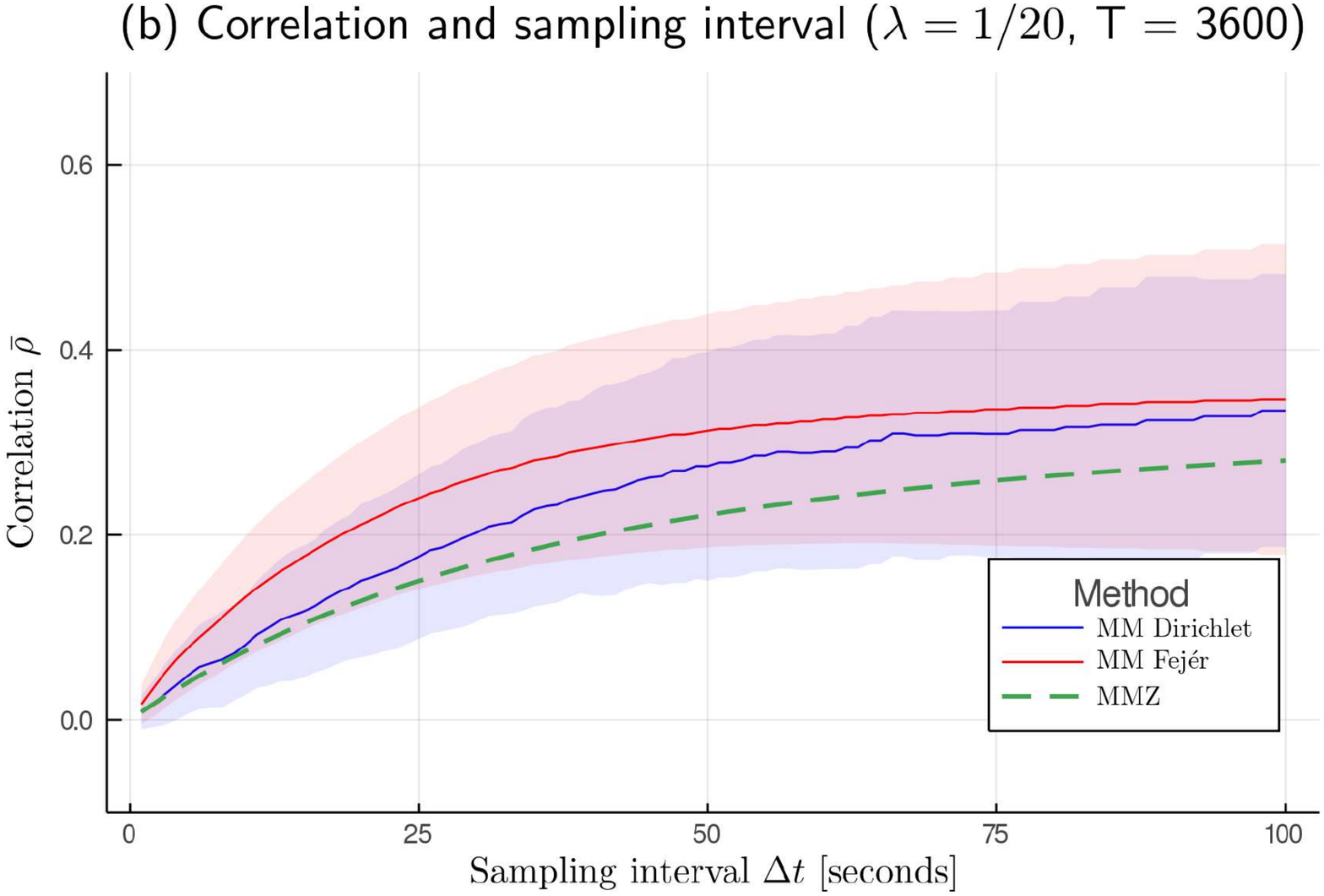}} \\
    \subfloat{\label{fig:avescale:c}\includegraphics[width=0.48\textwidth]{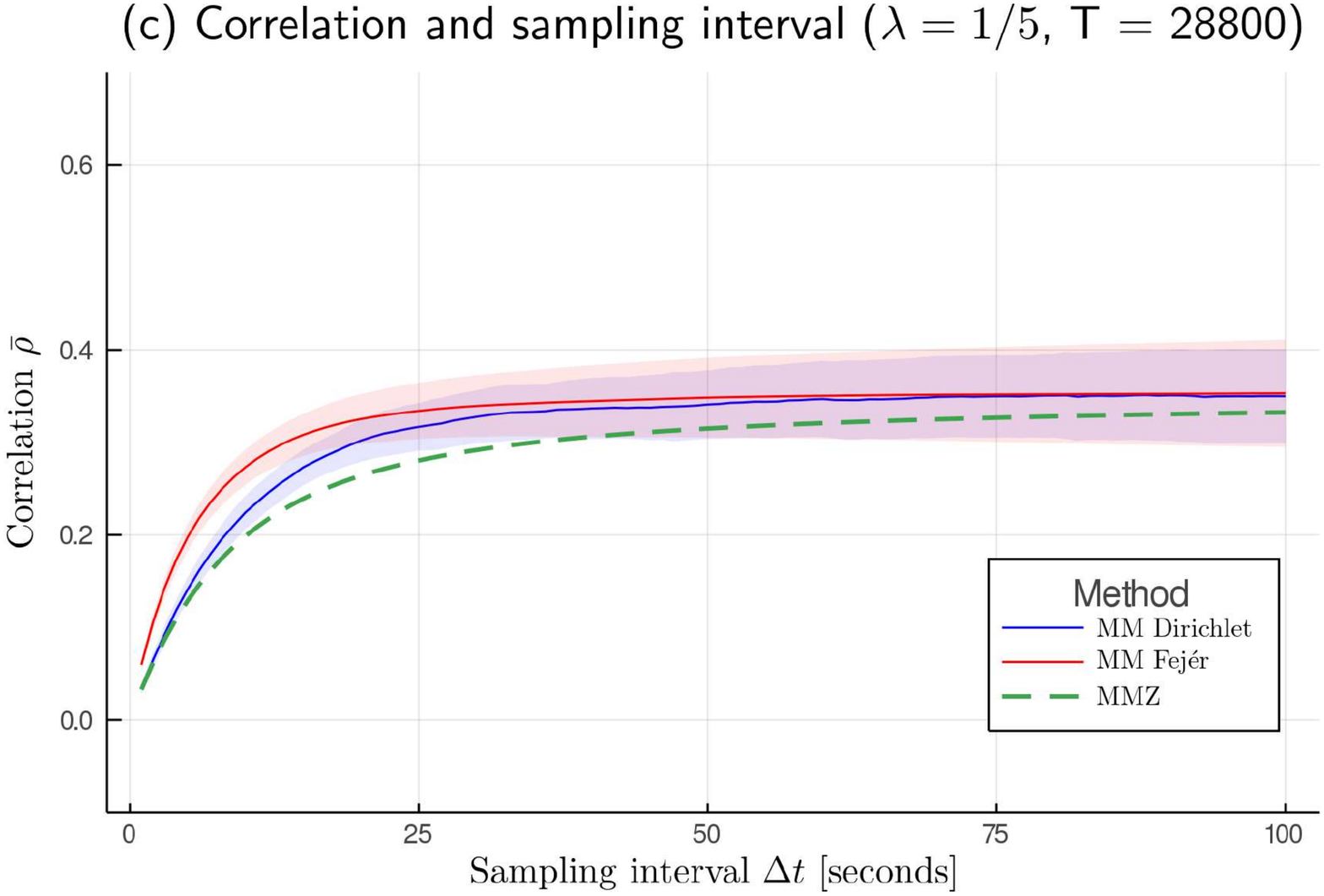}}
    \subfloat{\label{fig:avescale:d}\includegraphics[width=0.48\textwidth]{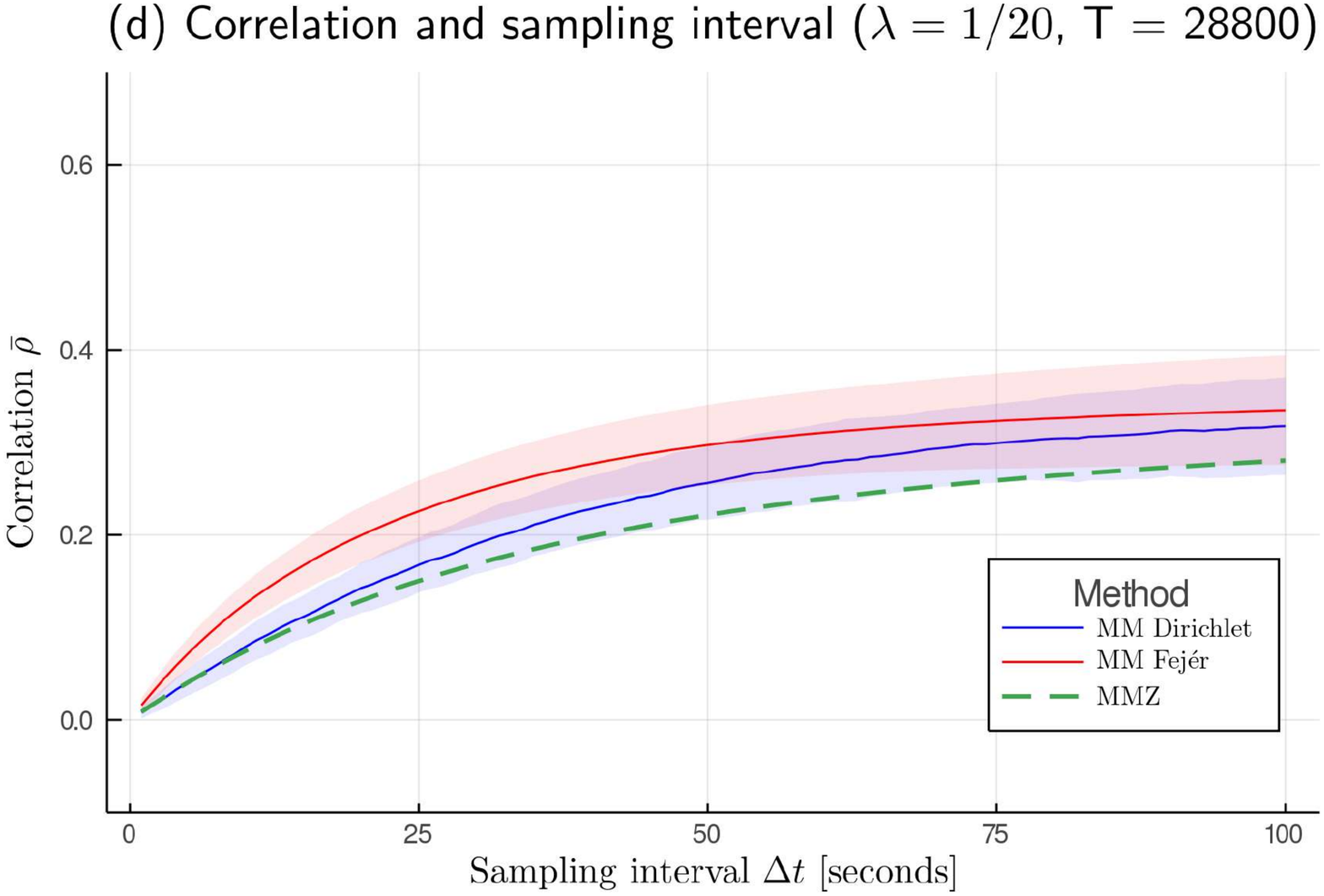}} \\
    \subfloat{\label{fig:avescale:e}\includegraphics[width=0.48\textwidth]{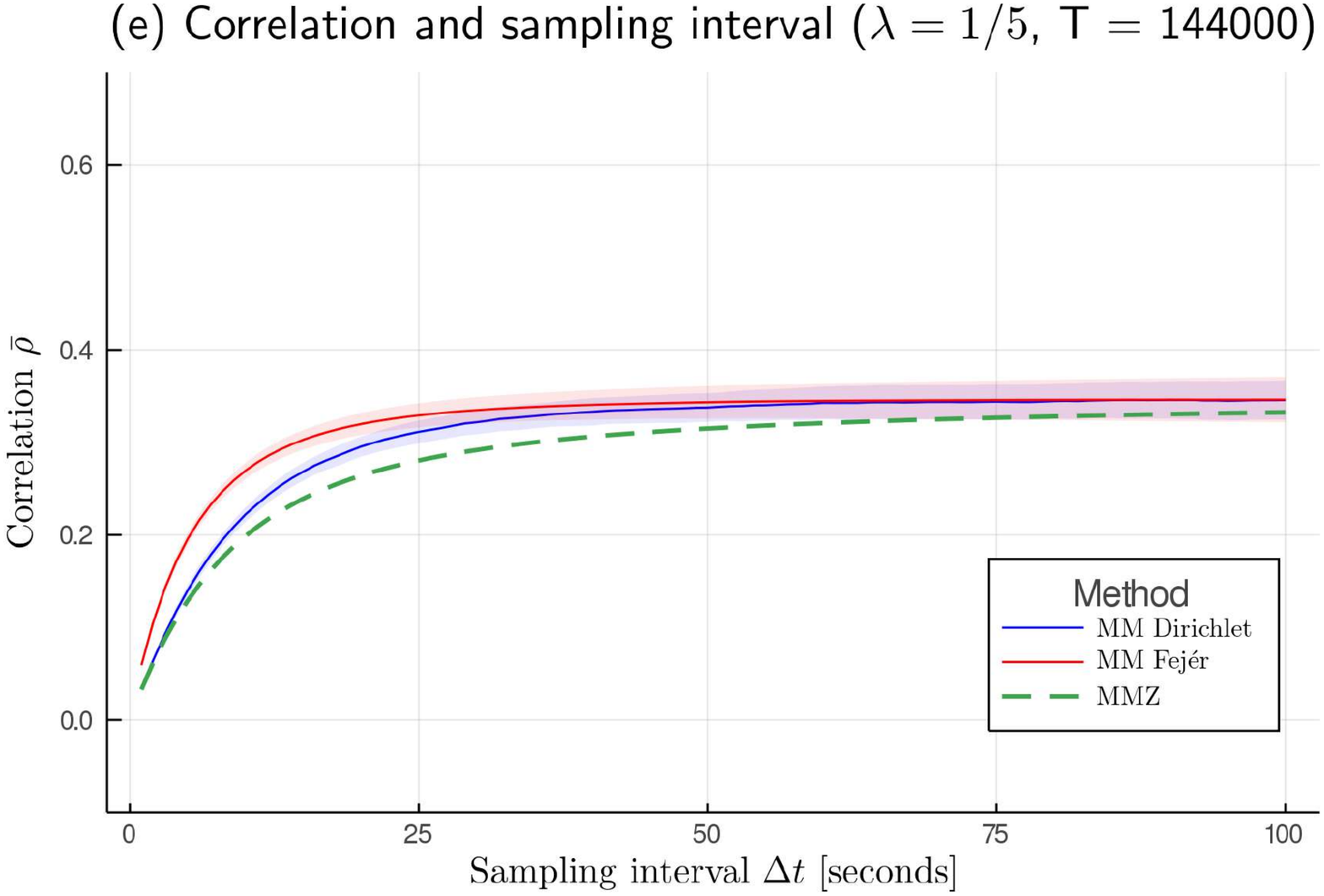}}
    \subfloat{\label{fig:avescale:f}\includegraphics[width=0.48\textwidth]{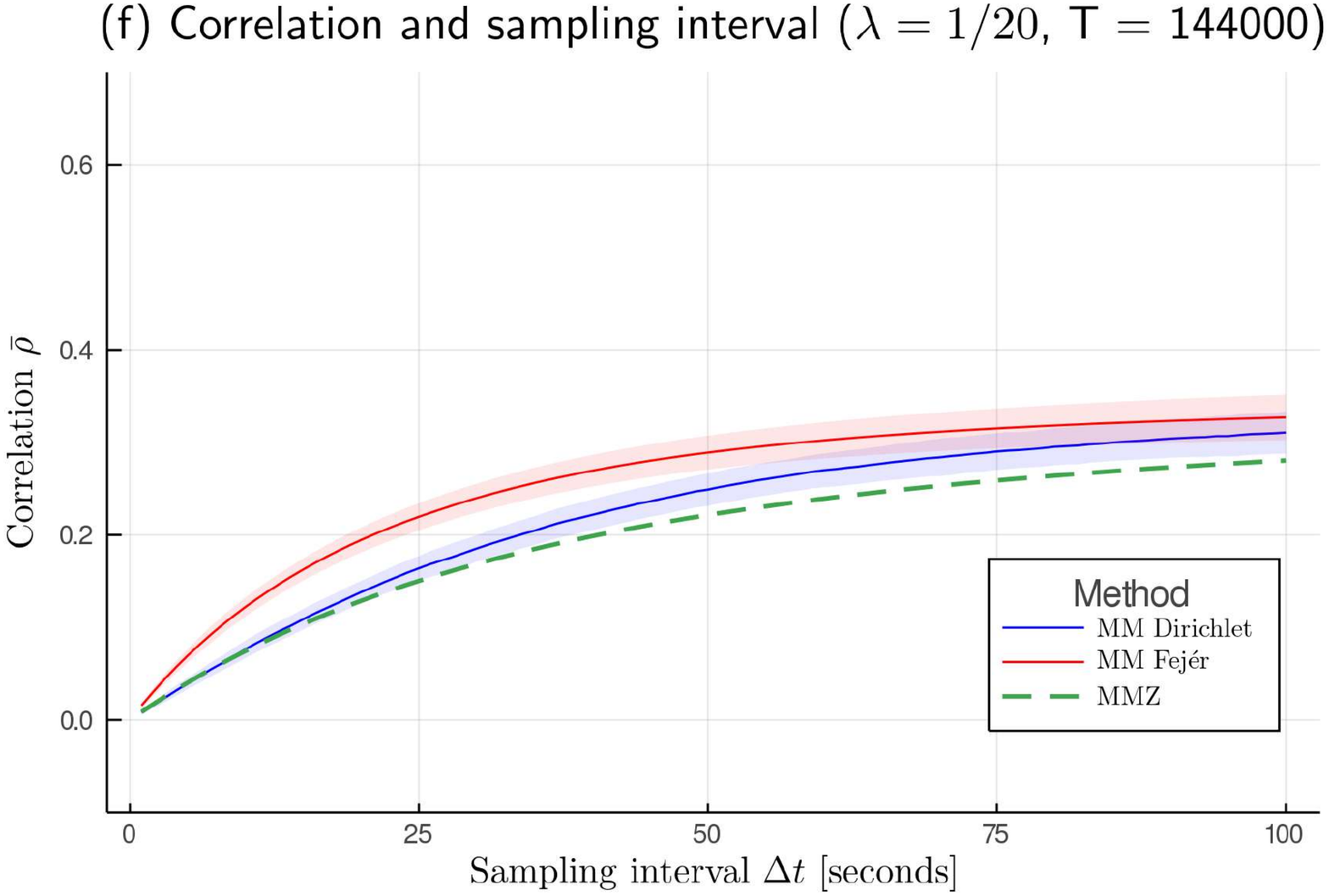}}
    \caption{\scriptsize{Here we demonstrate the relationship between the time-scale averaging in the Malliavin-Mancino estimator (as determined by the number of Fourier coefficients) and the analytic formula characterising the Epps effect arising from Poissonian sampling using \cref{eq:Der:21}. The conversion from $\Delta t$ to $N$ in the estimators is given by \cref{eq:Der:22}. The arrival time representation samples $T$ synchronous data points (each point representing a second in a day) with an inter-arrival time with rates $\lambda = \frac{1}{5}$ and $\lambda = \frac{1}{20}$ for the first and second columns respectively. The induced correlation across the replications is $0.35$. Here the green dashes (``MMZ'') is the plot of \cref{eq:Der:21}. The Malliavin-Mancino estimates for each $\Delta t$ are obtained using the Dirichlet and Fej\'{e}r basis kernels; these are respectively the blue line (``MM Dirichlet'') and red line (``MM Fej\'{e}r''). Here we repeat this process 100 times to obtain 100 arrival time representation paths, based on 100 different GBM paths to obtain 100 estimates for each $\Delta t$. The average correlation estimate at each $\Delta t$ is then plotted with error bars (computed using a t-distribution with 99 degrees of freedom and the sample standard deviation) representing 68\% of the variability between the estimated paths. The Dirichlet basis kernel (plausibly) recovers the Epps effect given by \cref{eq:Der:21}, while the Fej\'{e}r basis kernel is biased upwards with respect to the Dirichlet basis kernel (and the theoretical Epps effect) because there is induced averaging. For the same reason the estimate curves using the Dirichlet basis kernel are more volatile than those estimated using Fej\'{e}r kernel basis (see \Cref{fig:EmpMMZAll}). The figures can be recovered using the Julia script file \href{https://github.com/CHNPAT005/PCEPTG-MM-NUFFT/blob/master/Scripts/Time\%20Scales/MMZandMM}{MMZandMM} on the GitHub resource \cite{PCEPTG2020CODE}.}}
\label{fig:avescale}
\end{figure*}
Following the work of \cite{PI2007,RENO2001}, T\'{o}th and Kert\'{e}sz \cite{TK2007} and Mastromatteo et al. \cite{MMZ2011} were able to analytically quantify the Epps effect arising from asynchrony under an arrival time representation as:
\begin{equation} \label{eq:Der:21}
  \tilde{\rho}_{\Delta t}^{ij} = c \left( 1 + \frac{1}{\lambda \Delta t} \left( e^{-\lambda \Delta t} - 1 \right) \right).
\end{equation}
Here $c$ is the induced correlation and the sampling intensity is $\lambda$; the same for the price paths \cite{MMZ2011}. This will serve as our base-line theoretical Epps effect. 

We compare the relationship between the time-scale averaging in the Malliavin-Mancino estimator used by \cite{PI2007,RENO2001} to the analytic formula characterising the Epps effect arising from Poissonian sampling in \cref{eq:Der:21}. This is done by simulating $T$ data points from a bivariate Geometric Brownian Motion with the same parameters as \Cref{subsec:accuracy}. We consider one hour, one trading day and one trading weeks' worth of simulated data with a price realisation sampled each second. Thus, assuming that each trading day is 8 hours in Calendar time, we have $T = 3600, 28800$ and $144000$ synchronous data points for the various cases. The synchronous price paths are then sampled using an exponential inter-arrival time with the same rate $\lambda$ to create the arrival time representation of the asynchronous price paths.

\Cref{fig:avescale} plots \cref{eq:Der:21} as a function of $\Delta t$ (MMZ) ranging from $1$ to $100$ seconds and compared against the estimated correlations. The corresponding $N$\footnote{We note that \cref{eq:Der:22} may not always be a perfect conversion due to the range of Fourier modes in \cref{eq:Der:1}.} for the Malliavin-Mancino estimator is given by:
\begin{equation} \label{eq:Der:22}
  N = \left\lfloor \frac{1}{2} \left( \frac{T}{\Delta t} - 1 \right) \right\rfloor.
\end{equation}

The Malliavin-Mancino correlation estimates are computed using the fast Gaussian gridding implementation of the non-uniform fast Fourier transform with $\epsilon = 10^{-12}$ for the various choices of $\Delta t$. This is done for the Dirichlet basis kernel (MM Dirichlet) and the Fej\'{e}r basis kernel (MM Fej\'{e}r). Furthermore, this process is repeated 100 times so that the variability between the measured estimates can be investigated for various $n_i$ with $i=1,2$ and $N$. Here, $n_i \approx T / \lambda$ on average based on the Poissonian sampling.

\Cref{fig:avescale} plots the average correlation estimates over the various replications with the error bars representing 68\% of the variability between the estimated paths.\footnote{We use the sample standard deviation and a t-distribution with 99 degrees of freedom for 100 replications.} First, the precision of the estimates improves as $n_i$ and $N$ increase {\it i.e.} for decreasing time-scales. The exact contributions of $n_i$ and $N$ leading to the increased precision for larger $T$ is unclear, as larger $T$ implies larger $n_i$ and $N$. However, for a fixed $T$ we see the effect that larger $N$ has on the precision (ignoring the variability from $n_i$ changing from the replications). Second, the Dirichlet kernel can plausibly recover the theoretical Epps curve; while the Fej\'{e}r is biased upwards with respect to the Dirichlet basis and the theoretical Epps effect. This is because the Fej\'{e}r kernel places more weight on the lower frequencies and less weight on the higher frequencies, which makes it more stable under microstructure noise \cite{MS2011}. Furthermore, due to the weighting of frequencies in the Fej\'{e}r kernel we get smoother estimates compared to the Dirichlet kernel (see \Cref{fig:EmpMMZAll} for indication of individual realisations).

Deciding which basis kernel to use depends on how one wants to treat the Epps effect. The Epps effect is well known and has many factors contributing to it \cite{MMZ2011, MSG2010, MSG2011, PI2007, RENO2001, SS2014, TK2009, TK2007}. The effects include statistical causes that require correction such as asynchrony \cite{MSG2011,PI2007,RENO2001,TK2007} and tick-size \cite{MSG2010,MSG2011}, but also genuine effects such as lead-lag \cite{MMZ2011,RENO2001} and sampling interval dependent correlations \cite{BDHM2013a}. Therefore, the Dirichlet kernel will be more appropriate if one is interested in recovering the empirical nature of correlation dynamics at various time-scales as it (plausibly) recovers the theoretical Epps effect. However, the Fej\'{e}r kernel is more effective if one is interested in correcting the Epps effect as more weight is placed on lower order frequencies to avoid market microstructure noise. Furthermore, the Fej\'{e}r kernel can be coupled with $N$ which minimises the MSE.

There have been many methods proposed to correct the Epps effect arising from asynchrony, such as the estimator proposed by Hayashi and Yoshida \cite{HY2005}, a correction based on the distortion caused by asynchrony \cite{MSG2011}, or picking a smaller $N$ with the Malliavin-Mancino estimator \cite{MRS2017}. The implication of picking $N$ to minimise MSE is that control over the time-scale of interest is relinquished (as with the Hayashi-Yoshida estimator). It is clear in \Cref{fig:avescale} that in order for MSE to be minimised $N$ must be small, meaning that larger time intervals are investigated (which is a simple method to remove the Epps effect). This can be problematic if one is interested in disentangling the genuine causes of the Epps effect from the statistical causes at various time-scales \cite{MMZ2011}. The better approach here would be to measure the observed correlation dynamics (the Malliavin-Mancino estimator using NUFFTs provides a quick method to so) then correct for the statistical causes \cite{PCEPTG2020b,MSG2011} so that genuine causes in the decay of correlations at various time-scales can be investigated.

\subsection{Real-world data}

The estimated correlations at various high-frequency time-scales for 10 equity assets listed on the Johannesburg Stock Exchange (JSE) are given as a real-world example. The correlations are estimated using Trade and Quote (TAQ) event data for the 10 equities extracted from Bloomberg Pro and processed to remove repeated time stamps by aggregating trades with the same time stamp using a volume weighted average. The processed TAQ data can be found in \cite{PCEPTG2020DATA}. The 10 equities considered are: FirstRand Limited (FSR), Shoprite Holdings Ltd (SHP), Absa Group Ltd (ABG), Nedbank Group Ltd (NED), Standard Bank Group Ltd (SBK), Sasol Ltd (SOL), Mondi Plc (MNP), Anglo American Plc (AGL), Naspers Ltd (NPN) and British American Tobacco Plc (BTI). The period considered is the week from 24/06/2019 to 28/06/2019. The data is for a 5 day period with equities trading 8 hours a day. This yields $T = 5 \times 28,800 = 144,000$ seconds in the period of consideration. The TAQ data is discrete and asynchronous with different rates of trading for different stocks.

\begin{table}[h!]
\centering
\begin{tabular}{|rrrr|}
  \hline
Tickers & Vol. Traded & Unique Trades & \small $1/\hat{\lambda}$ [sec] \normalsize  \\ 
  \hline
  BTI & 3143263 & 7893 & 17.83 $\pm$ 0.64  \\ 
  NPN & 2791054 & 12378 & 11.38 $\pm$ 0.32 \\ 
  AGL & 5751811 & 9091 & 15.49 $\pm$ 0.50  \\ 
  MNP & 1701907 & 6562 & 21.43 $\pm$ 0.93  \\ 
  SOL & 6048773 & 10343 & 13.62 $\pm$ 0.43  \\ 
  SBK & 9427755 & 7441 & 18.93 $\pm$ 0.65  \\ 
  NED & 4518354 & 7090 & 19.85 $\pm$ 0.69  \\ 
  ABG & 6607644 & 6572 & 21.36 $\pm$ 0.78  \\ 
  SHP & 3758655 & 5549 & 25.35 $\pm$ 1.01  \\ 
  FSR & 38493240 & 10412 & 13.53 $\pm$ 0.39  \\ 
   \hline
\end{tabular}
\caption{The table provides a summary of the 10 equities considered for the week from 24/06/2019 to 28/06/2019. The table indicates the volume traded, the number of unique trades and the mean inter-arrival time between trades measured in seconds with the 95\% confidence interval provided.}
\label{tab:JSE10summary}
\end{table}

\begin{figure*}[hbt]
    \centering
    \subfloat{\label{fig:EmpMMZ2:a}\includegraphics[width=0.5\textwidth]{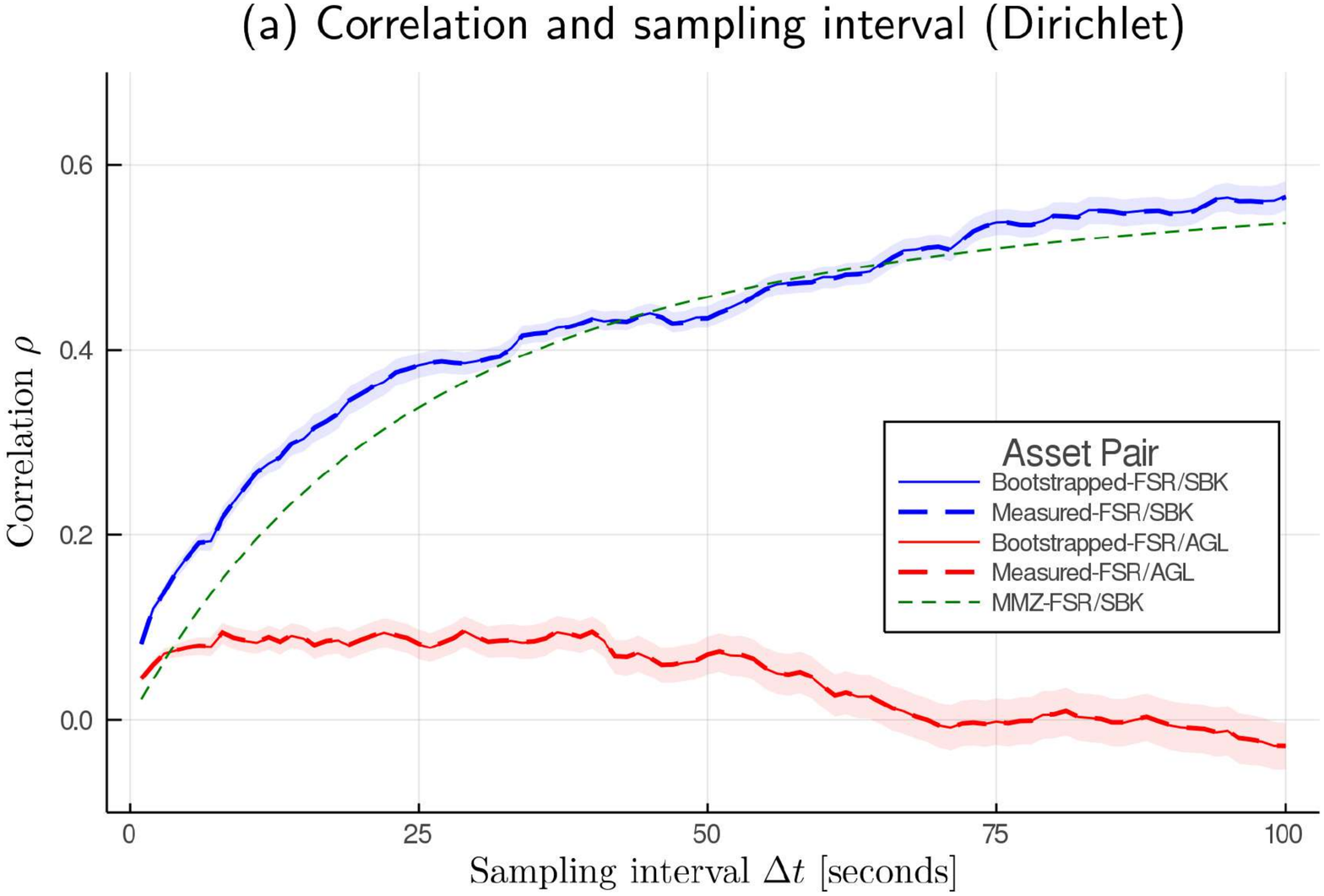}}
    \subfloat{\label{fig:EmpMMZ2:b}\includegraphics[width=0.5\textwidth]{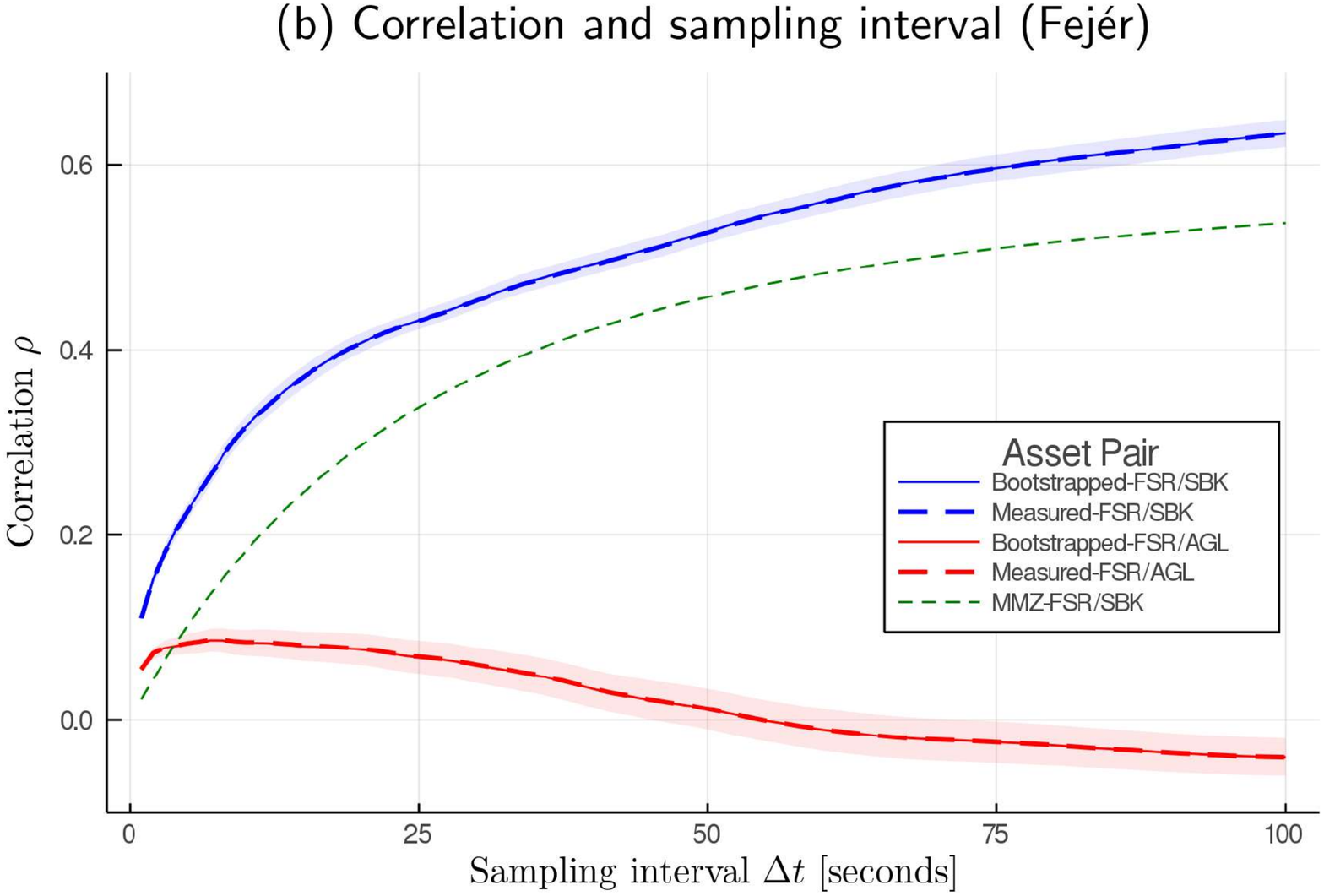}}
    \caption{The two most interesting correlation pairs from the 10 available equity data is shown by plotting the correlation as a function of the sampling interval $\Delta t$ measured in seconds  (the complete set is in \Cref{fig:EmpMMZAll} of \ref{app:EppsEDA}). The conversion between $\Delta t$ and $N$ is given by \cref{eq:Der:22} assuming $T = 144,000$. The correlation pairs considered are FSR/SBK (blue) and FSR/AGL (red). Here the lines with error bars are the mean estimates and 95\% variability between the paths obtained from block bootstrap, while the dashes are the measured estimates from the complete time series. Furthermore, we plot a simple theoretical Epps effect arising from asynchrony for the FSR/SBK (green dashes) pair (see \cref{eq:Der:21}). This is done by assuming the inter-arrival time of trades follow an exponential distribution with the same rate $\hat{\lambda} = 1/ 13.5$. We found that $c = 0.621$ provided a relatively good fit. The theoretical Epps is not plotted for the FSR/AGL pair because the correlation dynamics do not exhibit the Epps effect. The empirical reality of curves such as the upper blue curve for the FSR/SBK pair suggest that current theoretical Epps effect models can plausibly model the correlation dynamics for some asset pairs; while curves such as the lower red curve for the FSR/AGL pair suggest that the current theoretical explanations for the Epps effect are possibly insufficient. The figures can be recovered using the Julia script file \href{https://github.com/CHNPAT005/PCEPTG-MM-NUFFT/blob/master/Scripts/Time\%20Scales/Empirical}{Empirical} on the GitHub resource \cite{PCEPTG2020CODE}.}
\label{fig:EmpMMZ2}
\end{figure*}

\Cref{tab:JSE10summary} provides the volume traded, the number of unique trades, and the mean inter-arrival times between the trades for the 10 equities used in the analysis.\footnote{Here measured in seconds with a 95\% confidence interval provided computed using a t-distribution and the standard errors.} It is important to notice that the measured intensities $\hat{\lambda}$'s\footnote{The $\hat{\lambda}$'s are indicative, estimated from the TAQ data.} are not the same across the assets. In order to use \cref{eq:Der:21}, we make the simplifying assumption that the $\hat{\lambda}$'s are approximately the same and take on the larger intensity of the two {\it i.e.} $\hat{\lambda} = \max( \hat{\lambda}_i, \hat{\lambda}_j)$. We highlight that an extension of \cref{eq:Der:21} to model different intensities and lead-lags is provided by Mastromatteo et al. \cite{MMZ2011}. They considered multiple intensities $\lambda_i \neq \lambda_j$ in order to decouple effects from asynchronous sampling and the effects from lead-lag. We are interested in the general concave shape of the theoretical model (see \cite{MMZ2011,TK2007} and the plots therein) and do not use the extended model for the theoretical Epps effect. However, we point out that not all of the measured Epps curves conform to the shape of the known theoretical models; irrespective of whether one can conflate the lead-lag formulation for some lag $\tau$ with the asynchronous versions with different intensities $\lambda_i \neq \lambda_j$.

Before comparing the theoretical Epps effect against the measured Epps effect, we perform some Exploratory Data Analysis to identify the interesting correlation pairs out of the 45 available pairs. We investigate the 45 correlation pairs as a function of the sampling interval $\Delta t$ for the Dirichlet and Fej\'{e}r basis kernel (see \Cref{fig:EmpMMZAll}). The conversion for $\Delta t$ to $N$ is given by \cref{eq:Der:22}, assuming $T = 144,000$. The correlation estimates are estimated using the fast Gaussian gridding implementation of the non-uniform fast Fourier transform with $\epsilon = 10^{-12}$. We obtain correlation matrices for $100$ $\Delta t$'s ranging from 1 to 100. The compute time for $100$ different $N$'s took a total of $7.28$ seconds using the Dirichlet basis and $9.10$ seconds using the Fej\'{e}r basis --- demonstrating the efficacy of our fast Fourier method.\footnote{In this instance both the Dirichlet and Fej\'{e}r kernel produced positive semi-definite covariance matrices.} Snapshots of the correlations structures are then plotted as heat-maps for $\Delta t = 1, 30, 60$ and $100$ seconds for easier identification of the pairs (see \Cref{fig:EmpHM}). From considering all the correlation pairs as a function of the sampling intervals and the reduced heat maps (see \Cref{fig:EmpMMZAll,fig:EmpHM}), we were able to make two initial observations. First, the Fej\'{e}r kernel produces smoother estimates compared to the Dirichlet kernel. Second, nearly all the correlations pairs exhibit the Epps effect where the correlations rise as $\Delta t$ increases and conforming to the theoretical models in the literature \cite{MMZ2011,TK2009,TK2007}. However, there are exceptions where correlation pairs do not exhibit the behaviours easily accounted for by the prevailing models such as the FSR/AGL pair. Rather, it seems that the correlation drops as $\Delta t$ increases to the point where the sign of the correlation switches. 

\Cref{fig:EmpMMZ2} investigates this in more detail by plotting the correlation as a function of $\Delta t$ for two particular asset pairs. The indicative sample error bars are obtained through block bootstrap quoted to indicate 95\% of the variability between the estimates for each $\Delta t$.\footnote{This is achieved by splitting the data into 100 calendar time blocks and estimating correlations at various $\Delta t$'s with 1 block removed each time. $T$ remains the same across the various replications, so the missing block is treated as missing data. Standard deviations are obtained from the block bootstrap and error-bars computed using a t-distribution with 99 degrees of freedoms. The errors are overlaid on the mean estimates from the block bootstrap.} The first pair FSR/SBK (blue line/dashes) is a clear demonstration of the Epps effect. Thus, a simple theoretical Epps effect arising from Poissonian sampling is plotted for the pair (green dashes). This is done by assuming the inter-arrival time of trades follow an exponential distribution with larger of the two rates: $\max(\hat{\lambda}_{_{FSR}},\hat{\lambda}_{_{SBK}}) = \hat{\lambda} \approx 1/ 13.5$. We found that $c=0.621$ produced a relatively good fit for the measured correlations using the Dirichlet basis. The second pair FSR/AGL (red line/dashes) does not behave in accordance to the Epps effect, so no theoretical Epps curve is plotted for the pair as the correlation dynamics do not meaningfully fit the functional form of the model under estimation. One could possibly try argue that this is because there are stocks in the set with low relative correlations with respect to other stocks (see \ref{app:Ave3Asset} for a simple simulated 3-asset example) where the sample error can generate measured sign changes. However, we argue that this is insufficient as it does not recover the pathological behaviour seen in the FSR/AGL pair.

\Cref{fig:EmpMMZ2} illustrates a case where the Epps effect can be plausibly modelled and a case where it cannot be easily modelled. The majority of the correlation pairs fit into the more notable Epps effect models \cite{MMZ2011,MSG2011,TK2009,TK2007} which account for a drop in magnitude with a concave decay in correlations (see \Cref{fig:EmpMMZAll}). However, Mastromatteo et al. caution that a significant portion of the measured Epps effect cannot be completely accounted for by current models of the Epps effect. Meaning there are other factors which can affect the dynamics of the observed correlation \cite{MMZ2011}. The FSR/AGL pair is one such example. This suggests that either: (i) current theoretical explanations for the Epps effect are possibly insufficient, or (ii) the correlation dynamics under market-microstructure cannot be explained with only the Epps effect \cite{PCRBTG2019}.

\section{Conclusions} \label{sec:conclude}

We provide a fast novel implementation of the Malliavin-Mancino Fourier estimators using non-uniform fast Fourier transforms and promote the use of fast Gaussian gridding with the Fej\'{e}r basis function as our preferred implementation. 

First, we compared three averaging kernels: the Gaussian, Kaiser-Bessel, and exponential of semi-circle kernel. Based on the like-for-like algorithmic comparison, the fast Gaussian gridding is the fastest out of the three non-uniform fast Fourier methods. However, with appropriate low-level implementation techniques the exponential of semi-circle kernel can be made to be faster than the fast Gaussian gridding \cite{BMK2018}. All three non-uniform fast Fourier method significantly outperform the naive implementations of the Malliavin-Mancino estimators. 

Second, we demonstrate the requirement for using the non-uniform fast Fourier methods as motivated by the failure of the zero-padded fast Fourier transform using the arrival time representation of asynchrony. 

Third, we demonstrate that there is no adverse interplay between the kernel averaging and the time-scale averaging (arising from the choice of $N$); provided there is sufficient spreading to enough nearby grid points. Concretely, the requested tolerance $\epsilon$ must be less than $10^{-4}$ under our choice of $M_{sp}$. We show when this is the case, the non-uniform fast Fourier methods can recover the estimates of \cref{eq:Der:2,eq:Der:3} to machine precision. Moreover, we show that the NUFFT methods recover the same bias and MSE results as the direct evaluation of \cref{eq:Der:FC} and correctly recover the target integrated covariance.

Fourth, we provide the link between the work done by Ren\`{o} \cite{RENO2001} and Precup and Iori \cite{PI2007} with the work from T\'{o}th and Kert\'{e}sz \cite{TK2007} and Mastromatteo, Marsili and Zoi \cite{MMZ2011}. Moreover, we argue the Dirichlet kernel is the better choice if one wants to recover the empirical nature of correlation dynamics at various time-scales while the Fej\'{e}r kernel is more appropriate if one wants to correct the Epps effect (with appropriate choice of $N$).

Finally, we demonstrate the efficacy of our non-uniform fast Fourier methods with one week of Trade and Quote data from the JSE. We argue that the current theoretical explanations for the Epps effect are possibly insufficient in explaining the entirety of the empirical correlation dynamics under specific market microstructures.

Future work aims to expand our empirical understanding of correlation dynamics on various streaming event-data sources using this convenient estimation tool and to incorporate the estimators NUFFT implementation with an extension that makes the estimate robust to samples that explicitly incorporate jumps \cite{CT2015}.

\section*{Acknowledgements}
We would like to thank Melusi Mavuso and Roger Bukuru for comments. We would also like to thank the reviewers for helpful critique and suggestions. The data was sourced from Bloomberg Professional via the University of Cape Town Library service. Patrick Chang would like to acknowledge the support of the Manuel \& Luby Washkansky Scholarship and the South African Statistical Association [grant number 127931]. 

\balance

\bibliographystyle{elsarticle-harv} 
\bibliography{PCEPTG-NUFFT-001.bib}

\onecolumn
\appendix

\section{Additional figure sets}

\begin{figure}[h!]
    \centering
    \subfloat{\includegraphics[width=0.5\textwidth]{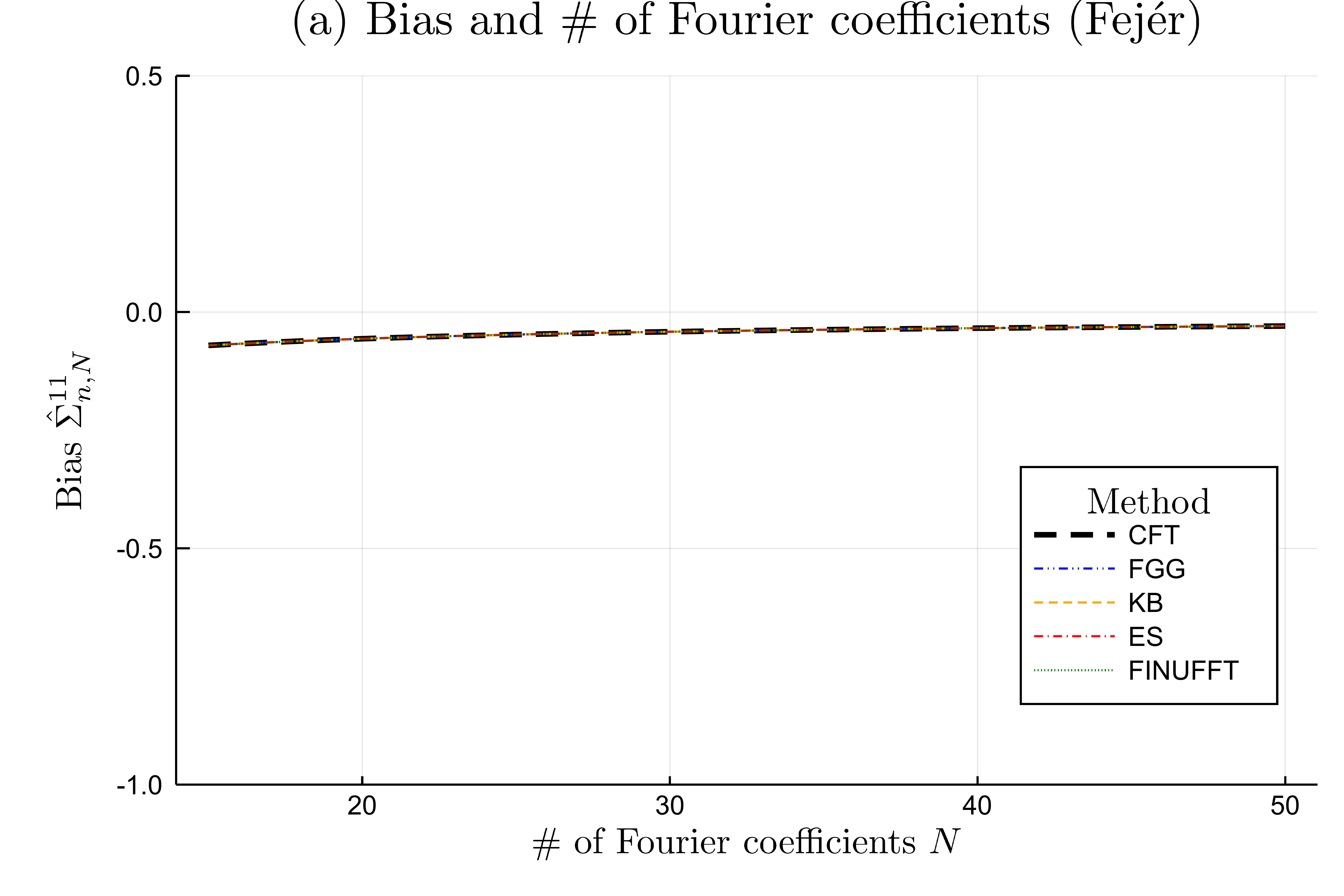}}
    \subfloat{\includegraphics[width=0.5\textwidth]{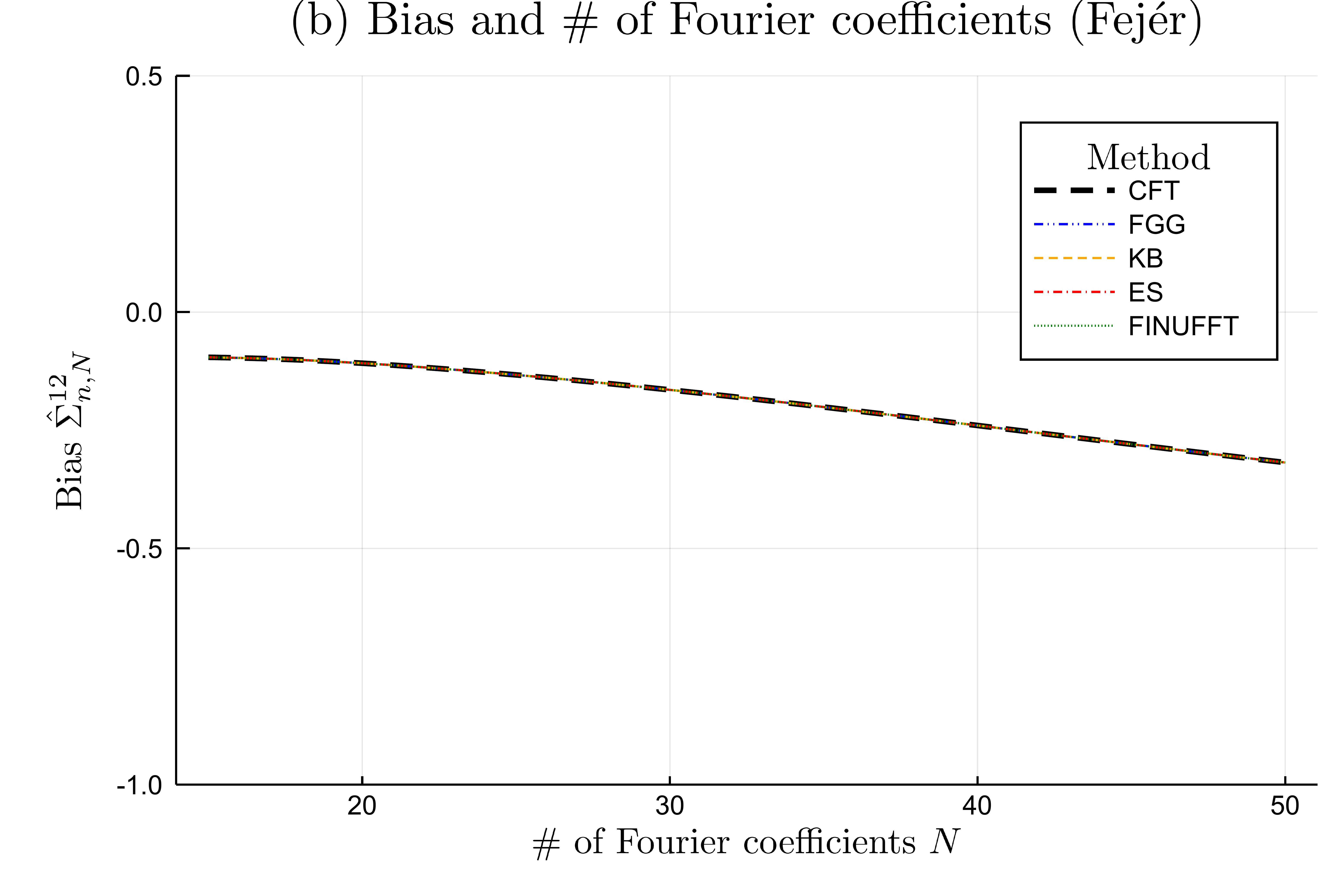}} \\
    \subfloat{\includegraphics[width=0.5\textwidth]{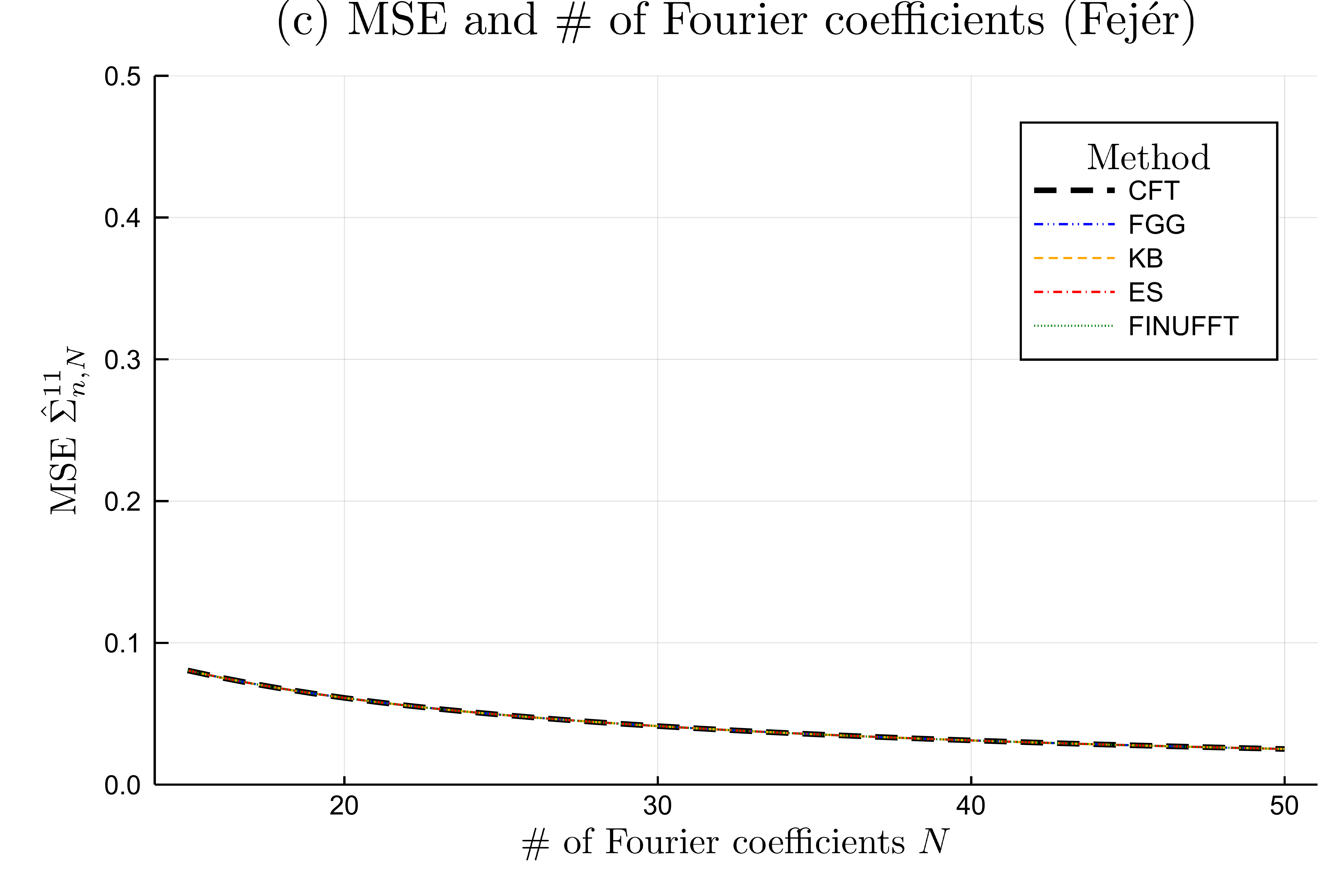}}
    \subfloat{\includegraphics[width=0.5\textwidth]{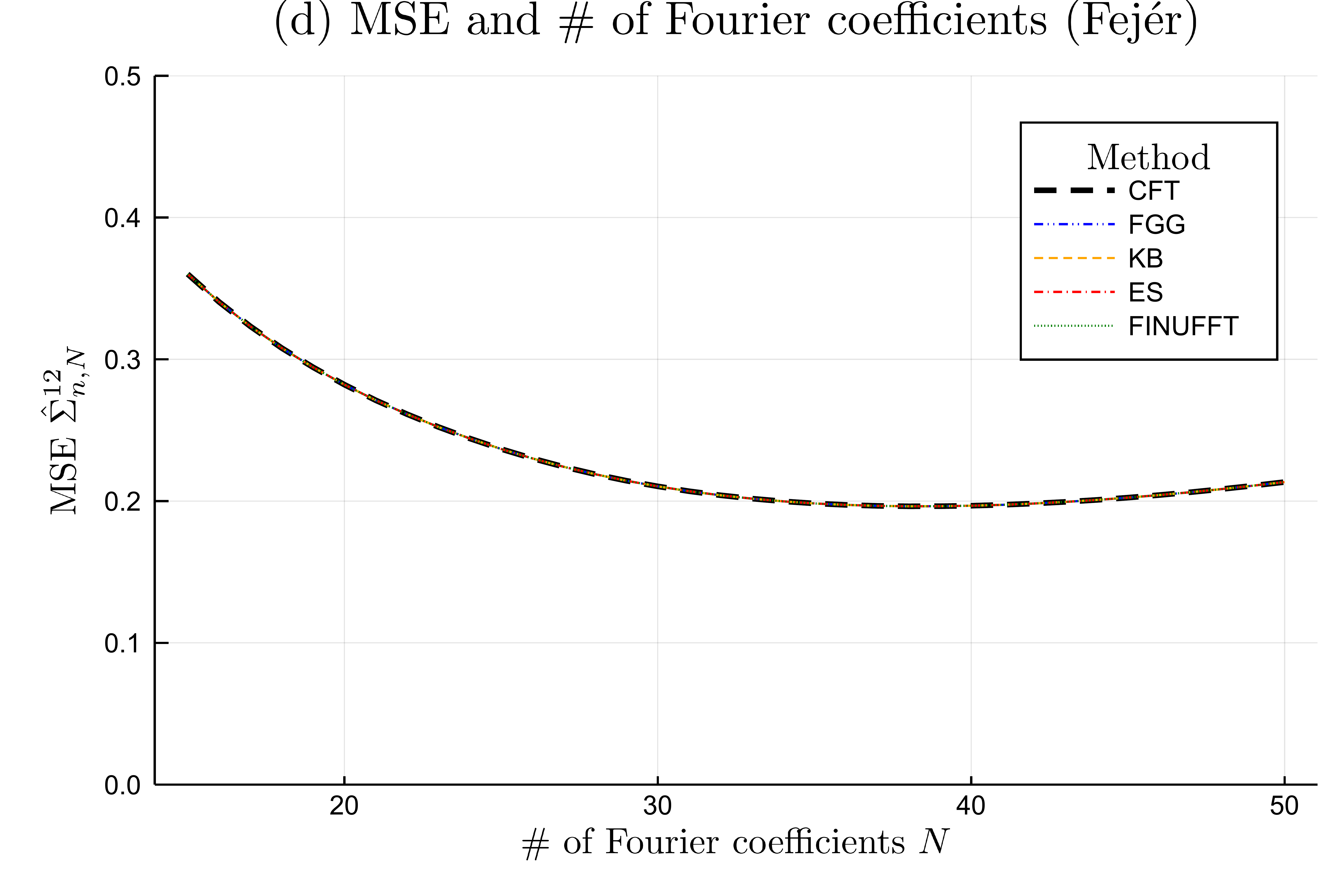}}
    \caption{The figures investigate the bias and MSE of the integrated covariance as a function of the number of Fourier coefficients using the Fej\'{e}r representation. The base-line price process is a synchronous GBM with $n=100$ data points. The missing data representation is induced here using the regular non-synchronous trading \cite{MRS2017}, where the second asset is observed at every second trade of asset one. The methods investigated are: the vectorised implementation (CFT - black dashes), the fast Gaussian gridding (FGG - blue dash-dot-dot), the Kaiser-Bessel kernel (KB - orange dashes), the exponential of semi-circle with our naive implementation (ES - red dash-dots) and the FINUFFT implementation (FINUFFT - green dots). The NUFFT methods are computed using the default $\epsilon = 10^{-12}$. We see that the fast Fourier methods recover the same bias and MSE results as the vectorised implementation and behave similarly to \cref{fig:MSEBias}. The figures can be recovered using the Julia script file \href{https://github.com/CHNPAT005/PCEPTG-MM-NUFFT/blob/master/Scripts/Accuracy/MSEBias}{MSEBias} on the GitHub resource \cite{PCEPTG2020CODE}.}
\label{fig:MSEBias_Fej}
\end{figure}

\section{Sensitivity test}

\begin{figure}[H]
    \centering
    \subfloat{\label{fig:Sensitivity:a}\includegraphics[width=0.5\textwidth]{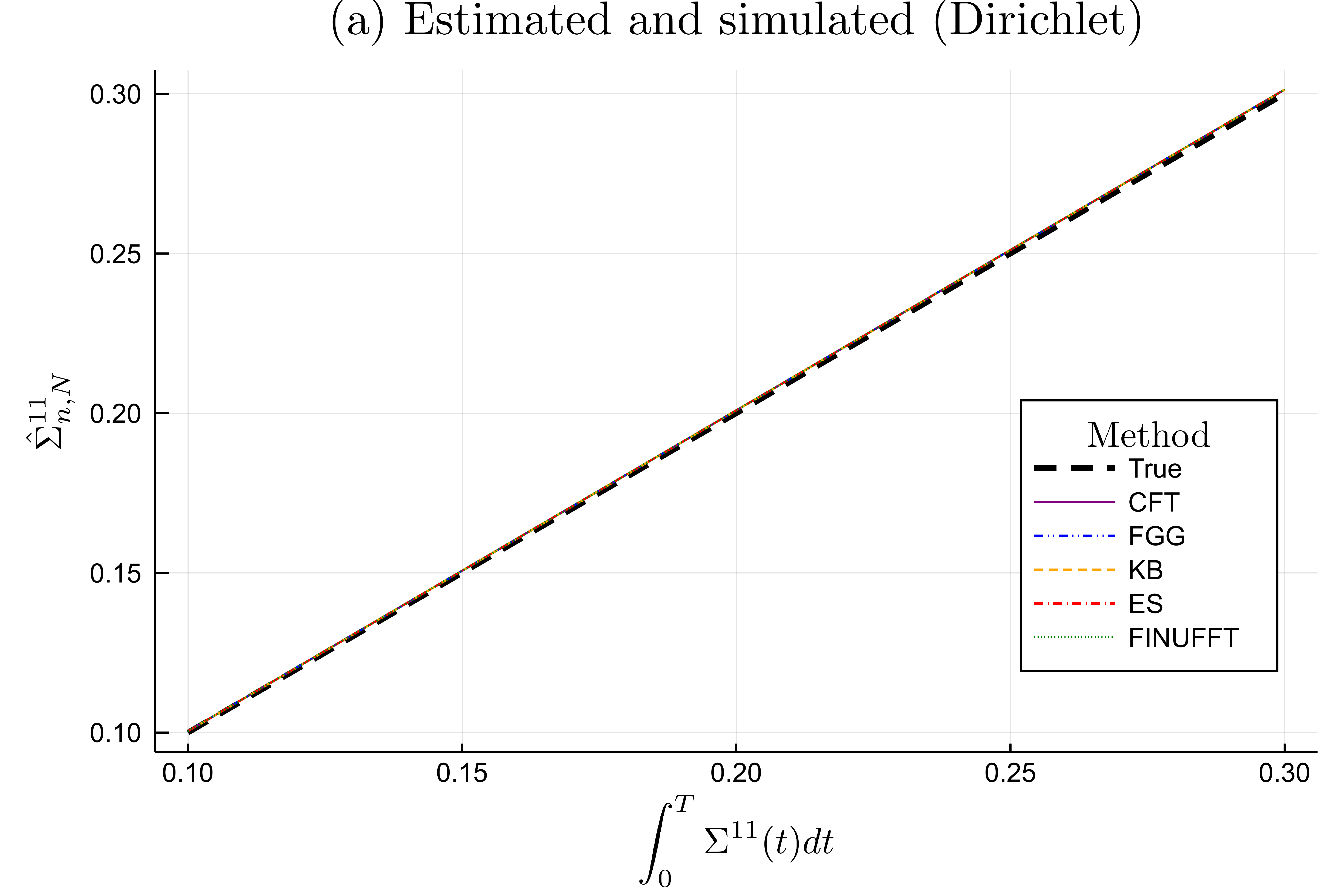}}
    \subfloat{\label{fig:Sensitivity:b}\includegraphics[width=0.5\textwidth]{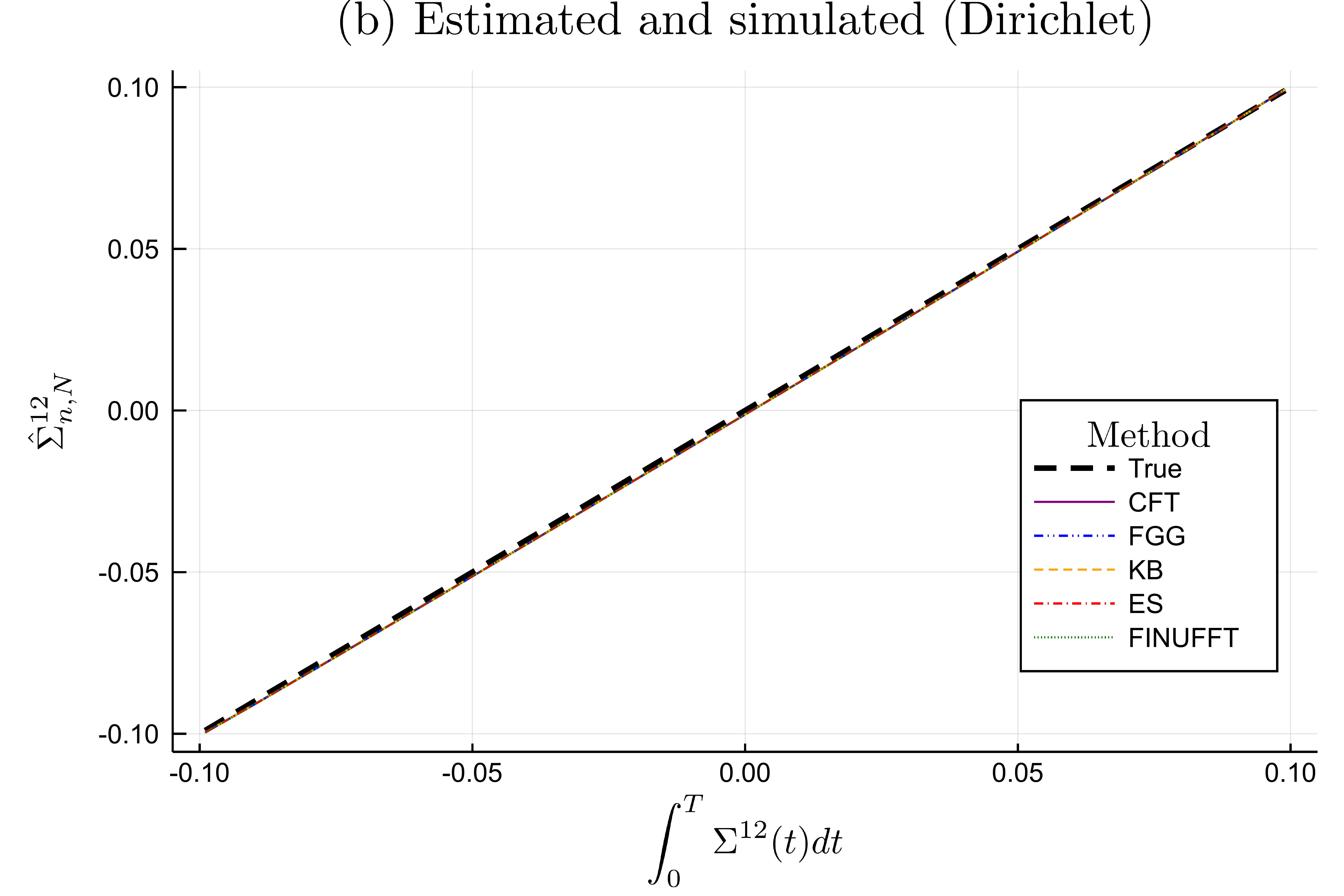}} \\
    \subfloat{\label{fig:Sensitivity:c}\includegraphics[width=0.5\textwidth]{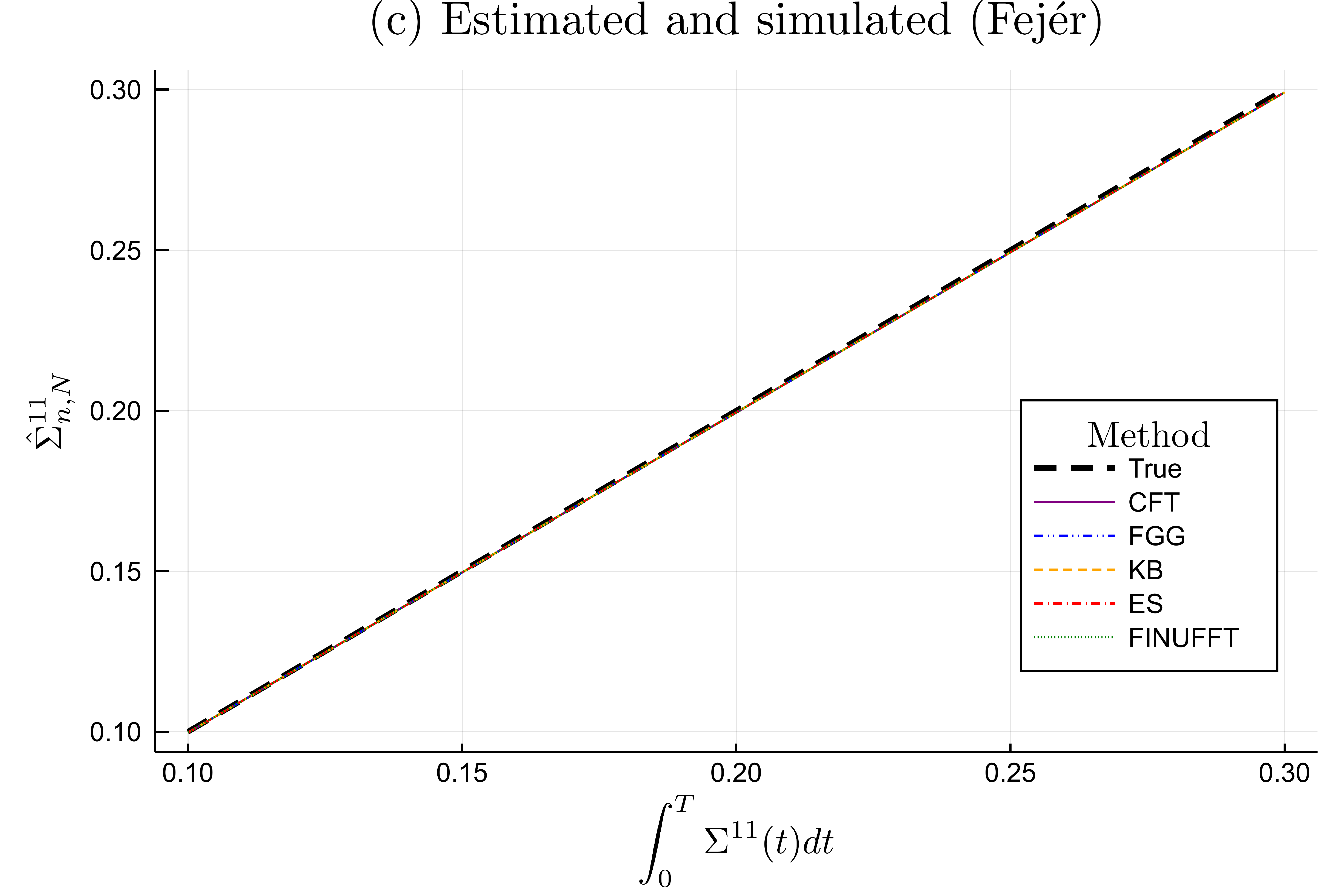}}
    \subfloat{\label{fig:Sensitivity:d}\includegraphics[width=0.5\textwidth]{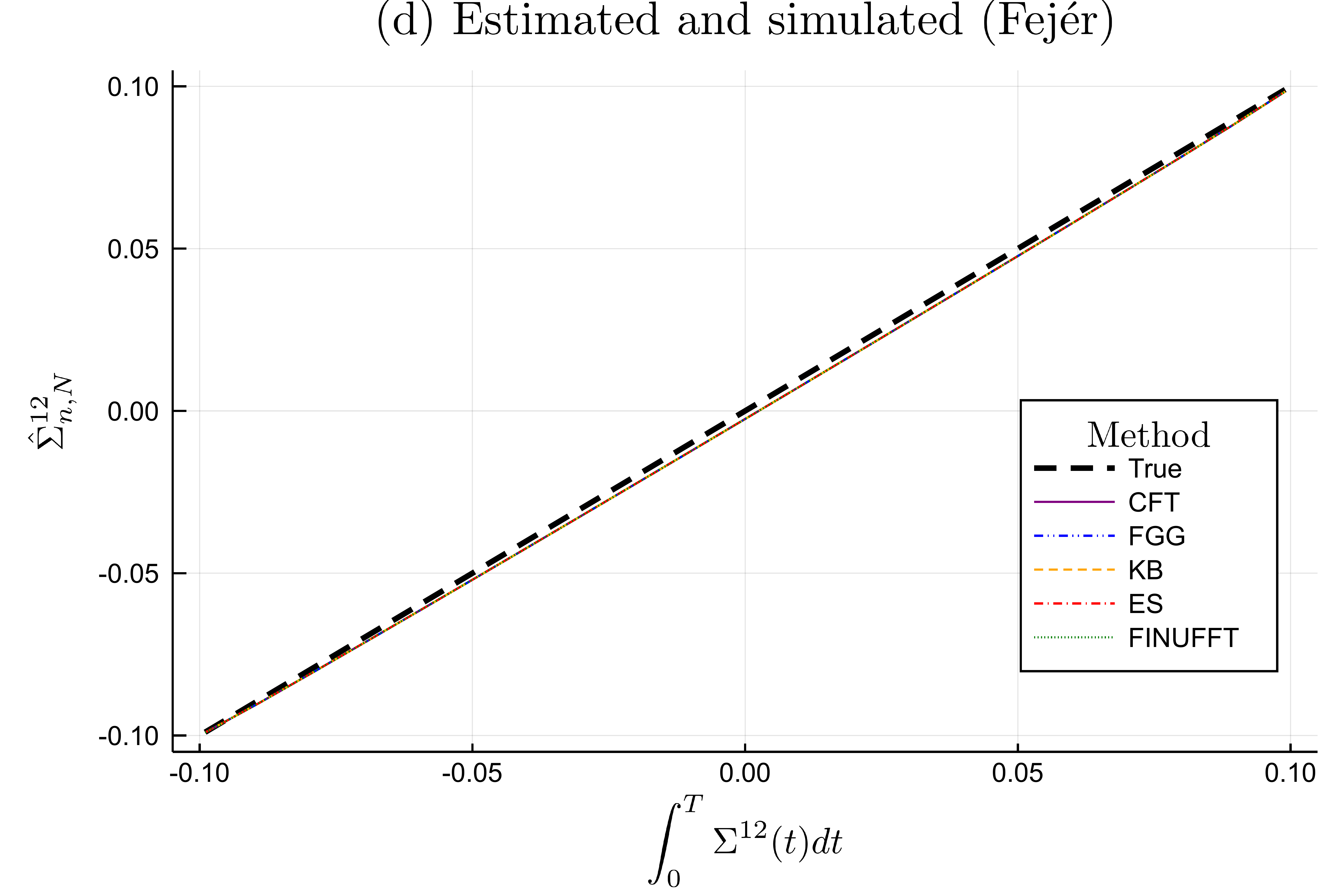}}
    \caption{The figures investigate if the NUFFT methods an correctly recover the target integrated covariance, independent of the specific parameters chosen. This is achieved by simulating a synchronous bivariate GBM with $n=10^4$ data points with the true value of $\int_0^T \Sigma^{11}(t)dt$ ranging from 0.1 to 0.3 and the true value of $\int_0^T \Sigma^{12}(t)dt$ ranging from -0.1 to 0.1. The methods investigated are: the vectorised implementation (CFT - purple line), the fast Gaussian gridding (FGG - blue dash-dot-dot), the Kaiser-Bessel kernel (KB - orange dashes), the exponential of semi-circle with our naive implementation (ES - red dash-dots) and the FINUFFT implementation (FINUFFT - green dots). The NUFFT methods are computed using the default $\epsilon = 10^{-12}$. The estimates recover a linear relationship between the estimated integrated covariance and the true integrated covariance, confirming that the NUFFT methods can correctly recover the target estimates. The figures can be recovered using the Julia script file \href{https://github.com/CHNPAT005/PCEPTG-MM-NUFFT/blob/master/Scripts/Accuracy/Sensitivity}{Sensitivity} on the GitHub resource \cite{PCEPTG2020CODE}.}
\label{fig:Sensitivity}
\end{figure}

We perform a sensitivity analysis to ensure that the NUFFT implementation can correctly recover the target integrated covariance, not depending on the parameters chosen in the main document. Here we simulate a synchronous bivariate GBM with $n=n_1=n_2=10^4$ data points with discretisation size $\Delta t = 1/n$. \Cref{fig:Sensitivity:a,fig:Sensitivity:b} is the Dirichlet representation, while \Cref{fig:Sensitivity:c,fig:Sensitivity:d} is the Fej\'{e}r representation. For the case of the integrated variance $\int_0^T \Sigma^{11}(t)dt$ the true value ranges from 0.1 to 0.3, while for the case of the integrated covariance $\int_0^T \Sigma^{12}(t)dt$ the true value ranges from -0.1 to 0.1.

\Cref{fig:Sensitivity} plots the estimated integrated covariance as a function of the true integrated covariance. The methods investigated are: the vectorised implementation (CFT), the fast Gaussian gridding (FGG), the Kaiser-Bessel kernel (KB), the exponential of semi-circle with our naive implementation (ES) and the FINUFFT implementation (FINUFFT). The NUFFT methods are computed using the default $\epsilon = 10^{-12}$. We see a linear relationship between the estimated integrated covariance and the true integrated covariance. This confirms that the NUFFT methods can correctly recover the target estimates. Moreover, the estimates for the various methods are exactly the same as each other. This confirms that the NUFFT methods can recover the same estimates as the traditional implementation of the estimator.

\section{Epps effect EDA for 10 JSE stocks} \label{app:EppsEDA}

\begin{figure*}[p]
\centering
    \subfloat{\label{fig:EmpMMZAll:a}\includegraphics[width=0.48\textwidth]{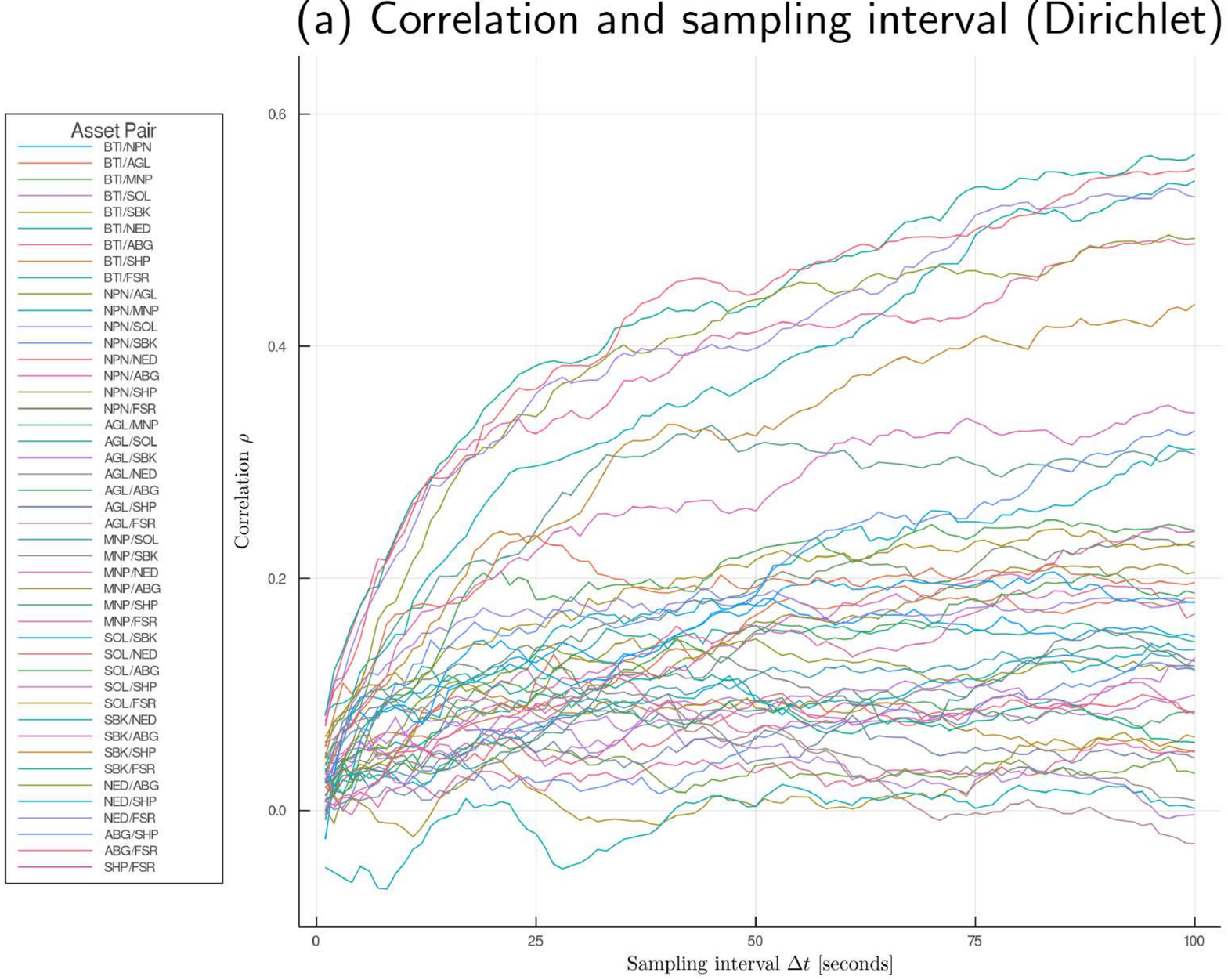}}
    \subfloat{\label{fig:EmpMMZAll:b}\includegraphics[width=0.48\textwidth]{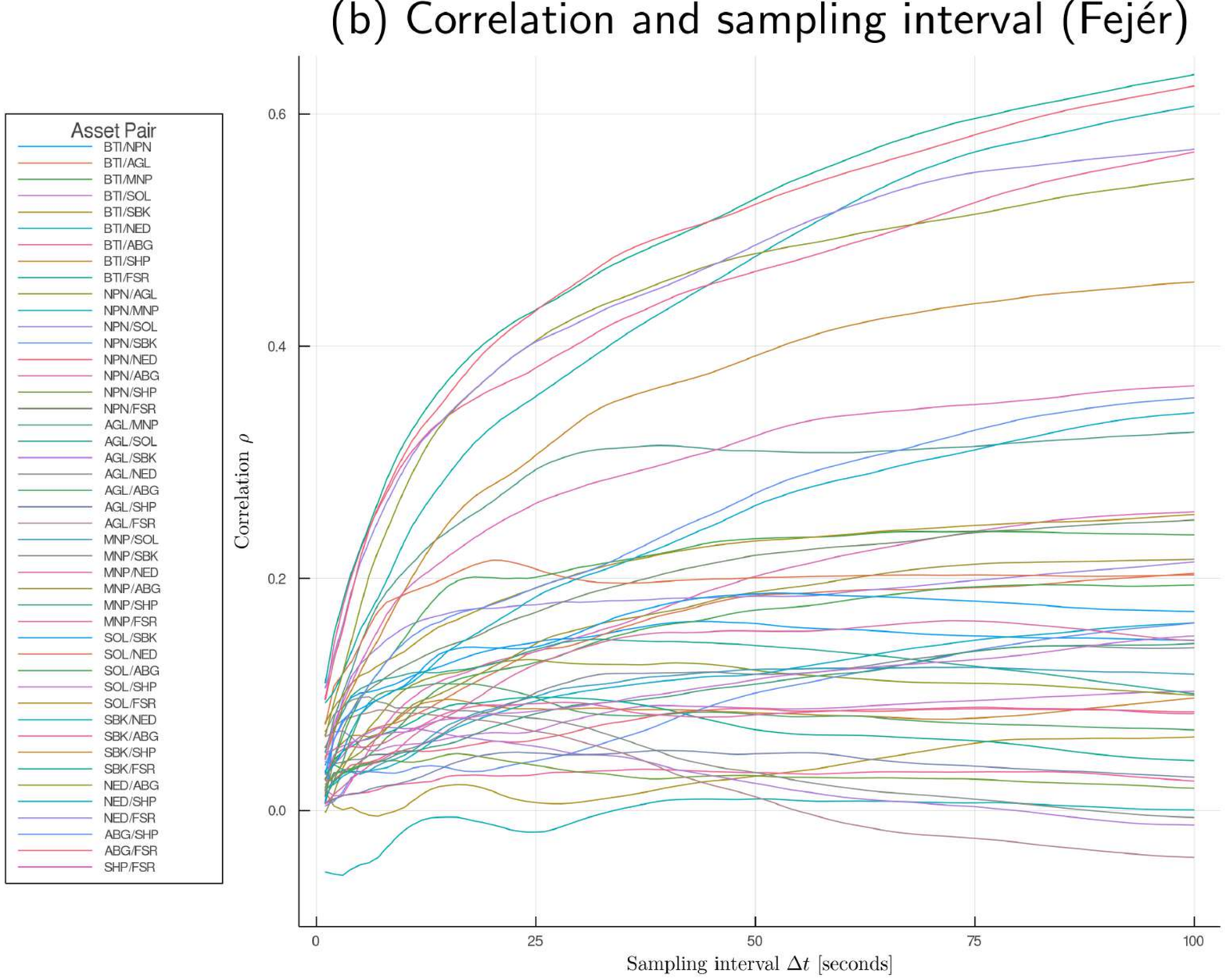}}
\caption{We investigate the Epps effect on the JSE by plotting the correlation estimate from the Malliavin-Mancino estimator as a function of the sampling interval $\Delta t$. The conversion between $\Delta t$ and $N$ is given by \eqref{eq:Der:22}, assuming $T = 144,000$ seconds in the 5 day period. The correlation pairs are plotted for all 10 equities. Sub-figure (a) is the estimates using the Dirichlet basis kernel and (b) the Fej\'{e}r basis kernel. It is clear that the Fej\'{e}r kernel produces smoother estimates compared to the Dirichlet kernel due to the induced averaging. Finally, we see that most of the equity pairs produce the Epps effect demonstrated in Figure \ref{fig:avescale}, but more interesting is that there is an equity pair where the correlation switches signs for different $\Delta t$. The figures can be recovered using the Julia script file \href{https://github.com/CHNPAT005/PCEPTG-MM-NUFFT/blob/master/Scripts/Time\%20Scales/Empirical}{Empirical} on the GitHub resource \cite{PCEPTG2020CODE}.}
\label{fig:EmpMMZAll}
\end{figure*}

\begin{figure*}[p]
\centering
    \subfloat{\label{fig:EmpHM:a}\includegraphics[width=0.245\textwidth]{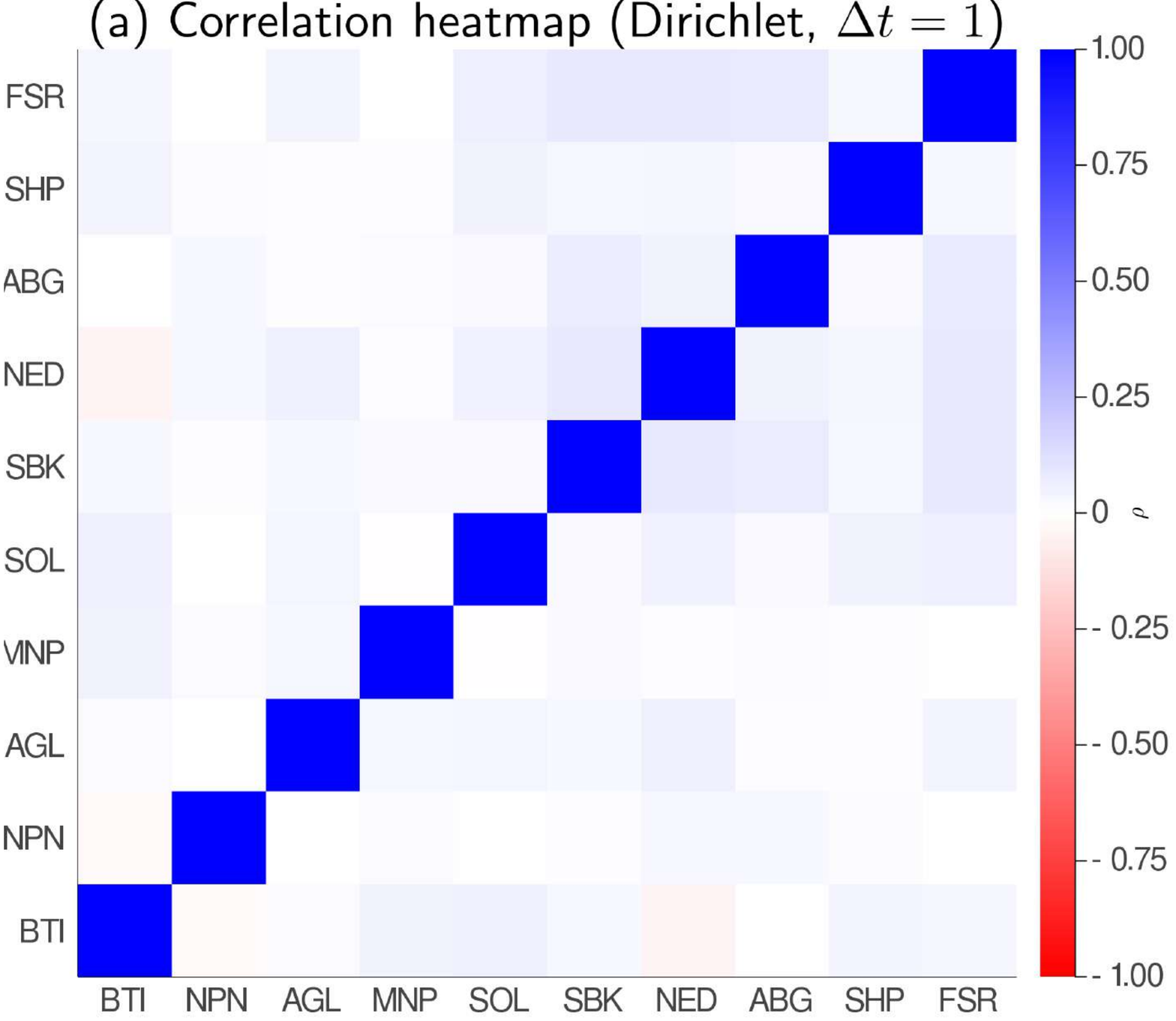}}
    \subfloat{\label{fig:EmpHM:b}\includegraphics[width=0.245\textwidth]{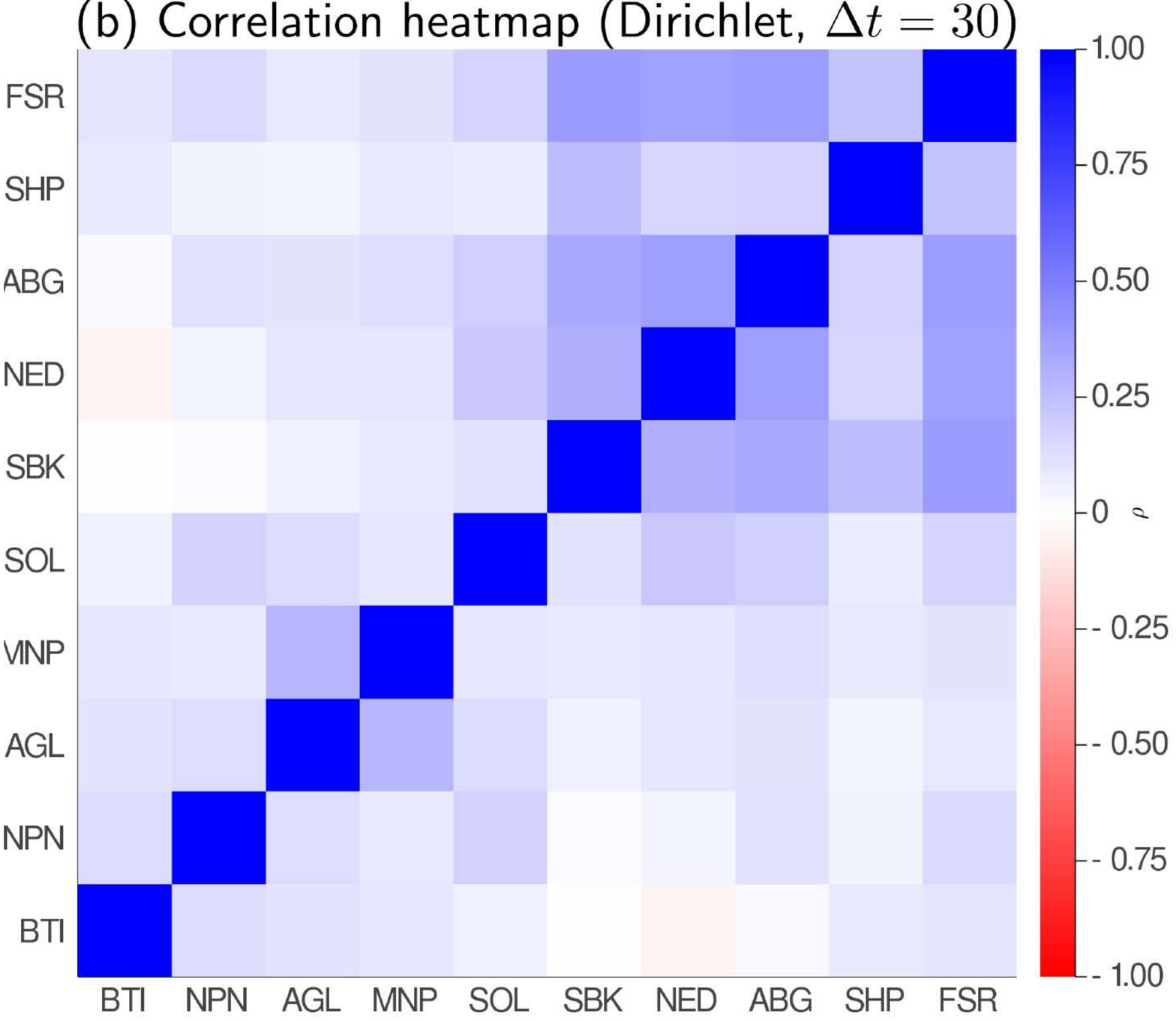}}
    \subfloat{\label{fig:EmpHM:c}\includegraphics[width=0.245\textwidth]{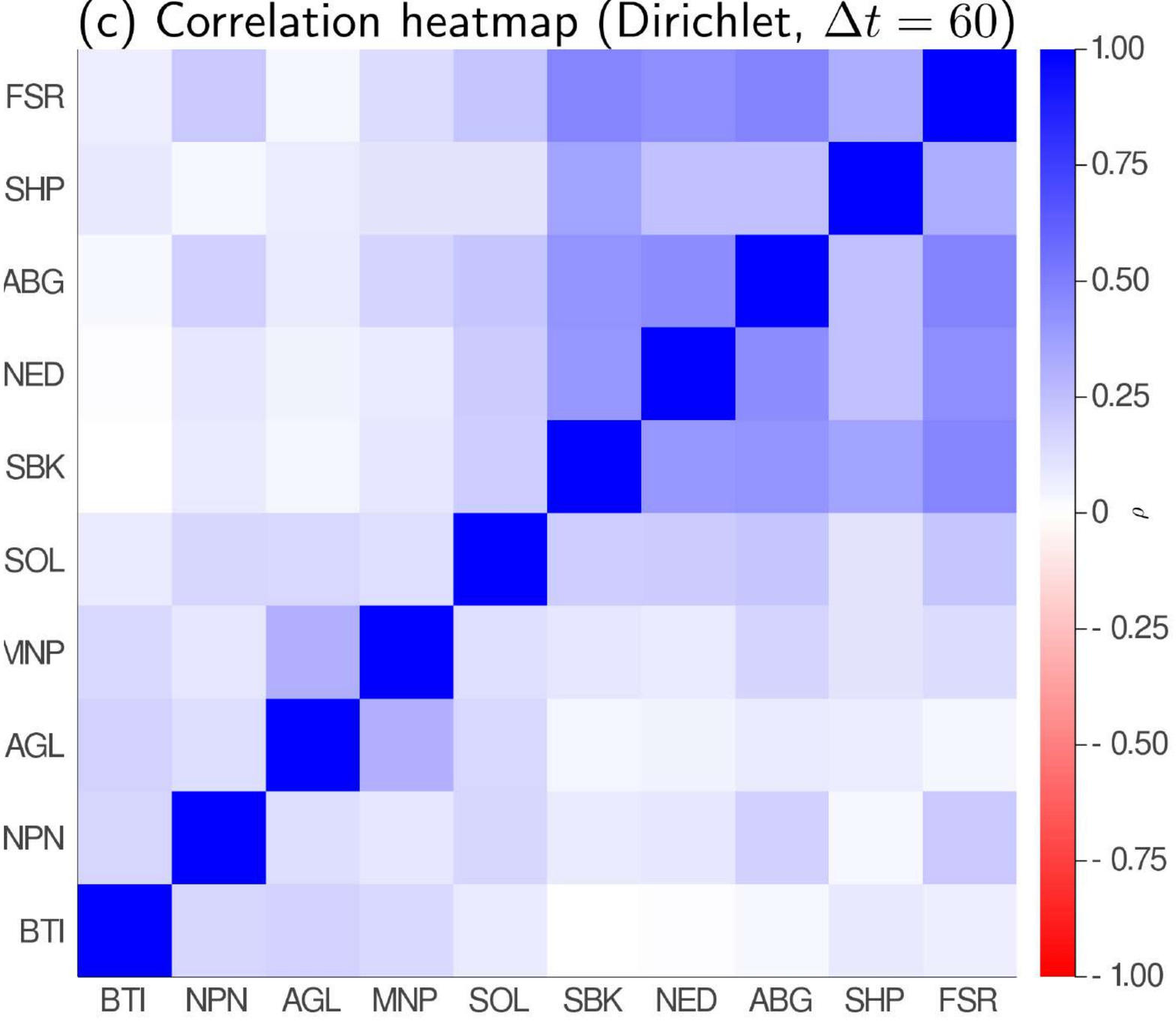}}
    \subfloat{\label{fig:EmpHM:d}\includegraphics[width=0.245\textwidth]{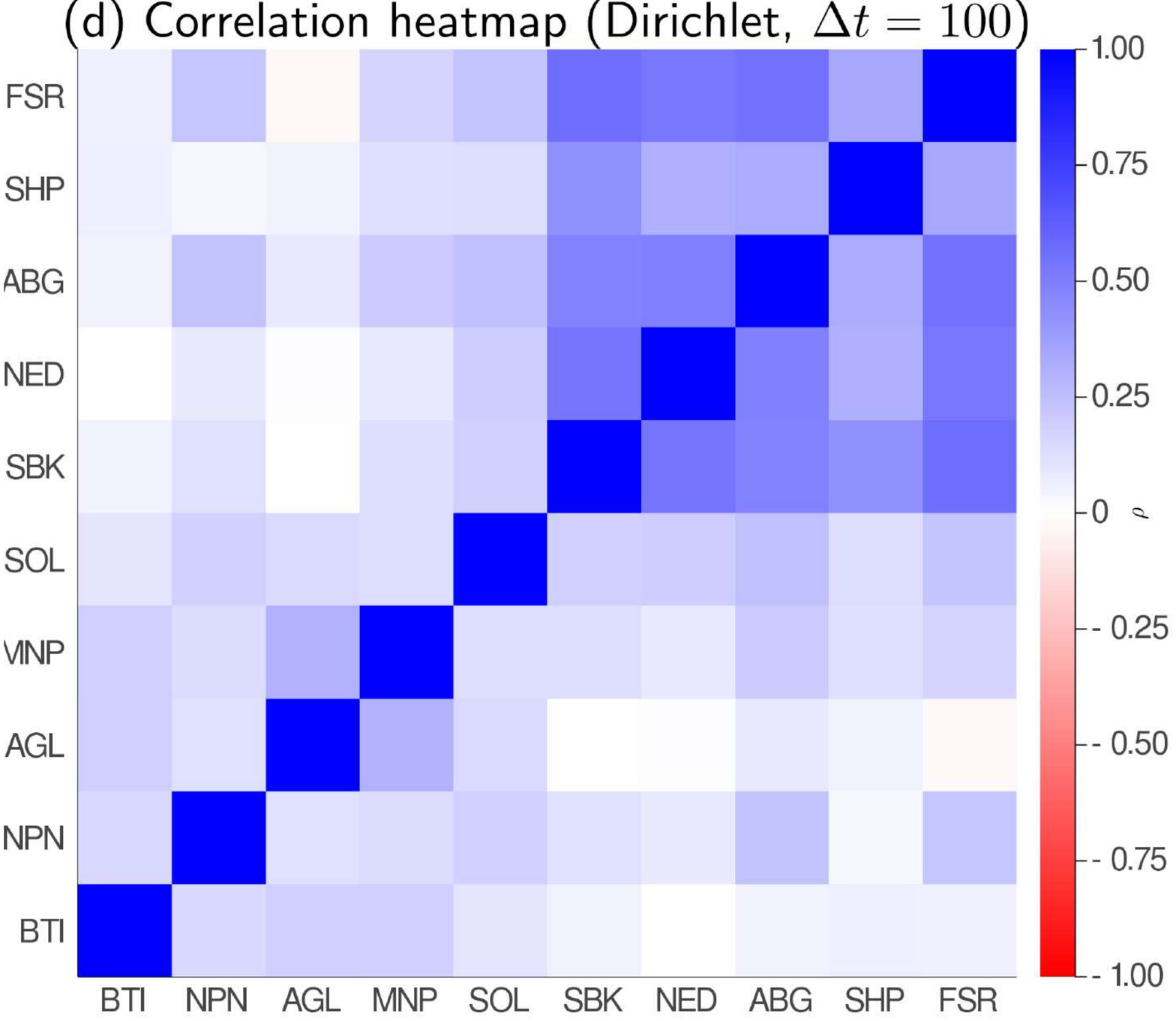}}  \\
    \subfloat{\label{fig:EmpHM:e}\includegraphics[width=0.245\textwidth]{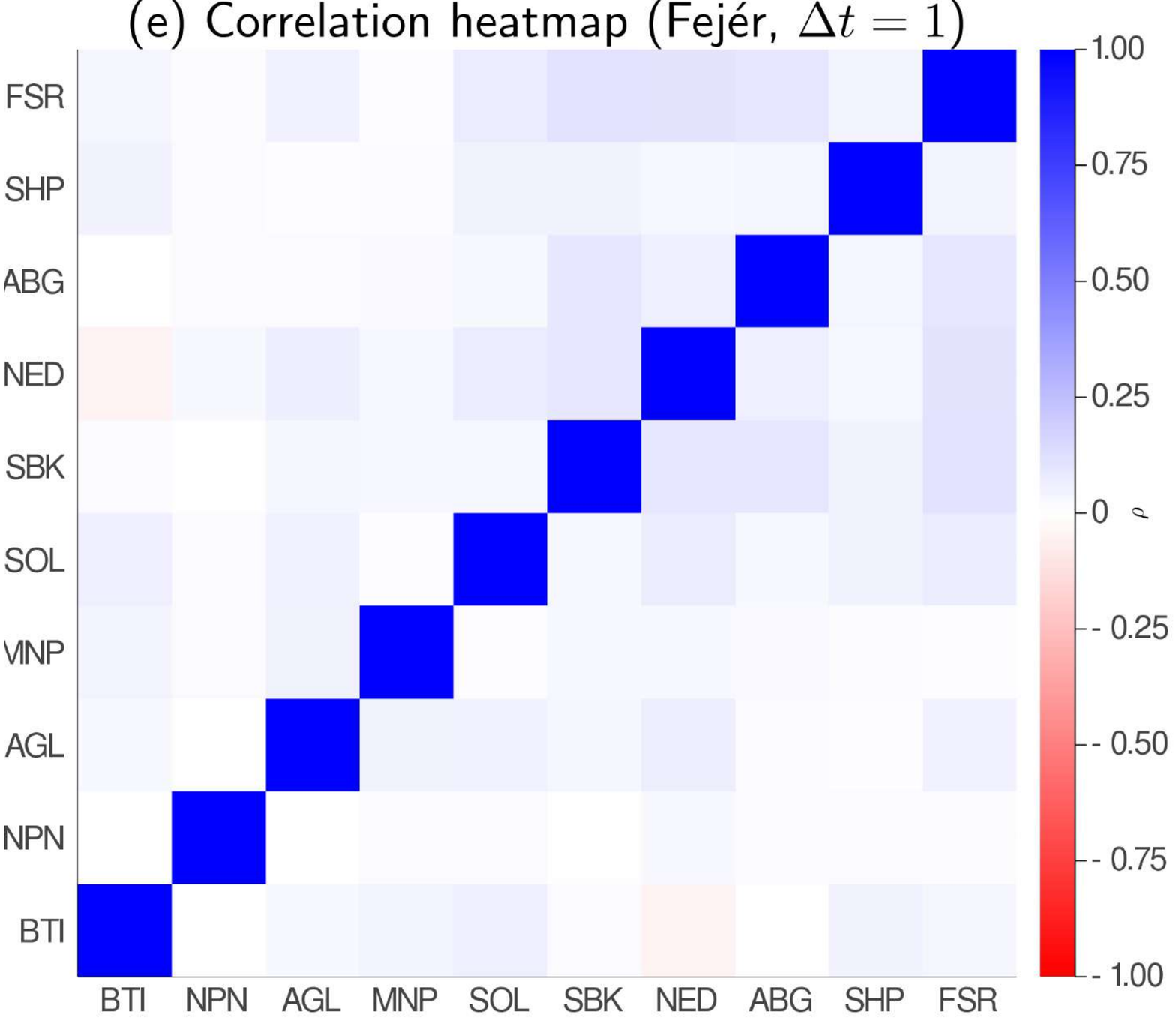}}
    \subfloat{\label{fig:EmpHM:f}\includegraphics[width=0.245\textwidth]{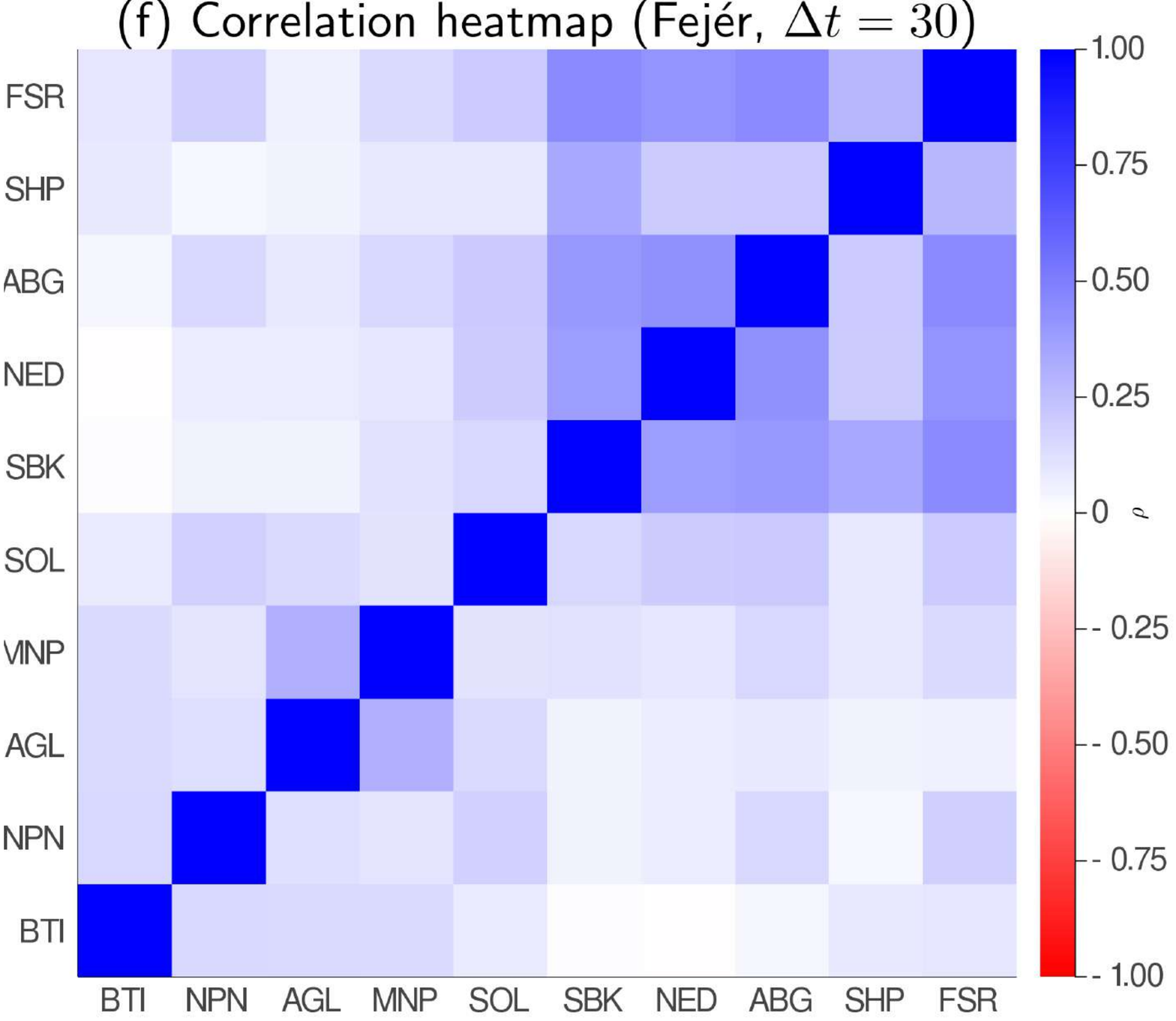}}
    \subfloat{\label{fig:EmpHM:g}\includegraphics[width=0.245\textwidth]{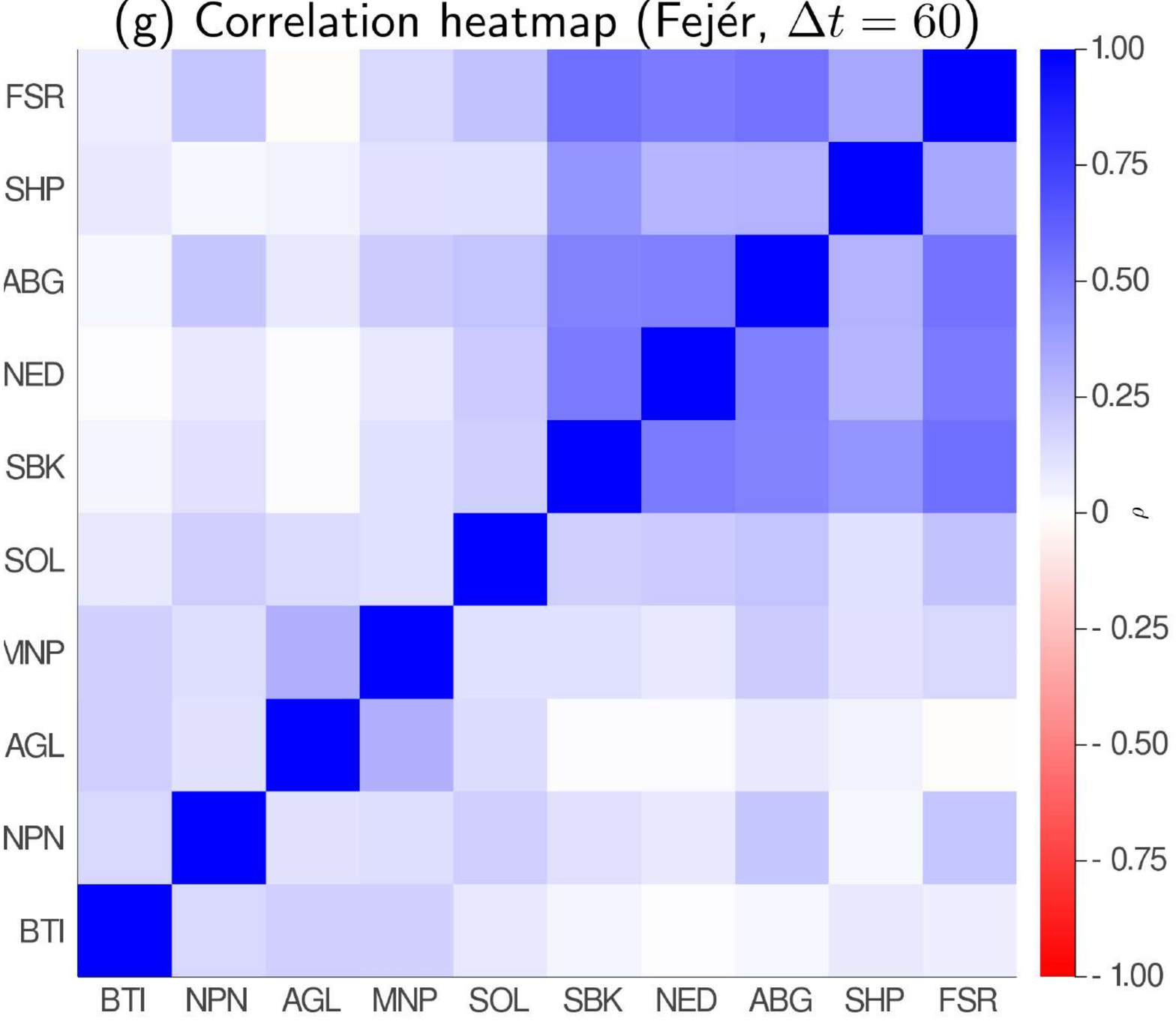}}
    \subfloat{\label{fig:EmpHM:h}\includegraphics[width=0.245\textwidth]{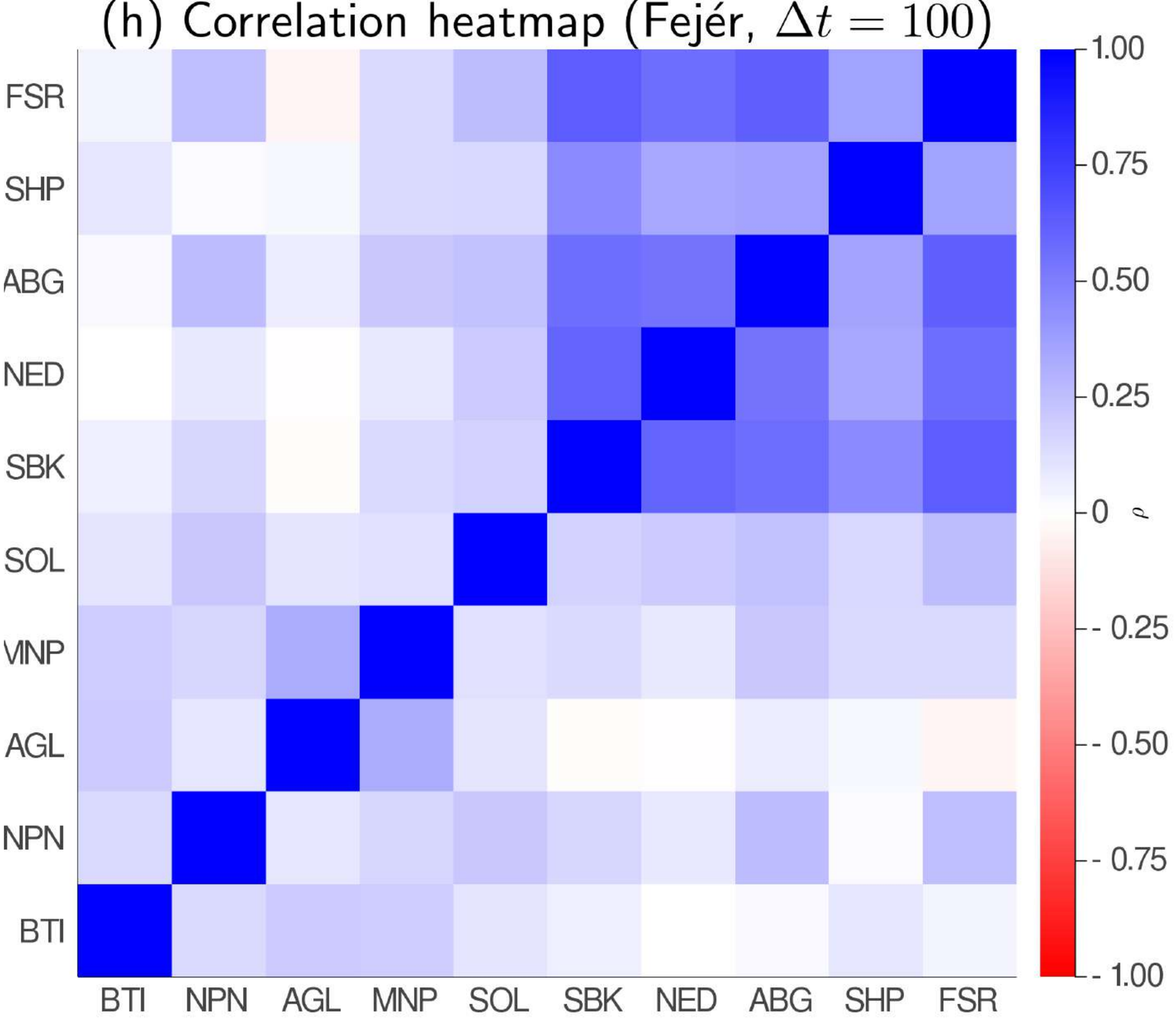}}
\caption{We investigate the Epps effect on the JSE by plotting correlation structure as heat-maps for snapshots from Figure \ref{fig:EmpMMZAll}. The snapshots are taken for $\Delta t = 1, 30, 60$ and $100$ seconds for (a) to (d) and (e) to (h) respectively. (a) to (d) plots the correlation structure using the Dirichlet basis and (e) to (h) the Fej\'{e}r basis. We see that the correlations at short time scales are generally positively correlated - with the top right quadrant being the most positively correlated as they are from the banking sector. More interestingly, we see that the correlation pair FSR/AGL goes from positively correlated to negatively correlated as $\Delta t$ increases. The figures can be recovered using the Julia script file \href{https://github.com/CHNPAT005/PCEPTG-MM-NUFFT/blob/master/Scripts/Time\%20Scales/Empirical}{Empirical} on the GitHub resource \cite{PCEPTG2020CODE}.}
\label{fig:EmpHM}
\end{figure*}

Here we consider the 10 stocks as described in Table \ref{tab:JSE10summary} and provide the correlation heat-maps in Figure \ref{fig:EmpHM} and Epps effect plots in Figure \ref{fig:EmpMMZAll}. 

\newpage

\section{Simulated 3-asset case} \label{app:Ave3Asset}

\begin{figure*}[hbt!]
\centering
    \subfloat{\label{fig:Ave3Asset:a}\includegraphics[width=0.48\textwidth]{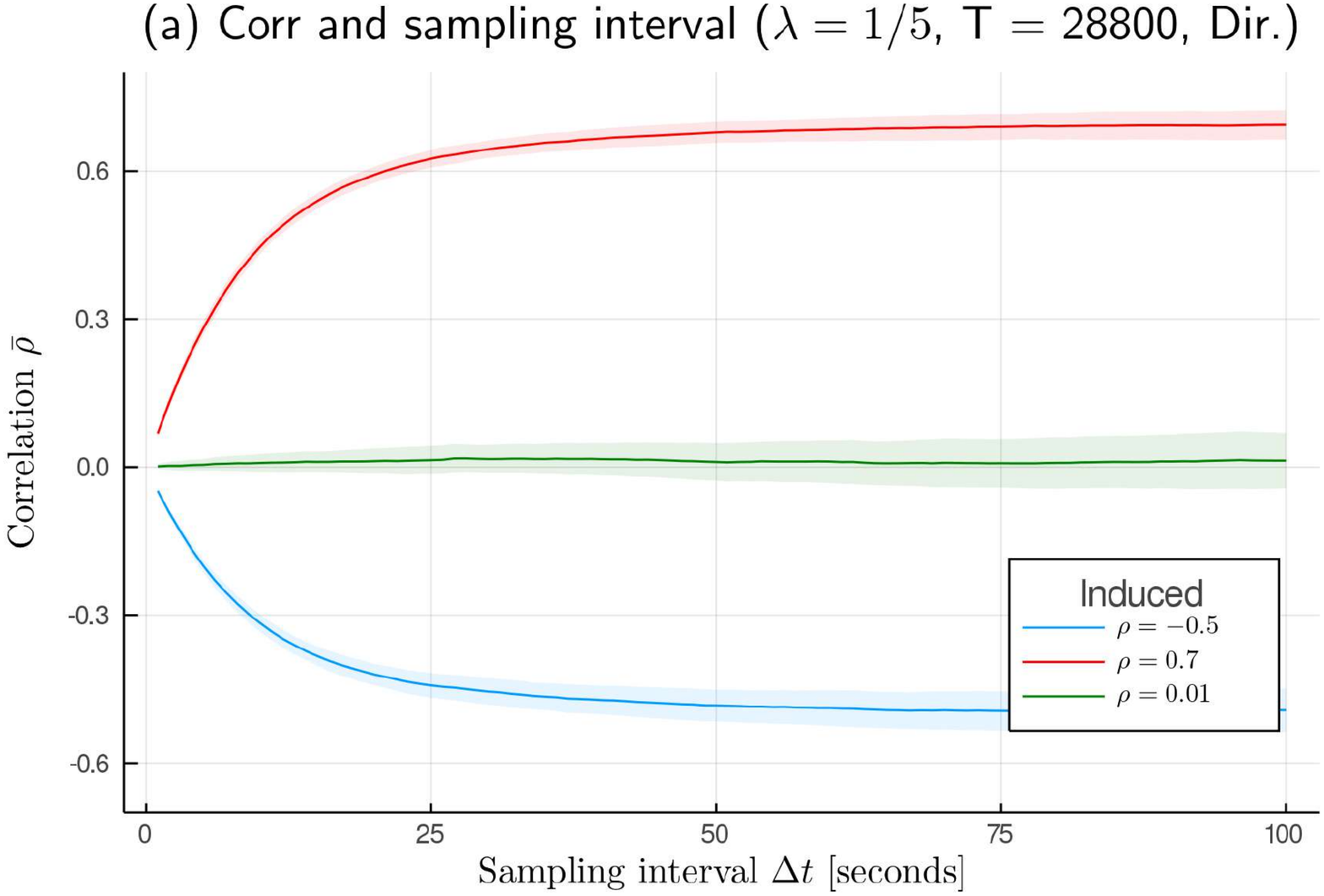}}
    \subfloat{\label{fig:Ave3Asset:b}\includegraphics[width=0.48\textwidth]{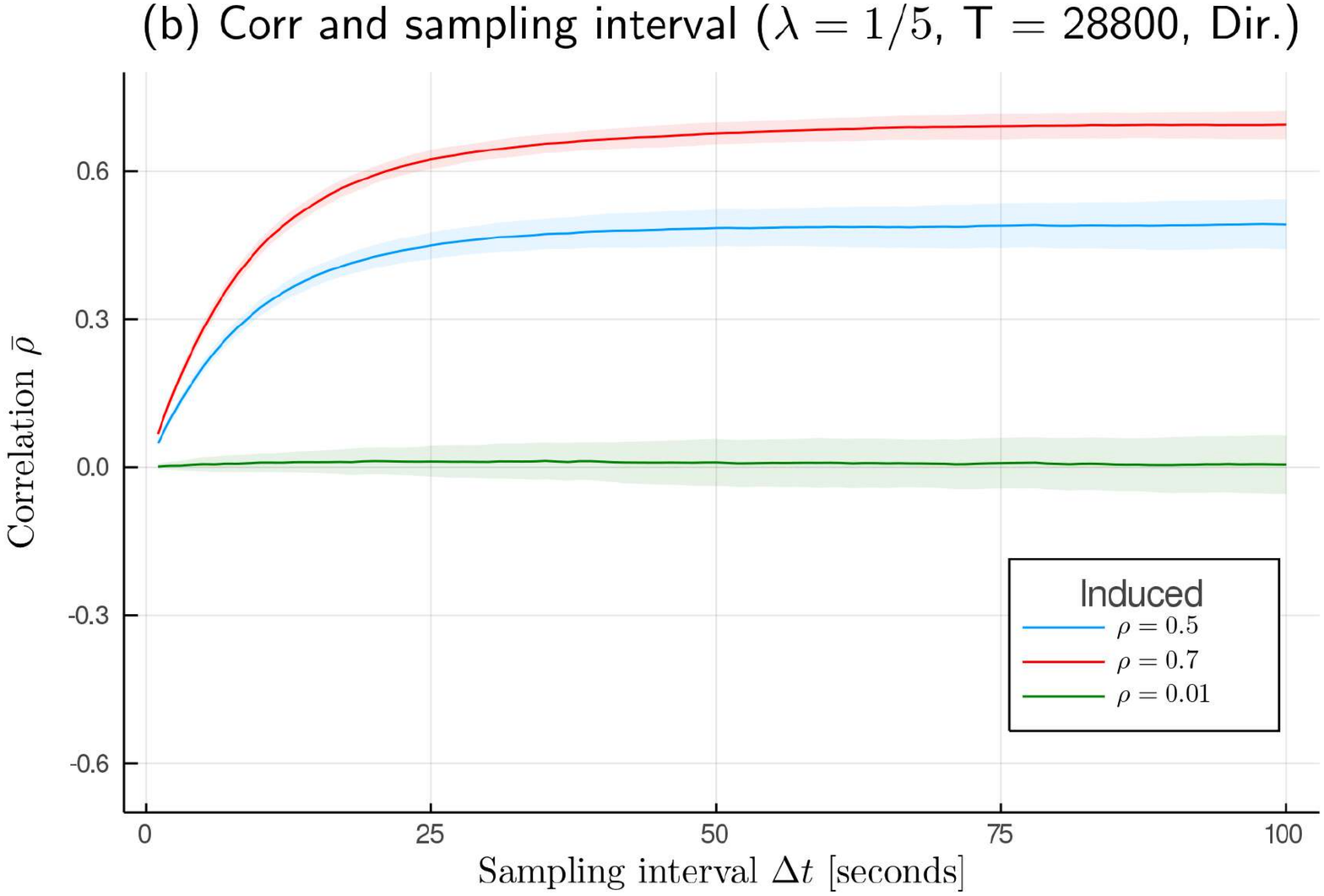}}
\caption{We investigate whether the interplay from correlation combinations or estimation uncertainty can explain the correlation dynamics found in FSR/AGL from Figure \ref{fig:EmpMMZ2}. We simulate $T=28800$ data points for a three feature Geometric Brownian Motion with induced correlation choices $\rho_{12}=-0.5$ (blue line), $\rho_{13}=0.7$ (red line) and $\rho_{23}=0.01$ (green line) for (a) and $\rho_{12}=0.5$ (blue line), $\rho_{13}=0.7$ (red line) and $\rho_{23}=0.01$ (green line) for (b). The synchronous case is then sampled with an exponential inter-arrival process with rate $\lambda = 1/5$. The Dirichlet estimates are obtained for $\Delta t$ ranging from 1 to 100 with the conversion to $N$ given in \cref{eq:Der:22}. This process is repeated 100 times and the average correlation estimate at each $\Delta t$ is then plotted with error bars (computed using a t-distribution with 99 degrees of freedom and the sample standard deviation) representing 68\% of the variability between the estimation paths. We see that for $\rho \approx 0$, when $n$ and $N$ is not large enough, estimation uncertainty arises, explaining the switching of signs; but it does not account for the correlation magnitude dropping as $\Delta t$ increases. The figures can be recovered using the Julia script file \href{https://github.com/CHNPAT005/PCEPTG-MM-NUFFT/blob/master/Scripts/Time\%20Scales/3Asset}{3Asset} on the GitHub resource \cite{PCEPTG2020CODE}.}
\label{fig:Ave3Asset}
\end{figure*}

Here we consider 3 simulated correlated stocks to demonstrate the interplay of the combination of negative and positive correlations in Figure \ref{fig:Ave3Asset}. The change of scale can lead to spurious negative and positive correlations when there is insufficient data. 

\section{Algorithms} \label{app:algo}
Algorithm outline for the various implementation methods.

\begin{table}[H]
\begin{algorithm}[H]
\begin{algorithmic}
\Require {\\ \begin{enumerate} 
\item n: number of price points to simulate.
\item $\boldsymbol{\mu}$: (D x 1) vector of drift parameters.
\item $\Vec{\Sigma}$: (D x D) covariance matrix.
\item start price: (D x 1) vector of $S(0)$.
\end{enumerate}}
\State Procedure for the $i^{th}$ feature:
\begin{enumerate} 
\item Generate: $\Vec{Z} \sim \mathcal{N}_D(\Vec{0}, \Vec{I}_{DxD})$.
\small
\item Set: $S_i(t_{k+1}) = S_i(t_k) \exp\big[(\mu_i - \frac{1}{2} \sigma^2_i)(t_{k+1} - t_k) + \sqrt{t_{k+1} - t_k} \sum_{k=1}^d A_{ik} Z_k \big]$.
\end{enumerate}
\State \Return{($\Vec{S}$)}
\end{algorithmic} \caption{GBM Algorithm} \label{algo:GBM}
\end{algorithm}
\caption{The Geometric Brownian Motion (GBM) Algorithm (see \cref{algo:GBM}) simulates a correlated multivariate GBM using the Euler–Maruyama scheme. It is subject to the initial condition $S(0) =$ start price. $\Vec{A}$ is the Cholesky decomposition of $\Sigma$. The Julia implementation can be found at \href{https://github.com/CHNPAT005/PCEPTG-MM-NUFFT/blob/master/Functions/Monte\%20Carlo\%20Simulation\%20Algorithms/GBM}{GBM} in the GitHub resource \cite{PCEPTG2020CODE} and was provided by \cite{GLASSERMAN2004}.}
\end{table}

\begin{table}
\begin{algorithm}[H]
\begin{algorithmic}
\Require {\\ \begin{enumerate} 
\item $\Vec{\tilde{t}}$: (n x D) of re-scaled sampled times. Non-trade times are represented using \textit{NaNs} or \textit{NAs}.
\end{enumerate}}
\State Set: $\Delta t_0^i$ = minimum distance between sampled times for asset $i$.
\State Set: $N_0^i = \frac{2 \pi}{\Delta t_0^i}$.
\State Set: $N_i = \lfloor \frac{N_0^i}{2} \rfloor$, where $\lfloor x \rfloor$ denotes the floor of $x$.
\State Set: $N = \min_i \{N_i\}$.
\State \Return(N)
\end{algorithmic} \caption{Nyquist Frequency} \label{algo:nyquist}
\end{algorithm}
\caption{The Nyquist frequency Algorithm (see \cref{algo:nyquist}) computes the Nyquist cutoff. The Julia implementation can be found in any of the \href{https://github.com/CHNPAT005/PCEPTG-MM-NUFFT/tree/master/Functions/Correlation\%20Estimators/Dirichlet}{Dirichlet} or \href{https://github.com/CHNPAT005/PCEPTG-MM-NUFFT/tree/master/Functions/Correlation\%20Estimators/Fejer}{Fej\'{e}r} implementations in the GitHub resource \cite{PCEPTG2020CODE} and is an auxiliary function based on the MATLAB implementation from \cite{TGDWCMDH2005, DHTGDW2017}.}
\end{table}

\begin{table}
\begin{algorithm}[H]
\begin{algorithmic}
\Require {\\ \begin{enumerate} 
\item $\Vec{P}$: (n x D) matrix of sampled prices. Non-trade times are represented using \textit{NaNs} or \textit{NAs}.
\item $\Vec{T}$: (n x D) matrix of sampled times. Non-trade times are represented using \textit{NaNs} or \textit{NAs}.
\item N (Optional): cutoff frequency (Integer) used in the convolution. Default is set to be the Nyquist cutoff.
\item tol (Optional): error tolerance for NUFFTs. Determines the number of grid points to spread. Default is set to $10^{-12}$.
\end{enumerate}}
\State Step I. Initialisation.
\begin{itemize}
    \item[I.1.] Re-scale the sampled times ($T$) (see \cref{algo:rescale}).
    \item[I.2.] Compute the Nyquist cutoff ($N$) --- unless specified otherwise through input parameter (see \cref{algo:nyquist}).
\end{itemize}
\State Step F: Compute the Fourier coefficients, $k \in \{-N, ..., N\}$.
\For {$i$ = 1 to D} 
\begin{itemize}
    \item[F.1.] Extract the re-scaled sampled times for the $i^{th}$ object: $\Vec{\tilde{t}}^i = \Vec{T}(i)$, excluding any \textit{NaNs} or \textit{NAs}.
    \item[F.2.] Extract and compute the logarithm of the sampled prices for the $i^{th}$ object: $\Vec{\tilde{p}}_i = \ln\left(\Vec{p}_i\left(\Vec{\tilde{t}}^i\right)\right)$, excluding any \textit{NaNs} or \textit{NAs}.
    \item[F.3.] Compute the returns: $\delta_{i}(I_h) = \tilde{p}_i(\tilde{t}_{h+1}^i) - \tilde{p}_i(\tilde{t}_{h}^i)$ 
    \item[F.4.] Compute the Fourier coefficients: \small $$c^+_k(i) = \sum_{h=1}^{n_i-1} e^{\iu k\tilde{t}_h^i} \delta_{i}(I_h); \quad c^-_k(i) = \sum_{h=1}^{n_i-1} e^{-\iu k\tilde{t}_h^i} \delta_{i}(I_h)$$ \normalsize
\end{itemize}
\EndFor
\State Step C: Convolution.
\begin{itemize}
    \item[C.1.] The Dirichlet implementation: \small $$\hat{\Sigma}_{ij} = \frac{1}{2N+1} \sum_{k = -N}^N [c^+_k(i) c^-_k(j)]$$ \normalsize
    \item[C.2.] The Fej\'{e}r implementation: \small $$\hat{\Sigma}_{ij} = \frac{1}{N+1} \sum_{k = -N}^N \left( 1 - \frac{|k|}{N} \right) [c^+_k(i) c^-_k(j)]$$ \normalsize
\end{itemize}
\State Correlation: $R_{ij} = \frac{\Sigma_{ij}}{\sqrt{\Sigma_{ii}} \sqrt{\Sigma_{jj}}}$
\State \Return{($\Vec{\Sigma}$, $\Vec{R}$)}
\end{algorithmic} \caption{Malliavin-Mancino Estimators} \label{algo:MM}
\end{algorithm}
\caption{The Malliavin-Mancino estimators (see \cref{algo:MM}) computes the Dirichlet or Fej\'{e}r implementation of the Malliavin-Mancino estimator \cite{MM2002,MM2009} using a complex exponential formulation of the Fourier transform. The algorithm is a mere sketch provided by \cite{DHTGDW2017} and is based on their MATLAB implementation \cite{TGDWCMDH2005}.}
\end{table}

\begin{table}
\begin{algorithm}[H]
\begin{algorithmic}
\Require {\\ \begin{enumerate} 
\item $\Vec{T}$: (n x D) matrix of sampled times. Non-trade times are represented using \textit{NaNs} or \textit{NAs}.
\end{enumerate}}
\State Set: {$t_{\text{min}} =$ minimum value of $\Vec{T}$}
\State Set: {$t_{\text{max}} =$ maximum value of $\Vec{T}$}
\For{$i$ = 1 to D}
\For{$h$ = 1 to $n_i$}
\State $$\tilde{t}_{h}^i = \frac{2 \pi (t_h^i - t_{\text{min}})}{t_{\text{max}} - t_{\text{min}}}$$
\EndFor
\EndFor
\State \Return{($\Vec{\tilde{t}}$)}
\end{algorithmic} \caption{Time-rescaling Algorithm} \label{algo:rescale}
\end{algorithm}
\caption{The Time-rescaling Algorithm (see \cref{algo:rescale}) re-scales the trading times from $[0, T]$ to $[0, 2 \pi]$. The Julia implementation can be found in any of the \href{https://github.com/CHNPAT005/PCEPTG-MM-NUFFT/tree/master/Functions/Correlation\%20Estimators/Dirichlet}{Dirichlet} or \href{https://github.com/CHNPAT005/PCEPTG-MM-NUFFT/tree/master/Functions/Correlation\%20Estimators/Fejer}{Fej\'{e}r} implementations in the GitHub resource \cite{PCEPTG2020CODE} and is an auxiliary function based on the MATLAB implementation from \cite{TGDWCMDH2005, DHTGDW2017}.}
\end{table}

\begin{table}
\begin{algorithm}[H]
\begin{algorithmic}
\Require {\\ \begin{enumerate} 
\item $\boldsymbol{\delta}_i = (\delta_{i}(I_h))_{h=1}^{n_i-1}$: vector of source strengths for asset $i$.
\item $\Vec{\tilde{t}}^i = (\tilde{t}^i_h)_{h=1}^{n_i-1}$: vector of re-scaled sample times for asset $i$.
\item N: the cutoff frequency.
\end{enumerate}}
\For{$s$ =1 to $2N + 1$}
\State $k = s-N-1$
\State \small $$c^+_s = \sum_{h=1}^{n_i-1} e^{\iu k \tilde{t}^i_h} \delta_{i}(I_h)$$ 
\State \small $$c^-_s = \sum_{h=1}^{n_i-1} e^{-\iu k \tilde{t}^i_h} \delta_{i}(I_h)$$ 
\normalsize
\EndFor
\State \Return($c^+_k, c^-_k$)
\end{algorithmic} \caption{``for-loop'' implementation} \label{algo:MS}
\end{algorithm}
\caption{The for-loop implementation \cite{MRS2017} (see \cref{algo:MS}) computes the Fourier coefficients using for-loops. The Julia implementation can be found in  \href{https://github.com/CHNPAT005/PCEPTG-MM-NUFFT/blob/master/Functions/Correlation\%20Estimators/Dirichlet/MScorrDK}{MScorrDK} or \href{https://github.com/CHNPAT005/PCEPTG-MM-NUFFT/blob/master/Functions/Correlation\%20Estimators/Fejer/MScorrFK}{MScorrFK} on the GitHub resource \cite{PCEPTG2020CODE} and correspond to the Dirichlet and Fej\'{e}r representation respectively. The implementation is based on the MATLAB implementation from \cite{MRS2017}.}
\end{table}
\begin{table}
\begin{algorithm}[H]
\begin{algorithmic}
\Require {\\ \begin{enumerate}
\item $\boldsymbol{\delta}_i = (\delta_{i}(I_h))_{h=1}^{n_i-1}$: vector of source strengths for asset $i$.
\item $\Vec{\tilde{t}}^i = (\tilde{t}^i_h)_{h=1}^{n_i-1}$: vector of re-scaled sample times for asset $i$.
\item N: the cutoff frequency.
\end{enumerate}}
\State Set: $\Vec{k} = (1,2,\ldots,N)^{_T}$ a column vector 1 to N.
\State Compute: $\Vec{c}_{1:N} = \boldsymbol{\delta}_i^\intercal \exp(-\iu \ \Vec{\tilde{t}}^i \ \Vec{k}^\intercal)$
\State Compute: $c_0 = \sum_{h=1}^{n_i-1} \delta_{i}(I_h)$
\State Piece $\Vec{c}_{1:N}$, $\overline{\Vec{c}_{1:N}}$ and $c_0$ together to obtain $c^+_k$ and $c^-_k$
\State \Return($c^+_k, c^-_k$)
\end{algorithmic} \caption{Vectorised code implementation} \label{algo:legacy}
\end{algorithm}
\caption{The vectorised code implementation \cite{DHTGDW2017, TGDWCMDH2005} (see \cref{algo:legacy}) replaces for-loops to vectorise the computation of the Fourier coefficients and exploits the Hermitian symmetry of the real source strengths. The Julia implementation can be found in \href{https://github.com/CHNPAT005/PCEPTG-MM-NUFFT/blob/master/Functions/Correlation\%20Estimators/Dirichlet/CFTcorrDK}{CFTcorrDK} or \href{https://github.com/CHNPAT005/PCEPTG-MM-NUFFT/blob/master/Functions/Correlation\%20Estimators/Fejer/CFTcorrFK}{CFTcorrFK} on the GitHub resource \cite{PCEPTG2020CODE} and correspond to the Dirichlet and Fej\'{e}r representation respectively.}
\end{table}
\begin{table}
\begin{algorithm}[H]
\begin{algorithmic}
\Require {\\ \begin{enumerate}
\item $\boldsymbol{\delta}_i = (\delta_{i}(I_h))_{h=1}^{n_i-1}$: vector of source strengths for asset $i$.
\item $\Vec{\tilde{t}}^i = (\tilde{t}^i_h)_{h=1}^{n_i-1}$: vector of re-scaled sample times for asset $i$ ($\tilde{t}^i_h \in [0, 2 \pi]$).

\item $N^* = \lfloor \frac{2 \pi}{\Delta t} \rceil$, where $\lfloor x \rceil$ denotes rounding $x$ to the nearest Integer and $\Delta t$ is the minimum distance between sampled times from \cref{algo:nyquist}.
\end{enumerate}}
\State Initialise: $(\tilde{f}_{\ell})_{\ell=1}^{N^*} = \Vec{0}$, a zero vector of length $N^*$.
\For{$h$ = 1 to $n_i - 1$}
\State $\ell = \lfloor \frac{\tilde{t}^i_h N^*}{2 \pi} \rceil + 1$
\State $\tilde{f}_{\ell} = \delta_{i}(I_h)$
\EndFor
\State \Return($\tilde{f}_{\ell}$ for FFT computation)
\end{algorithmic} \caption{Zero-padded FFT implementation} \label{algo:ZPFFT}
\end{algorithm}
\caption{The zero-padded FFT implementation (see \cref{algo:ZPFFT}) creates a uniform grid and shifts the non-uniform source points to the nearest grid point on an up-sampled uniform grid. The Julia implementation can be found in  \href{https://github.com/CHNPAT005/PCEPTG-MM-NUFFT/blob/master/Functions/Correlation\%20Estimators/Dirichlet/FFTZPcorrDK}{FFTZPcorrDK} or \href{https://github.com/CHNPAT005/PCEPTG-MM-NUFFT/blob/master/Functions/Correlation\%20Estimators/Fejer/FFTZPcorrFK}{FFTZPcorrFK} on the GitHub resource \cite{PCEPTG2020CODE} and correspond to the Dirichlet and Fej\'{e}r representation respectively. Note that the index $\ell$ is set for languages with array indices starting from 1.}
\end{table}
\begin{table}[H]
\begin{algorithm}[H]
\begin{algorithmic}
\Require {\\ \begin{enumerate}
\item $\boldsymbol{\delta}_i = (\delta_{i}(I_h))_{h=1}^{n_i-1}$: vector of source strengths for asset $i$.
\item $\Vec{\tilde{t}}^i = (\tilde{t}^i_h)_{h=1}^{n_i-1}$: vector of re-scaled sample times for asset $i$ ($\tilde{t}^i_h \in [0, 2 \pi]$).
\item $M = 2N + 1$: the number of Fourier modes computed.
\item $\epsilon$: error tolerance. 
\end{enumerate}}
\State Step I. Initialisation:
\begin{itemize}
    \item[I.1.] Set: $\sigma=2$.
    \item[I.2.] Set: $M_r = \sigma M$; $M_{sp} = \lfloor \frac{-\ln(\epsilon) (\sigma-1/2)}{(\pi (\sigma-1))} + \frac{1}{2} \rfloor$.
    \item[I.3.] Set: $\lambda = \frac{\sigma^2 M_{sp}}{\sigma(\sigma-0.5)}$; $h_x = \frac{2 \pi}{M_r}$; $t_1 = \frac{\pi}{\lambda}$.
    \item[I.4.] Set: $\tau = \frac{\pi \lambda}{M_r^2}$.
    \item[I.5.] Initialise: $(\tilde{f}_{\ell})_{\ell=1}^{M_r} = \Vec{0}$, a zero vector of length $M_r$.
\end{itemize}
\For{$k$ = 1 to $M_{sp}$}
\State $E_{3, k} = \exp(-t_1 k^2)$
\EndFor
\State Step C: Convolution (see \cref{eq:Der:15}).
\For{$h$ = 1 to $n_i - 1$}
\State $b_0 = \lfloor \frac{\tilde{t}^i_h}{h_x} \rfloor$, index of nearest up-sampled grid $\xi_{b_0} \leq \tilde{t}^i_h$.
\State d = $\frac{\tilde{t}^i_h}{h_x} - b_0$.
\State $E_0 = \Vec{0}$, a zero vector of length $2M_{sp}$.
\State $E_1 = e^{-t_1 \text{d}^2}$; $E_{0, M_{sp}} = E_1$; $E_2 = e^{2t_1 \text{d}}$.
\For{$k$ = 1 to $M_{sp}$}
\State $E_{0, M_{sp}+k} = E_{3, k} E_1 E_2^k$.
\EndFor
\For{$k$ = 1 to $M_{sp}-1$}
\State $E_{0, M_{sp}-k} = E_{3, k} E_1 E_2^{-k}$.
\EndFor
\State $b_d = \min(M_{sp}-1, b_0)$; $b_u = \min(M_{sp}, M_r - b_0 - 1)$.
\For{$k$ = $-M_{sp}+1$ to $-b_d-1$}
\State $\tilde{f}_{b_0 + k + M_r + 1} = \tilde{f}_{b_0 + k + M_r + 1} + \delta_{i}(I_h) E_{0, M_{sp}+k}$.
\EndFor
\For{$k$ = $-b_d$ to $b_u$}
\State $\tilde{f}_{b_0 + k + 1} = \tilde{f}_{b_0 + k + 1} + \delta_{i}(I_h) E_{0, M_{sp}+k}$.
\EndFor
\For{$k$ = $b_u+1$ to $M_{sp}$}
\State $\tilde{f}_{b_0 + k - M_r + 1} = \tilde{f}_{b_0 + k - M_r + 1} + \delta_{i}(I_h) E_{0, M_{sp}+k}$.
\EndFor
\EndFor
\State Step F: Compute FFT on over-sampled grid (see \cref{eq:Der:16}).
\begin{itemize}
    \item[F.1.] Find Fourier coefficients $F_G(dp_i)(k)$ via FFT on the grid $\tilde{f}_{\ell}$.
\end{itemize}
\State Step D: Deconvolution (see \cref{eq:Der:17}).
\begin{itemize}
    \item[D.1.] Compute: $F(dp_i)(k) = \sqrt{\frac{\pi}{\tau}} e^{k^2 \tau} F_G(dp_i)(k) \frac{1}{M_r}$.
\end{itemize}
\State \Return($F(dp_i)(k), k \in \{-N, ..., N\}$)
\end{algorithmic} \caption{Fast Gaussian Gridding NUFFT implementation} \label{algo:FGG}
\end{algorithm}
\caption{The non-uniform FFT implementation (see \cref{algo:FGG}) creates an up-sampled uniform grid and convolves the non-uniform source points onto the uniform grid, applies the FFT on the up-sampled grid and deconvolves the convolution effects in the Fourier space. \Cref{algo:FGG} is specific for the Gaussian kernel as it uses the fast Gaussian gridding implementation by \cite{GL2004} to reduce the number of exponential evaluations. The Julia implementation can be found in  \href{https://github.com/CHNPAT005/PCEPTG-MM-NUFFT/blob/master/Functions/NUFFT/NUFFT-FGG}{NUFFT-FGG} on the GitHub resource \cite{PCEPTG2020CODE}. The Algorithm is a replication of the \href{https://cims.nyu.edu/cmcl/nufft/nufft.html}{FORTRAN} source code from \cite{GL2004}, but adjusted for languages with array indices starting from 1.}
\end{table}
\begin{table}[H]
\begin{algorithm}[H]
\begin{algorithmic}
\Require {\\ \begin{enumerate} 
\item $\boldsymbol{\delta}_i = (\delta_{i}(I_h))_{h=1}^{n_i-1}$: vector of source strengths for asset $i$.
\item $\Vec{\tilde{t}}^i = (\tilde{t}^i_h)_{h=1}^{n_i-1}$: vector of re-scaled sample times for asset $i$ ($\tilde{t}^i_h \in [0, 1]$).
\item $M = 2N + 1$: the number of Fourier modes computed.
\item $\epsilon$: error tolerance. 
\end{enumerate}}
\State Step I. Initialisation:
\begin{itemize}
    \item[I.1.] Set: $\sigma=2$.
    \item[I.2.] Set: $M_r = \sigma M$; $M_{sp} =  \lfloor \frac{1}{2} (\lceil \log_{10}(\frac{1}{\epsilon}) \rceil + 2) \rfloor$.
    \item[I.3.] Initialise: $(\tilde{f}_{\ell})_{\ell=1}^{M_r} = \Vec{0}$, a zero vector of length $M_r$.
\end{itemize}
\State Step C: Convolution (see \cref{eq:Der:15}).
\For{$h$ = 1 to $n_i - 1$}
\State $b_0 = \lfloor \tilde{t}^i_h M_r \rfloor$, index of nearest up-sampled grid $\xi_{b_0} \leq \tilde{t}^i_h$.
\State d = $\tilde{t}^i_h - \frac{b_0}{M_r}$.
\State $b_d = \min(M_{sp}-1, b_0)$; $b_u = \min(M_{sp}, M_r - b_0 - 1)$.
\For{$k$ = $-M_{sp}$ to $-b_d-1$}
\State $\tilde{f}_{b_0 + k + M_r + 1} = \tilde{f}_{b_0 + k + M_r + 1} + \delta_{i}(I_h) \varphi_{_{KB}}(d - \frac{k}{M_r})$.
\EndFor
\For{$k$ = $-b_d$ to $b_u$}
\State $\tilde{f}_{b_0 + k + 1} = \tilde{f}_{b_0 + k + 1} + \delta_{i}(I_h) \varphi_{_{KB}}(d - \frac{k}{M_r})$.
\EndFor
\For{$k$ = $b_u+1$ to $M_{sp}$}
\State $\tilde{f}_{b_0 + k - M_r + 1} = \tilde{f}_{b_0 + k - M_r + 1} + \delta_{i}(I_h) \varphi_{_{KB}}(d - \frac{k}{M_r})$.
\EndFor
\EndFor
\State Step F: Compute FFT on over-sampled grid (see \cref{eq:Der:16}).
\begin{itemize}
    \item[F.1.] Find Fourier coefficients $F_{{KB}}(dp_i)(k)$ via FFT on the grid $\tilde{f}_{\ell}$.
\end{itemize}
\State Step D: Deconvolution (see \cref{eq:Der:17}).
\begin{itemize}
    \item[D.1.] Compute: $F(dp_i)(k) = \frac{1}{M_r} F_{{KB}}(dp_i)(k) / \hat{\varphi}_{_{KB}}(k)$.
\end{itemize}
\State \Return($F(dp_i)(k), k \in \{-N, ..., N\}$)
\end{algorithmic} \caption{Kaiser-Bessel NUFFT implementation} \label{algo:KB}
\end{algorithm}
\caption{The non-uniform FFT implementation (see \cref{algo:KB}) creates an up-sampled uniform grid and convolves the non-uniform source points onto the uniform grid, applies the FFT on the up-sampled grid and deconvolves the convolution effects in the Fourier space. \Cref{algo:KB} uses the Kaiser-Bessel kernel here, but the algorithm can be applied to any kernel that is 1-periodic. The algorithm is structured without the pre-computation step used in \cite{PS2003} and evaluates the algorithm ``on-the-fly''. The Julia implementation can be found in  \href{https://github.com/CHNPAT005/PCEPTG-MM-NUFFT/blob/master/Functions/NUFFT/NUFFT-KB}{NUFFT-KB} on the GitHub resource \cite{PCEPTG2020CODE}. The algorithm is set for languages with array indices starting from 1.}
\end{table}
\begin{table}[H]
\begin{algorithm}[H]
\begin{algorithmic}
\Require {\\ \begin{enumerate}
\item $\boldsymbol{\delta}_i = (\delta_{i}(I_h))_{h=1}^{n_i-1}$: vector of source strengths for asset $i$.
\item $\Vec{\tilde{t}}^i = (\tilde{t}^i_h)_{h=1}^{n_i-1}$: vector of re-scaled sample times for asset $i$ ($\tilde{t}^i_h \in [0, 2 \pi]$).
\item $M = 2N + 1$: the number of Fourier modes computed.
\item $\epsilon$: error tolerance. 
\end{enumerate}}
\State Step I. Initialisation:
\begin{itemize}
    \item[I.1.] Set: $\sigma=2$.
    \item[I.2.] Set: $M_r = \sigma M$; $M_{sp} =  \lfloor \frac{1}{2} (\lceil \log_{10}(\frac{1}{\epsilon}) \rceil + 2) \rfloor + 2$.
    \item[I.3.] Initialise: $(\tilde{f}_{\ell})_{\ell=1}^{M_r} = \Vec{0}$, a zero vector of length $M_r$.
\end{itemize}
\State Step C: Convolution (see \cref{eq:Der:15}).
\For{$h$ = 1 to $n_i - 1$}
\State $b_0 = \lfloor \tilde{t}^i_h M_r \rfloor$, index of nearest up-sampled grid $\xi_{b_0} \leq \tilde{t}^i_h$.
\State d = $\tilde{t}^i_h - \frac{b_0}{M_r}$.
\State $b_d = \min(M_{sp}-1, b_0)$; $b_u = \min(M_{sp}, M_r - b_0 - 1)$.
\For{$k$ = $-M_{sp}$ to $-b_d-1$}
\State $\tilde{f}_{b_0 + k + M_r + 1} = \tilde{f}_{b_0 + k + M_r + 1} + \delta_{i}(I_h) \varphi_{_{ES}}(d - \frac{2 \pi k}{M_r})$.
\EndFor
\For{$k$ = $-b_d$ to $b_u$}
\State $\tilde{f}_{b_0 + k + 1} = \tilde{f}_{b_0 + k + 1} + \delta_{i}(I_h) \varphi_{_{ES}}(d - \frac{2 \pi k}{M_r})$.
\EndFor
\For{$k$ = $b_u+1$ to $M_{sp}$}
\State $\tilde{f}_{b_0 + k - M_r + 1} = \tilde{f}_{b_0 + k - M_r + 1} + \delta_{i}(I_h) \varphi_{_{ES}}(d - \frac{2 \pi k}{M_r})$.
\EndFor
\EndFor
\State Step F: Compute FFT on over-sampled grid (see \cref{eq:Der:16}).
\begin{itemize}
    \item[F.1.] Find Fourier coefficients $F_{ES}(dp_i)(k)$ via FFT on the grid $\tilde{f}_{\ell}$.
\end{itemize}
\State Step D: Deconvolution (see \cref{eq:Der:17}).
\begin{itemize}
    \item[D.1.] Compute: $\hat{\varphi}_{_{ES}}(k) = \alpha \int_{-\infty}^{\infty} {\phi}_{_{ES}}(x) e^{\iu \alpha k x} dx$ using numerical integration.
    \item[D.2.] Compute: $F(dp_i)(k) = \frac{2 \pi}{M_r} F_{ES}(dp_i)(k) / \hat{\varphi}_{_{ES}}(k)$.
\end{itemize}
\State \Return($F(dp_i)(k), k \in \{-N, ..., N\}$)
\end{algorithmic} \caption{Exponential of semi-circle NUFFT implementation} \label{algo:ES}
\end{algorithm}
\caption{The non-uniform FFT implementation (see \cref{algo:ES}) creates an up-sampled uniform grid and convolves the non-uniform source points onto the uniform grid, applies the FFT on the up-sampled grid and deconvolves the convolution effects in the Fourier space. \Cref{algo:ES} uses the exponential of semi-circle kernel. The algorithm is a naive implementation based on the steps provided in \cite{BMK2018} and does not exploit the piecewise polynomial kernel approximation nor the Gauss-Legendre quadrature for implementation acceleration used in FINUFFT \cite{BMK2018}. The implementation relies on the \href{https://github.com/JuliaMath/QuadGK.jl}{QuadGK} package to perform the numerical integration using adaptive Gauss-Kronrod quadrature. The Julia implementation can be found in  \href{https://github.com/CHNPAT005/PCEPTG-MM-NUFFT/blob/master/Functions/NUFFT/NUFFT-ES}{NUFFT-ES} on the GitHub resource \cite{PCEPTG2020CODE}. The Algorithm is set for languages with array indices starting from 1.}
\end{table}

\end{document}